%% file: main.tex
\documentclass[a4paper,11pt]{scrartcl}

\usepackage{fullpage}

\usepackage[latin1]{inputenc}
\usepackage[T1]{fontenc}
\usepackage{scrtime}

\usepackage{url}
\usepackage{makeidx}

\usepackage{enumerate}
\usepackage[all]{xy}
\usepackage{array}
\usepackage{hhline}

\usepackage{amsmath}
\usepackage{amsfonts}
\usepackage{amsthm}
\usepackage{amssymb}
\usepackage{stmaryrd}
\usepackage{latexsym}
\usepackage{euler}

\usepackage{bm} % Bold mathematical symbols
\usepackage{ifpdf}
\usepackage[%
  draft,
  appendix,
  short,
]{optional}

\usepackage[pagebackref]{hyperref}
  
\ifpdf
  \hypersetup{
    pdftex,
    colorlinks=true,
    urlcolor=blue,        % \href{...}{...} external (URL)
    filecolor=blue,       % \href{...} local file
    linkcolor=blue,       % \ref{...} and \pageref{...}
    citecolor=blue,       % \cite{...}
    anchorcolor=blue
  }
\fi

\usepackage{tabularx}
  \newcolumntype{L}{>{\raggedright\arraybackslash}X}
  \newcolumntype{C}{>{\centering\arraybackslash}X}
  \newcolumntype{R}{>{\raggedleft\arraybackslash}X}

\input{Macros/main}

\input{figures}

\title{On the confluence of lambda-calculus\\ with conditional rewriting}

\author{
Frédéric Blanqui (INRIA)
\footnote{FIT 3-604, Tsinghua University,
Haidian District, Beijing 100084, China}
\and Claude Kirchner (INRIA)
\footnote{INRIA Bordeaux - Sud-Ouest,
B{\^a}t. A29, 351 cours de la Lib{\'e}ration, 33405 Talence, France}
\and Colin Riba (LIP - ENS Lyon)
\footnote{UMR 5668 CNRS ENS-Lyon UCBL INRIA,
  46 all{\'e}e d'Italie, 69364 Lyon Cedex 7, France}
}

\date{}

\begin{document}
%%%%%%%%%%%%%%%%%%%%%%%%%%%%%%%%%%%%%%%%%%%%%%%%%%%%%%%%%%%%%%%%%%%%%%%%%%%

\maketitle

\input{abstract}

\newpage\tableofcontents\newpage

\input{intro}

\input{calc}

\input{ex}

\input{confl}

\input{over}

\input{Aconfl}

\input{Bconfl}

\input{orth}

\input{conclu}

\bibliographystyle{smfalpha}
\bibliography{bibliographie}

\appendix{
}

\end{document}

%% file: Macros/main.tex
\input{Macros/general}

\input{Macros/theoremes}

\input{Macros/alphabetiques}

\input{Macros/notations}

\input{Macros/symbols}

%% file: Macros/general.tex
\newcommand{\comment}[1]{} % Comments
\newcommand{\xyS}{20pt}   % Standard
\newcommand{\xyTS}{10pt}  % Third Standard
\newcommand{\xyL}{50pt}   % Large

\newcommand{\imp}{\Longrightarrow}
\newcommand{\tq}{~|~}
\newcommand{\gs}{\quad|\quad}

\newcommand{\esp}{{\;}}
\newcommand{\dbesp}{\esp\esp}
\newcommand{\ptesp}{{\,}}
\newcommand{\mbin}[1]{\mathbin{#1}}

\newcommand{\deq}{\mathbin{=_{\text{def}}}}

\newcommand{\dom}{\mathcal{D}om}

\newcommand{\N}{\mathbb{N}}

\newcommand{\mul}{\text{mul}}
\renewcommand{\o}[1]{{\overline{#1}}}

%% file: Macros/theoremes.tex
\newcommand{\abovethm}{\topsep}
\newcommand{\belowthm}{\topsep}

\newtheoremstyle{myplain}
  {\abovethm}{\belowthm}  % Espacement dessus et dessous
  {\itshape}
  {}          % Indentation (standard: pas d'indentation)
  {\bfseries} % Texte du titre
  {}
  {.5em}      % head-after-space
  {\thmname{#1} \thmnumber{#2} \thmnote{\textup{(#3)}}}

\newtheoremstyle{myremark}
  {\abovethm}{\belowthm}
  {\itshape}
  {}{\bfseries}
  {}
  {.5em}
  {\thmname{#1} \thmnumber{#2} \thmnote{\textup{(#3)}}}

\theoremstyle{myplain}
  \newtheorem{theorem}{Theorem}[section]
  \newtheorem{lemma}[theorem]{Lemma}
  \newtheorem{proposition}[theorem]{Proposition}
  \newtheorem{definition}[theorem]{Definition}

\theoremstyle{myremark}
  \newtheorem{remark}[theorem]{Remark}
  \newtheorem{example}[theorem]{Example}

\theoremstyle{myplain}
  \newtheorem{theorem*}{Theorem}
  \newtheorem{lemma*}{Lemme}
  \newtheorem{proposition*}{Proposition}
  \newtheorem{definition*}{Definition}
  \newtheorem{corollary*}{Corollary}
\theoremstyle{myremark}
  \newtheorem{remark*}{Remark}
  \newtheorem{example*}{Example}

%% file: Macros/alphabetiques.tex
\newcommand{\sle}{\subseteq}

\newcommand{\sgt}{\supset}

\renewcommand{\a}{\rightarrow}

\newcommand{\al}{\leftarrow}
\newcommand{\ad}{\downarrow}

\newcommand{\alr}{\leftrightarrow}

\newcommand{\at}{\mapsto}

\renewcommand{\b}{\beta}

\newcommand{\e}{\eta}
\renewcommand{\t}{\theta}

\newcommand{\la}{\lambda}

\renewcommand{\r}{\rho}
\newcommand{\s}{\sigma}
\newcommand{\sfa}{\mathsf{a}}
\renewcommand{\sfb}{\mathsf{b}}
\newcommand{\sfc}{\mathsf{c}}
\newcommand{\sfd}{\mathsf{d}}

\newcommand{\sff}{\mathsf{f}}
\newcommand{\sfg}{\mathsf{g}}
\newcommand{\sfh}{\mathsf{h}}

\newcommand{\va}{{\vec{a}}}
\newcommand{\vb}{{\vec{b}}}
\newcommand{\vc}{{\vec{c}}}
\newcommand{\vd}{{\vec{d}}}

\newcommand{\vl}{{\vec{l}}}

\newcommand{\vt}{{\vec{t}}}
\newcommand{\vu}{{\vec{u}}}
\newcommand{\vv}{{\vec{v}}}
\newcommand{\vw}{{\vec{w}}}
\newcommand{\vx}{{\vec{x}}}

\comment{ % Useless
} % Useless

%% file: Macros/notations.tex
\newcommand{\Te}{\Lambda}       % Lambda-Terms
\newcommand{\Vte}{\mathcal{X}}  % Term variables
\newcommand{\FV}{{FV}}

\newcommand{\wth}[2]{[#2/#1]}
\newcommand{\wthdots}[4]{[#2/#1,\dots,#4/#3]}

\newcommand{\ap}{{a}}             % Arité applicative
\newcommand{\AN}{{\mathcal{AN}_{\!\ap}}}
\newcommand{\Occ}{\mathcal{O}cc}

\newcommand{\bnf}{\mathop{\beta\text{nf}}}
\newcommand{\re}{{=}}

\newcommand{\R}{\mathcal{R}}
\newcommand{\rpr}{\rhd}
\newcommand{\rpl}{\lhd}

\newcommand{\SP}{{SP}}

\newcommand{\Si}{\Sigma}       % Signature

\newcommand{\SN}{\mathcal{SN}}
\newcommand{\WN}{\mathcal{WN}}

\newcommand{\abs}{\textsc{Abs}}

\newcommand{\appl}{\textsc{App}_\textsc{L}}
\newcommand{\appr}{\textsc{App}_\textsc{R}}
\newcommand{\subst}{\textsc{Subst}}

\newcommand{\pvar}{\rpr\textsc{Var}}
\newcommand{\papp}{\rpr\textsc{App}}
\newcommand{\psymb}{\rpr\textsc{Symb}}

\newcommand{\pBeta}{\rpr\beta}
\newcommand{\pRule}{\rpr\R}

%% file: Macros/symbols.tex
\newcommand{\id}{\mathsf{id}}
\newcommand{\err}{\mathsf{err}}

\newcommand{\true}{\mathsf{true}}
\newcommand{\false}{\mathsf{false}}

\newcommand{\minus}{\mathsf{minus}}

\newcommand{\zero}{\mathsf{zero}}
\newcommand{\suc}{\mathsf{succ}}

\newcommand{\nil}{\mathsf{nil}}
\newcommand{\cons}{\mathsf{cons}}
\newcommand{\car}{\mathsf{car}}
\newcommand{\cdr}{\mathsf{cdr}}
\newcommand{\get}{\mathsf{get}}
\newcommand{\length}{\mathsf{length}}
\newcommand{\filter}{\mathsf{filter}}

\newcommand{\apply}{\mathsf{apply}}
\newcommand{\appaux}{\mathsf{app}}

\newcommand{\node}{\mathsf{node}}

\newcommand{\occ}{\mathsf{occ}}
\newcommand{\replace}{\mathsf{replace}}
\newcommand{\rep}{\mathsf{rep}}

\newcommand{\pair}{\mathsf{pair}}
\newcommand{\fst}{\mathsf{fst}}
\newcommand{\snd}{\mathsf{snd}}

\newcommand{\W}{\Omega}
\newcommand{\Zero}{{Zero}}
\newcommand{\Succ}{{Succ}}

\newcommand{\Iter}{{Iter}}
\newcommand{\Rec}{{Rec}}

\comment{ % Case change
\newcommand{\Zero}{{zero}}
\newcommand{\Succ}{{succ}}

\newcommand{\Iter}{{iter}}
\newcommand{\Rec}{{rec}}

} % Case change

\comment{ % Old Fonts

\newcommand{\W}{\Omega}
\newcommand{\Zero}{\text{Zero}}
\newcommand{\Succ}{\text{Succ}}

\newcommand{\Rec}{\text{Rec}}

} % Old Fonts

%% file: figures.tex
\newcommand\FigOverSimple{
\begin{tabularx}{\linewidth}
{| c | C | C | C | C | C |}
\hline
\S	& Left-Hand Sides & Right-Hand Sides & Conditions & Terms 	& Result \\
\hline
\hline
  \ref{sec-Aconfl-sc}	
& Algebraic \& Linear %(Def.~\ref{def-terms})
& Applicative %(Def.~\ref{def-terms})
& No equality tests between open terms
  (Semi-closed, Def.~\ref{def-sc-cond})
& All ($\Te(\Si)$)
& $\a_\R$ Confluent $\imp$ $\a_{\b \cup \R}$ Confluent
  (Thm.~\ref{thm-Aconfl-sc})
\\		
\hline		
  \ref{sec-Aconfl-wn}
& Algebraic %(Def.~\ref{def-terms})
  \& Respect an arity $\ap$ (Def.~\ref{def-arity-cond})
& Algebraic \&
  Respect the arity $\ap$ %(Def.~\ref{def-arity})
& Algebraic \& Respect the arity $\ap$
& Weakly $\b$-normalizing \&
  $\b$nf
  respect the arity $\ap$ %of rewrite rules
  ($\AN$, Def.~\ref{def-an})
& $\a_\R$ Confluent $\imp$ $\a_{\b \cup \R}$ Confluent
  (Thm.~\ref{thm-Aconfl-bnf})
\\
\hline
\hline
  \ref{sec-Bconfl-sc}
& Algebraic \& Linear %(Def.~\ref{def-terms})
  \& Respect an arity $\ap$ (Def.~\ref{def-arity-cond})
& Algebraic
  \& Respect the arity $\ap$
& Algebraic
  \& Respect the arity $\ap$\qquad\qquad
  \& No equality tests between open terms 
  (Semi-closed, Def.~\ref{def-sc-cond})
& Respect the arity $\ap$ %of rewrite rules
  (Conditionally $(\R,\ap)$-stable, Def.~\ref{def-rstable-cond})
& $\a_\R$ Confluent $\imp$ $\a_{\b \cup \R(\b)}$ Confluent 
  (Thm.~\ref{thm-Bconfl-sc})
\\
\hline
  \ref{sec-Bconfl-wn}	
& Algebraic %(Def.~\ref{def-terms})
  \& Respect an arity $\ap$ (Def.~\ref{def-arity-cond})
& Algebraic \& Respect the arity $\ap$
& Algebraic \&
  Respect the arity $\ap$
& Weakly $\b$-normalizing \&
  $\b$nf
  respect the arity $\ap$ %of rewrite rules
  ($\AN$, Def.~\ref{def-an})
& $\a_\R$ Confluent $\imp$ $\a_{\b \cup \R(\b)}$ Confluent
  (Thm.~\ref{thm-Bconfl-bnf})
\\
\hline
\hline
  \ref{sec-exclus}
& Algebraic \& Linear %(Def.~\ref{def-terms})
&
& Orthonormal (Def.~\ref{def-orth})
& All ($\Te(\Si)$)
& $\a_{\b \cup \R(\b)}$ Shallow Confluent (Thm.~\ref{thm:lcr:beta:cond})
\\
\hline
\end{tabularx}
} % FigOverview

%% file: abstract.tex
%%%%%%%%%%%%%%%%%%%%%%%%%%%%%%%%%%%%%%%%%%%%%%%%%%%%%%%%%%%%%%%%%%%%%%%%%%%
% Abstract
%%%%%%%%%%%%%%%%%%%%%%%%%%%%%%%%%%%%%%%%%%%%%%%%%%%%%%%%%%%%%%%%%%%%%%%%%%%
\begin{abstract}
The confluence of untyped $\la$-calculus with {\em unconditional}
re\-writing is now well understood. In this
paper, we investigate the confluence of $\la$-calculus with {\em
conditional} rewriting and provide general results in two directions.

First, when conditional rules are algebraic. This extends results of
M\"uller and Dougherty for unconditional rewriting. Two cases are
considered, whether beta-reduction is allowed or not in the evaluation
of conditions. Moreover, Dougherty's result is improved from the
assumption of strongly normalizing $\b$-reduction to weakly
normalizing $\b$-reduction. We also provide examples showing that
outside these conditions, modularity of confluence is difficult to
achieve.

Second, we go beyond the algebraic framework and get new
confluence results using a restricted notion of orthogonality that
takes advantage of the conditional part of rewrite rules.
\end{abstract}

%% file: intro.tex
%%%%%%%%%%%%%%%%%%%%%%%%%%%%%%%%%%%%%%%%%%%%%%%%%%%%%%%%%%%%%%%%%%%%%%%%%%%
\section{Introduction}
%%%%%%%%%%%%%%%%%%%%%%%%%%%%%%%%%%%%%%%%%%%%%%%%%%%%%%%%%%%%%%%%%%%%%%%%%%%

Rewriting~\cite{dj90book} and $\la$-calculus~\cite{barendregt84book}
are two universal computation models which are
both used, with their own advantages, in programming languages design
and implementation, as well as for the foundation of logical
frameworks and proof assistants. Among other things,
$\la$-calculus allows to manipulate abstractions and higher-order variables,
while rewriting is traditionally well suited for
defining functions over data-types and for dealing with equality.

Starting from Klop's work on higher-order rewriting and because of
their complementarity, many frameworks have been designed with a view
to integrate these two formalisms. This integration
has been handled
either by enriching first-order rewriting with higher-order
capabilities, by adding to $\la$-calculus algebraic features or, more
recently, by a uniform integration of both paradigms. In the first
case, we find the works on combinatory reduction systems~\cite{kor93tcs}
and other higher-order rewriting systems~\cite{wolfram93phd,nipkow91lics}
each of them subsumed by van Oostrom 
and van Raamsdonk's axiomatization of HORSs~\cite{or94lfcs},
and by the every expressive framework of CCERSs~\cite{gkk05ptc}.
The second case concerns the more atomic
combination of $\la$-calculus
with term rewriting~\cite{jo91lics,blanqui05mscs}
and the last category
the rewriting calculus~\cite{cirstea01igpal,bckl03popl}.

Despite this strong interest in the combination of both concepts, few works have
considered {\em conditional} higher-order rewriting with $\la$-calculus.
This is of particular interest for both computation and deduction.
Indeed, conditional rewriting appears to be very convenient
when programming with rewrite rules and its combination
with higher-order features provides a quite agile background
for the combination of algebraic and functional programming.
This is also of main use in proof assistants based
on the Curry-Howard-de Bruijn isomorphism where,
as emphasized in {\em deduction modulo}~\cite{dhk03jar,blanqui05mscs},
rewriting capabilities for defining functions and
proving equalities automatically is clearly of great interest when making large
proof developments.
Furthermore, while many confluence proofs
often rely on termination and local confluence, in some cases,
confluence may be necessary for proving termination (e.g.\ with
type-level rewriting or strong elimination~\cite{blanqui05mscs}). It
is therefore of crucial interest to have also criteria for the
preservation of confluence when combining conditional rewriting and
$\b$-reduction without assuming the termination of the combined
relation. In particular, assuming the termination of just one of the
two relations is already of interest.

The earliest work on preservation of confluence when combining
typed $\la$-calculus and first-order rewriting concerns the simple type
discipline~\cite{breazu88lics} and the result has been extended to
polymorphic $\la$-calculus in~\cite{bg94ic}. Concerning untyped
$\la$-calculus, the result was shown in~\cite{muller92ipl} for
left-linear rewriting.
It is extended as a modularity result for higher-order rewriting
in~\cite{or94lfcs}.
In~\cite{dougherty92ic}, it is shown that 
left-linearity is not necessary,
provided that terms considered are strongly
$\b$-normalizable and are well-formed with respect to the declared
arity of symbols, a property that we call here {\em arity compliance}.
Higher-order conditional rewriting is studied in~\cite{avenhaus94ccl}
and the confluence result relies on
joinability of critical pairs, hence on termination of the combined
rewrite relation.
An approach closer to ours is taken 
with a form of conditional $\la$-calculus in~\cite{takahashi93tlca},
and with CCERSs in~\cite{gkk05ptc}.
In both cases, confluence relies on a form of conditional orthogonality.
However, in these works, conditions are abstract predicates on terms,
and confluence is achieved by assuming that the satisfaction of these
predicates is preserved by reduction.
These results do not directly apply in our case,
since proving that the satisfaction of conditions
is preserved by reduction is actually 
the most difficult task for confluence, and this requires
a precise knowledge of the shape of the conditions.
%(see also Section~\ref{sec-exclus}).
These systems are related to those presented in Section~\ref{sec-exclus}.
Our results can rather be seen as a form of modularity properties.
Concerning confluence of unconditional term rewriting, the early work
of~\cite{toyama87jacm} has been extended to the higher-order
case in~\cite{or94lfcs}.
In the case of conditional rewriting,
if modularity properties have been investigated
in the pure first-order setting (e.g.~\cite{middeldorp91ctrs,gramlich96tcs}), 
to the best of our knowledge, there was up to now no result on the 
preservation of confluence for the {\em combination with $\b$-reduction}.

In this paper, we study the confluence property of the combination of
$\b$-reduction with a confluent conditional rewrite system. This of
course should rely on a clear understanding of the conditional rewrite
relation under use and, as usual, the way matching is performed
and conditions are checked is crucial.
We always consider left-hand sides without abstractions. So, rewriting
is not in need of higher-order pattern-matching but just relies on syntactic
matching.

We begin in Section~\ref{sec-calc} by presenting 
our notations and some basic facts on $\la$-calculus and conditional rewriting.
We start from $\la$-calculus and discuss, via Böhm's theorem,
the need of enriching its syntax with symbols defined by rewrite rules.
We then present the different kinds of conditional rewriting considered in this paper.
We are interested in {\em join} conditional rewriting: the conditions
of rewrite rules are evaluated by testing the {\em joinability} of terms.
Given a conditional rewrite system, we consider two conditional rewrite relations,
whether $\b$-reduction is allowed or not in the evaluation of conditions.
The case where $\b$-reduction is allowed in the conditions
is termed {\em $\b$-conditional} rewriting.
We also discuss the particular case of {\em normal} rewriting,
i.e.\ when one side of the conditions is made of terms in normal form.
We then give two examples of conditional rewrite system.
The first one recalls the use of conditional rewriting
in the study of $\la$-calculus with surjective pairing~\cite{vrijer89lics}.
The second one is a term manipulation system inspired from a program
of~\cite{huet86notes}.
We conclude this section by some basic material on confluence.

In Section~\ref{sec-over-gen} we state precisely
the known results from which this paper starts
and give a short overview of our results.
The general goal of this paper is to give sufficient conditions for the 
confluence of $\beta$-reduction with $\b$-conditional rewriting
(i.e.\ with $\b$-steps allowed in the evaluation of conditions).
Our main objective is the
{\em preservation} of confluence, that is, given a conditional
rewrite system, to get confluence
$\b$-conditional rewriting combined with $\b$-reduction
assuming the confluence of conditional rewriting.
Our approach is to generalize known results 
on the combination of $\b$-reduction with unconditional rewriting.
We present in Section~\ref{sec-rew} the two different cases
we start with:
M{\"u}ller's result~\cite{muller92ipl} for left-linear rewriting,
and 
Dougherty's result~\cite{dougherty92ic} for algebraic rewriting
on strongly $\b$-normalizing terms respecting some arity conditions
(called {\em arity compliance}).
In each case, we will first consider 
the case of $\b$-reduction with conditional rewriting
(when $\b$-reduction is not allowed in the evaluation of conditions)
and then extend these results to $\b$-conditional rewriting.
However, Example~\ref{ex-Bconfl} shows that for $\b$-conditional
rewriting, we can not go beyond algebraic rewriting with arity conditions.
In order to handle rewrite rules
which can contain active variables and abstractions
in right-hand sides or in conditions,
we build on orthogonal conditional rewriting.
Known results on the confluence of orthogonal for normal algebraic conditional rewriting
are discussed in Section~\ref{sec-orth-cond}.
We conclude this section by an informal overview of our results.
They are summarized in Figure~\ref{fig-oversimple},
page~\pageref{fig-oversimple}.
%See Figure~\ref{fig-summary} page~\pageref{fig-summary} for precise statements.

The last three sections contain the technical contributions of the paper.
We begin in Section~\ref{sec-Aconfl} by extending M{\"u}ller's
and Dougherty's result to conditional rewriting combined with
$\b$-reduction.
M{\"u}ller's result~\cite{muller92ipl} assumes the left-linearity
of rewrite rules.
Of course, with conditional rewriting,
non-linearity can be simulated by linear systems.
Extending the result of M{\"u}ller~\cite{muller92ipl},
we prove in Section~\ref{sec-Aconfl-sc} that the confluence of
conditional rewriting combined with $\b$-reduction
follows from the confluence of conditional rewriting when conditional rules are
applicative, left-linear and semi-closed,
which means that the conditions of rules cannot test for equality of open terms.
In Section~\ref{sec-Aconfl-wn} we adapt Dougherty's method~\cite{dougherty92ic}
to conditional rewriting and extend it to show
that for a large set of {\em weakly} $\b$-normalizing
terms, the left-linearity and semi-closure hypotheses can be dropped provided that
rules are algebraic and terms are arity compliant.

We then turn in Section~\ref{sec-Bconfl} to the confluence of
$\b$-conditional rewriting combined with $\b$-reduction.
We show in Example~\ref{ex-Bconfl} that confluence
is in general not preserved with non-algebraic rules.
When rules are algebraic, we show that arity compliance is a
sufficient condition to deduce the confluence of $\b$-conditional rewriting
combined with $\b$-reduction
from the confluence of conditional rewriting alone.
This is done first for left-linear semi-closed systems in Section~\ref{sec-Bconfl-sc},
a restriction that we also show to be superfluous when considering only 
{\em weakly} $\b$-normalizing terms (Section~\ref{sec-Bconfl-wn}).

The case of non-algebraic rules is handled in Section~\ref{sec-exclus}.
Such rules can contain active variables and abstractions
in right-hand sides or in conditions (but still not in left-hand sides).
In this case, the confluence of $\b$-conditional rewriting combined with $\b$-reduction
does not follow anymore
from the confluence of conditional rewriting.
We show that confluence
holds under a syntactic condition, called {\em orthonormality},
ensuring that if two rules overlap at a non-variable position,
then their conditions cannot be both satisfied.
An orthonormal system is therefore an orthogonal system
whose orthogonality follows from the confluence of the rewrite relation
(recall that with conditional rewriting critical pairs
contain conditions ; hence orthogonality depends on the rewrite relation
since it depends on the satisfiability of these conditions).

This paper is an extended version of~\cite{bkr06fossacs}.
We assume familiarity with
$\la$-calculus~\cite{barendregt84book} and conditional
rewriting~\cite{do90tcs,ohlebusch02book}.
We recall the main notions in the next section.

%%%%%%%%%%%%%%%%%%%%%%%%%%%%%%%%%%%%%%%%%%%%%%%%%%%%%%%%%%%%%%%%%%%%%%%%%%%
\begin{figure}
%%%%%%%%%%%%%%%%%%%%%%%%%%%%%%%%%%%%%%%%%%%%%%%%%%%%%%%%%%%%%%%%%%%%%%%%%%%
\begin{center} \FigOverSimple \end{center}
\caption{Overview of the results.
Algebraic and applicative terms are defined in Def.~\ref{def-terms}
\label{fig-oversimple}}
\end{figure}

%% file: calc.tex
%%%%%%%%%%%%%%%%%%%%%%%%%%%%%%%%%%%%%%%%%%%%%%%%%%%%%%%%%%%%%%%%%%%%%%%%%%%
\section{Lambda-calculus and conditional rewriting}
\label{sec-calc}
%%%%%%%%%%%%%%%%%%%%%%%%%%%%%%%%%%%%%%%%%%%%%%%%%%%%%%%%%%%%%%%%%%%%%%%%%%%

In this section we present the tools used in this paper and recall 
some well-known facts.

%%%%%%%%%%%%%%%%%%%%%%%%%%%%%%%%%%%%%%%%%%%%%%%%%%%%%%%%%%%%%%%%%%%%%%%%%%%
\subsection{Terms and rewrite relations}
%%%%%%%%%%%%%%%%%%%%%%%%%%%%%%%%%%%%%%%%%%%%%%%%%%%%%%%%%%%%%%%%%%%%%%%%%%%

We consider $\la$-terms with curried function symbols.
Among them we distinguish {\em applicative terms}
that do not contain abstractions, and {\em algebraic terms}
that are applicative terms with no variable in active position.

%%%%%%%%%%%%%%%%%%%%%%%%%%%%%%%%%%%%%%%%%%%%%%%%%%%%%%%%%%%%%%%%%%%%%%%%%%%
\begin{definition}[Terms]
\label{def-terms}
%%%%%%%%%%%%%%%%%%%%%%%%%%%%%%%%%%%%%%%%%%%%%%%%%%%%%%%%%%%%%%%%%%%%%%%%%%%
Let $\Si$ be a set of function symbols and $\Vte$ be a set of variables.
\begin{enumerate}
\item
The set $\Te(\Si)$ of {\em $\la$-terms} is defined by the grammar
\[
t,u \in \Te(\Si)
\quad::=\quad x
\gs \la x.t
\gs t \ptesp u
\gs \sff
~,
\]
where $x \in \Vte$ and $\sff \in \Si$.
We denote by $\Te$ the set $\Te(\emptyset)$ of {\em pure}
$\la$-terms.

\item
The set of {\em applicative terms} is defined by the grammar
\[
t,u \quad::=\quad x \gs t \ptesp u \gs \sff
~.
\]

\item
The set of {\em algebraic terms} is defined by the grammar
\[
t \quad::=\quad x \gs \sff \ptesp t_1 \dots t_n
~.
\]
\end{enumerate}
\end{definition}
\noindent
As usual, $\la$-terms are considered equal modulo $\alpha$-conversion.
We denote by $\FV(t)$ the set of variables occurring free in the term $t$.
A term is {\em closed} if it has no free variables and {\em open} otherwise,
it is {\em linear} if each of its free variables occurs at most once.
Given $h \in \Vte \cup \Si$, we write $h \ptesp \vt$
for $h \ptesp t_1 \dots t_n$ and let $|\vt| \esp\deq\esp n$.
Similarly, we write $\la \vx.t$ for
$\la x_1.\dots \la x_n.t$.
% and let $|\vx| \esp\deq\esp n$.

%%%%%%%%%%%%%%%%%%%%%%%%%%%%%%%%%%%%%%%%%%%%%%%%%%%%%%%%%%%%%%%%%%%%%%%%%%%
\begin{example}
%%%%%%%%%%%%%%%%%%%%%%%%%%%%%%%%%%%%%%%%%%%%%%%%%%%%%%%%%%%%%%%%%%%%%%%%%%%
Intuitively, an algebraic term is a curried first-order term with no
arity constraint on symbols. For instance the terms $\filter$ and
$\filter \ptesp p \ptesp x \ptesp l$ are algebraic,
as well as $\filter \ptesp p \ptesp x \ptesp l \ptesp y \ptesp z$.
An applicative term is an algebraic term which may contain
variables in head position, such as $x \ptesp \filter$.
The $\la$-term $\la x.x$ is not applicative
(and thus not algebraic).
\end{example}

A lot of proofs of this paper are made by induction on the structure
of $\la$-terms. However, it is often not convenient to reason
directly on their syntax as given by the productions of $\Te(\Si)$.
For instance, knowing that a term $t$ is an application, say $t = u \esp v$,
gives little information on its behavior:
we do not know whether $u$ is
an abstraction, in which case $t$ is a $\b$-redex,
or whether it is an algebraic term, in which case $t$ may be
the instantiated left-hand side of a rewrite rule.
It it is therefore useful to have an induction principle
on $\la$-terms which makes apparent more informations
on their structure.
This is provided by the following well-known lemma, due to
Wadsworth~\cite{wadsworth71phd}.

%The following well-known lemma is useful to reason on $\la$-terms.
%%%%%%%%%%%%%%%%%%%%%%%%%%%%%%%%%%%%%%%%%%%%%%%%%%%%%%%%%%%%%%%%%%%%%%%%%%%
\begin{lemma}[\cite{wadsworth71phd}]
\label{lem-wadsworth}
%%%%%%%%%%%%%%%%%%%%%%%%%%%%%%%%%%%%%%%%%%%%%%%%%%%%%%%%%%%%%%%%%%%%%%%%%%%
Any $\la$-term $t \in \Te(\Si)$
can be uniquely written in one of the following forms:
\begin{align}
\tag{a}                 \la x_1.\dots \la x_m.v\, &a_1 \dots a_n  \\
\tag{b} \text{or} \quad \la x_1.\dots \la x_m.(\la y.b) &a_0 \, a_1 \dots a_n 
\end{align}
where $n,m \geq 0$ and $v \in \Vte \cup \Si$.
\end{lemma}

A {\em substitution} is a map $\s : \Vte \a \Te(\Si)$ of finite domain.
We denote by $t\s$ the capture-avoiding application 
of the substitution $\s$ to the term $t$.
If $\s$ is the substitution which maps $x_i$ to $u_i$
for all $i \in \{1,\dots,n\}$, then we may write
$t\wthdots{x_1}{u_1}{x_n}{u_n}$ instead of $t\s$.

%%%%%%%%%%%%%%%%%%%%%%%%%%%%%%%%%%%%%%%%%%%%%%%%%%%%%%%%%%%%%%%%%%%%%%%%%%%
\begin{definition}[Rewrite Relations]
%%%%%%%%%%%%%%%%%%%%%%%%%%%%%%%%%%%%%%%%%%%%%%%%%%%%%%%%%%%%%%%%%%%%%%%%%%%
A rewrite relation is a binary relation $\a$ on $\Te(\Si)$
closed under the following rules,
where $\s$ is a substitution:
\[
(\abs)~
\dfrac{t \esp\a\esp u}
      {\la x.t \esp\a\esp \la x.u}
\qquad%\qquad
(\appl)~
\dfrac{t \esp\a\esp u}
      {t \ptesp v \esp\a\esp u \ptesp v}
\qquad%\qquad
(\appr)~
\dfrac{t \esp\a\esp u}
      {v \ptesp t \esp\a\esp v \ptesp u}
\qquad%\qquad
(\subst)
\dfrac{t \esp\a\esp u}
      {t\s \esp\a\esp u\s}
\]
\end{definition}

\noindent
We denote by $\a^+$ the transitive closure of $\a$,
by $\a^\re$ its reflexive closure,
by $\a^*$ its reflexive and transitive closure,
by $\al$ its inverse and by $\alr$ its reflexive symmetric and
transitive closure.
We write $t \ad u$ if there exists $v$ such that $t \a^* v \al^* u$
and $t \a^k u$ if $t \a^* u$ in at most $k$ steps.

Given two rewrite relations $\a_{A}$ and $\a_{B}$,
we let $\a_{A \cup B} \ptesp\mbin{\deq}\ptesp \a_A \cup \a_B$.
%Given a rewrite relation $\a_A$,
We say that a term $t$ is an {\em $A$-normal form}
if there is no $u$ such that $t \a_A u$.
We let $\SN_A$, {\em the set of strongly $A$-normalizing terms},
be the set of terms on which the relation $\a_A$ is well-founded
and we let $\WN_A$, {\em the set of weakly $A$-normalizing terms},
be the set of terms which rewrite to an $A$-normal form.

Rewrite relations $\a$ satisfy the following property:
for all $t,u,v \in \Te(\Si)$,
\[
\text{if}\qquad
t \a u
\qquad\text{then}\qquad
v\wth{x}{t} \esp\a^*\esp v\wth{x}{u}
~.
\]
In the following, we will often use a stronger property:
for all $t,u,v \in \Te(\Si)$,
\[
\text{if}\qquad
t \a u
\qquad\text{then}\qquad
v\wth{x}{t} \esp\a\esp v\wth{x}{u}
~.
\]
This is in general false with rewrite relations, but
this holds with {\em parallel} rewrite relations.

%%%%%%%%%%%%%%%%%%%%%%%%%%%%%%%%%%%%%%%%%%%%%%%%%%%%%%%%%%%%%%%%%%%%%%%%%%%
\begin{definition}[Parallel Rewrite Relations]
\label{def-par-rew}
%%%%%%%%%%%%%%%%%%%%%%%%%%%%%%%%%%%%%%%%%%%%%%%%%%%%%%%%%%%%%%%%%%%%%%%%%%%
A {\em parallel rewrite relation} is a rewrite relation $\rpr$
closed under the rules
\[
(\pvar)~
\dfrac{}
      {x \esp\rpr\esp x}
\qquad\qquad
(\psymb)~
\dfrac{}
      {\sff \esp\rpr\esp \sff}
\qquad\qquad
(\papp)~
\dfrac{t_1 \esp\rpr\esp u_1 \qquad t_2 \esp\rpr\esp u_2} 
      {t_1 \ptesp t_2 \esp\rpr\esp u_1 \ptesp u_2}
\]
\end{definition}

\noindent
Note that given a parallel rewrite relation $\rpr$,
we have $\la x.t \rpr \la x.u$ if $t \rpr u$, since by definition
parallel rewrite relations are rewrite relations.

Given a rewrite relation $\a$ and two substitutions $\s$ and $\s'$,
we write $\s \a \s'$ if $\s$ and $\s'$ have the same domain
and $\s(x) \a \s'(x)$ for all $x \in \dom(\s)$.

%%%%%%%%%%%%%%%%%%%%%%%%%%%%%%%%%%%%%%%%%%%%%%%%%%%%%%%%%%%%%%%%%%%%%%%%%%%
\begin{proposition}
\label{prop-par-rew}
%%%%%%%%%%%%%%%%%%%%%%%%%%%%%%%%%%%%%%%%%%%%%%%%%%%%%%%%%%%%%%%%%%%%%%%%%%%
If $\rpr$ is a parallel rewrite relation on $\Te(\Si)$
then $\s \rpr \s'$ implies $v\s \rpr v\s'$.
\end{proposition}

\begin{proof}
By induction on $v$.
\begin{description}
\item[$v \in \Vte \cup \Si$.]
If $v = x \in \dom(\s)$ then $v\s = \s(x) \rpr \s'(x) = v\s'$.
Otherwise, $v\s = v \rpr v = v\s'$
thanks to the rules $(\pvar)$
and $(\psymb)$.

\item[$v = v_1 \ptesp v_2$.]
By induction hypothesis we have $v_i\s \rpr v_i\s'$
for all $i \in \{1,2\}$, and we conclude by the rule $(\papp)$.

\item[$v = \la x.v_1$.]
By induction hypothesis.
\qedhere
\end{description}
\end{proof}

\noindent
In particular, if $\dom(\s) \cap \FV(v) = \emptyset$ then $v \rpr v$:
parallel rewrite relations are reflexive.

%%%%%%%%%%%%%%%%%%%%%%%%%%%%%%%%%%%%%%%%%%%%%%%%%%%%%%%%%%%%%%%%%%%%%%%%%%%
\subsection{Lambda-calculus}
%%%%%%%%%%%%%%%%%%%%%%%%%%%%%%%%%%%%%%%%%%%%%%%%%%%%%%%%%%%%%%%%%%%%%%%%%%%

$\la$-calculus is characterized by $\b$-reduction.
This is the smallest rewrite relation $\a_\b$ on $\Te(\Si)$
such that
\[
  (\la x.t)u
  \quad\a_\b\quad
  t\wth{x}{u}
~.
\]
In order to understand our motivations for studying the combination
of $\la$-calculus with (conditional) rewriting, let us recall
some facts about {\em pure} $\la$-calculus.
%We denote by $\Te$ the set $\Te(\emptyset)$ of {\em pure} $\la$-terms.
It is well-known that integers can be coded within pure $\la$-calculus.
An example of such coding is that of {\em Church's numerals}.
The $\Zero$ and $\Succ$ functions are represented by the following terms:
\[
  \Zero \quad\deq\quad \la x.\la f.x
  \qquad\text{and}\qquad
  \Succ \quad\deq\quad \la n.\la x.\la f.f \esp (n \esp x \esp f)
~.
\]
We can code {\em iteration} with the term
$\Iter \esp x \ptesp y \ptesp z \esp\deq\esp z \ptesp x \ptesp y$,
and for all $n,u,v \in \Te$ we have
\[
\begin{array}{l !{\quad=\quad} l !{\quad\a^2_\b\quad} l}
  \Iter \esp u \esp v \esp \Zero
& (\la xf.x) \esp u \esp v
& u \\
  \Iter \esp u \esp v \esp (\Succ \esp n)
& (\la xf.f \esp (n \esp x \esp f)) \esp u \esp v
& v \esp (n \esp u \esp v)
  \quad=\quad v \esp (\Iter \esp u \esp v \esp n)
~.
\end{array}
\]
However, {\em recursion} cannot be implemented in constant time
(see for instance~\cite{parigot89csl}):
there is no term $\Rec \esp x \ptesp y \ptesp z$ such that
there is $k \in \N$ such that  for all $u,v,n \in \Te$,
\[
  \Rec \esp u \esp v \esp \Zero \quad\a^k_\b\quad u
  \qquad\text{and}\qquad
  \Rec \esp u \esp v \esp (\Succ \esp n)
  \quad\a^k_\b\quad
  v \esp (\Rec \esp u \esp v \esp n) \esp n
  ~.
\]
In particular, there is no coding of the predecessor function
for Church's numerals in constant time.
This suggests that practical utilizations of the $\la$-calculus may require
extensions of $\b$-reduction.
At this point it is interesting to recall B{\"o}hm's theorem.
It states that any proper extension of $\b\eta$-conversion
on the set of weakly $\b$-normalizing pure $\la$-terms is inconsistent.

%%%%%%%%%%%%%%%%%%%%%%%%%%%%%%%%%%%%%%%%%%%%%%%%%%%%%%%%%%%%%%%%%%%%%%%%%%%
\begin{theorem}[B\"ohm~\cite{bohm68}]
%%%%%%%%%%%%%%%%%%%%%%%%%%%%%%%%%%%%%%%%%%%%%%%%%%%%%%%%%%%%%%%%%%%%%%%%%%%
Let $\a_\e$ be the smallest rewrite relation on $\Te$ such that
$\la x.t \ptesp x \a_\e t$ if $x \notin \FV(t)$.
If $\simeq$ is an equivalence relation on $\Te$ which is stable
by contexts, contains $\alr_{\b\e}$ and such that
$\simeq \setminus \alr_{\b\e}$ contains a pair of weakly $\b$-normalizing terms,
then for all $t,u \in \Te$ we have $t \simeq u$.
\end{theorem}

This theorem suggests to find extensions of $\b$-reduction
operating on extensions of the set of pure $\la$-terms $\Te$.
A possibility, that we consider in this paper, is to 
work with {\em function symbols} $\sff \in \Si$
defined by {\em rewrite rules}.

%%%%%%%%%%%%%%%%%%%%%%%%%%%%%%%%%%%%%%%%%%%%%%%%%%%%%%%%%%%%%%%%%%%%%%%%%%%
\subsection{Conditional rewriting}
%%%%%%%%%%%%%%%%%%%%%%%%%%%%%%%%%%%%%%%%%%%%%%%%%%%%%%%%%%%%%%%%%%%%%%%%%%%

In this paper, we are interested in {\em conditional} rewriting.
The following example introduces the main ideas.
Consider lists built
from the empty list $\nil$ and the constructor $\cons$.
We use the symbols $\true$ and $\false$ to represent the boolean values
"true" and "false".
We would like to define, via rewriting, a function $\filter$ such that
\begin{itemize}
\item
$\filter \ptesp p \ptesp \nil$ rewrites to $\nil$, 

\item
$\filter \ptesp p \ptesp (\cons \ptesp t \ptesp ts)$
rewrites to $\cons \ptesp t \ptesp (\filter \ptesp p \ptesp ts)$
if $p \ptesp t$ rewrites to $\true$,
and 

\item
$\filter \ptesp p \ptesp (\cons \ptesp t \ptesp ts)$
rewrites to $\filter \ptesp p \ptesp ts$
if $p \ptesp t$ rewrites to $\false$.
\end{itemize}
This specification can be written using {\em conditional rewrite rules}
($\sgt$ reads {\em implies}):
\begin{equation}
\begin{aligned}
\label{eqn-filter}
\begin{array}{l c l c l c l}
              &   &         &      & \filter \esp p \esp \nil
  & \at & \nil \\
    p \esp x  & = & \true   & \sgt & \filter \esp p \esp (\cons \esp x \esp xs)
  & \at & \cons \esp x \esp (\filter \esp p \esp xs) \\
    p \esp x  & = & \false  & \sgt & \filter \esp p \esp (\cons \esp x \esp xs)
  & \at & \filter \esp p \esp xs
\end{array}
\end{aligned}
\end{equation}
If we try to define a rewrite relation $\a$
that corresponds to our specification, we get that
\begin{equation}
%\label{eqn-ex-condrew}
  \filter \esp p \esp (\cons \esp t \esp ts)
\quad\a\quad
  \cons \esp t \esp (\filter \esp p \esp ts)
\qquad\text{if}\qquad
  p \esp t ~\a^*~ \true
  ~.
\end{equation}
In other words, to define $\a$ in the step
\[
  \filter \esp p \esp (\cons \esp t \esp ts)
\quad\a\quad
  \cons \esp t \esp (\filter \esp p \esp ts)
~,
\]
we need to test if $p \ptesp t \a^* \true$,
hence to use the relation $\a$.
This circularity can be broken off by using an inductive definition of
conditional rewriting:
the relation $\a$ is stratified in relations $(\a_i)_{i \in \N}$.
The correctness of the definition is ensured by Tarski's fixpoint
theorem,
which can be applied 
because, when replacing the symbol 
$=$ by $\a^*$ in (\ref{eqn-filter}),
the obtained formula is {\em positive} in $\a$ (it is in fact a Horn clause).

We now turn to formal definitions.

%%%%%%%%%%%%%%%%%%%%%%%%%%%%%%%%%%%%%%%%%%%%%%%%%%%%%%%%%%%%%%%%%%%%%%%%%%%
\begin{definition}[Conditional Rewrite Rules]
\label{def-cond-rules}
%%%%%%%%%%%%%%%%%%%%%%%%%%%%%%%%%%%%%%%%%%%%%%%%%%%%%%%%%%%%%%%%%%%%%%%%%%%
A conditional rewrite rule is an expression of the form
\[
  d_1 = c_1 \esp\land\esp \dots \esp\land\esp d_n = c_n
  \esp\sgt\esp l \dbesp\at\dbesp r
\]
where $d_1,\dots,d_n,c_1\dots,c_n,l,r \in \Te(\Si)$ and 
\begin{enumerate}
\item every variable of $\vd,\vc,r$ occurs also in $l$,
\item $l$ is an algebraic term which is not a variable.
\end{enumerate}
In conditional rewrite rules, we distinguish
\begin{itemize}
\item the left-hand side $l$, the right-hand side $r$ ;
\item the conditions $d_1 = c_1 \esp \land \dots \land \esp d_n = c_n$.
\end{itemize}
%We assume that the left-hand side $l$ is an algebraic term which is not a variable.
%
A rule $d_1 = c_1 \land \dots \land d_n = c_n \sgt l \at r$
is {\em unconditional} if $n = 0$.
It is {\em left-linear} if $l$ is linear.
\end{definition}

Since left-hand sides are algebraic terms,
rewriting is performed using syntactical first-order matching.
Note that the conditions of rewrite rules are not symmetric:
the condition $d = c$ is not the same as $c = d$.
%This asymmetry is used to define conditional rewrite relations
%that evaluate conditions in an asymmetric way, as in~(\ref{eqn-ex-condrew}).

Given a set $\R$ of conditional rewrite rules,
different conditional rewrite relations can be defined,
depending on the evaluation of the conditions:
by conversion, by joinability or by reduction.
This leads respectively to 
semi-equational, join and oriented conditional
rewriting.
In this paper, we focus on join conditional rewriting,
and call it simply {\em conditional rewriting}.
We also condsider the case of join condition rewriting
with $\beta$-reduction allowed in the evaluation of conditions,
and call it {\em $\beta$-conditional rewriting}.
%We consider two cases, whether $\b$-reduction
%is allowed or not the evaluation of conditions, we term them
%$\b$-conditional rewriting

%%%%%%%%%%%%%%%%%%%%%%%%%%%%%%%%%%%%%%%%%%%%%%%%%%%%%%%%%%%%%%%%%%%%%%%%%%%
\begin{definition}[Conditional Rewriting]
\label{def-cond-rew}
%%%%%%%%%%%%%%%%%%%%%%%%%%%%%%%%%%%%%%%%%%%%%%%%%%%%%%%%%%%%%%%%%%%%%%%%%%%
Let $\R$ be a set of conditional rewrite rules.
\begin{itemize}
\item
The {\em conditional rewrite relation}
$\a_{\R}$ is defined as
\[
\a_{\R} 
\quad\deq\quad \bigcup_{i \in \N} \a_{\R_i}
~,
\]
%$(\a_{\R_i})_{i \in \N}$,
where $\a_{\R_0} \ \deq\ \emptyset$ and for all $i \in \N$,
$\a_{\R_{i+1}}$ is the smallest rewrite relation
such that for every rule $\vd = \vc \sgt l \at_\R r$
and every substitution $\s$,
\[
  \text{if}\qquad
  \vd\s \quad\ad_{\R_{i}}\quad \vc\s
  \qquad\text{then}\qquad
  l\s \quad\a_{\R_{i+1}}\quad r\s
~.%\fn
\]

\item
The {\em $\b$-conditional rewrite relation}
$\a_{\R(\b)}$ is defined as
\[
\a_{\R(\b)} 
\quad\deq\quad \bigcup_{i \in \N} \a_{\R(\b)_i}
~,
\]
where $\a_{\R(\b)_0} \ \deq\ \emptyset$ and for all $i \in \N$,
$\a_{\R(\b)_{i+1}}$ is the smallest rewrite relation
such that for every rule $\vd = \vc \sgt l \at_\R r$
and every substitution $\s$,
\[
  \text{if}\qquad
  \vd\s \quad\ad_{\b \cup \R(\b)_{i}}\quad \vc\s
  \qquad\text{then}\qquad
  l\s \quad\a_{\R(\b)_{i+1}}\quad r\s
~.%\fn
\]
\end{itemize}
\end{definition}

\noindent
Hence, with conditional rewriting $\a_\R$, $\b$-reduction is not allowed
in the evaluation of conditions, while it is allowed with $\b$-conditional
rewriting $\a_{\R(\b)}$.
Note that $\a_{\R} \sle \a_{\R(\b)}$.
The converse is false, as shown by the following example.

%%%%%%%%%%%%%%%%%%%%%%%%%%%%%%%%%%%%%%%%%%%%%%%%%%%%%%%%%%%%%%%%%%%%%%%%%%%
\begin{example}
%%%%%%%%%%%%%%%%%%%%%%%%%%%%%%%%%%%%%%%%%%%%%%%%%%%%%%%%%%%%%%%%%%%%%%%%%%%
Consider the rule
\[
p\esp x \ =\ {\true}  \quad \sgt\quad {\filter}\esp p \esp (\cons \esp x \esp l)
  \quad \at\quad \cons \esp x \esp ({\filter}\esp p \esp l)
\]
issued from the 
conditional rewrite system~(\ref{eqn-filter}) and assume that
$\id \esp x \at x$.
With conditional rewriting we have
\[
  \filter\esp \id \esp (\cons \esp \true \esp ts)
  \quad\a_{\R}\quad
  \cons \esp \true \esp (\filter \esp \id \esp ts) 
\qquad\text{since}\qquad
  \id \esp \true \quad\a_{\R}\quad \true
~.
\]
%(This also holds for $\b$-conditional rewriting.)
With $\b$-conditional rewriting we also have
\[
  \filter\esp (\la x. x) \esp (\cons\esp \true \esp ts)
  \dbesp\a_{\R(\b)}\dbesp
  \cons \esp \true \esp (\filter \esp \la x.x \esp ts) 
\quad~\text{since}~\quad
  (\la x. x) \esp \true \dbesp\a_{\b}\dbesp \true
~,
\]
but 
the term
$\filter \esp (\la x.x) \esp (\cons \esp \true \esp ts)$
is a $\a_{\R}$-normal form.
\end{example}

An interesting particular case of join conditional rewriting is 
{\em normal} rewriting.

%%%%%%%%%%%%%%%%%%%%%%%%%%%%%%%%%%%%%%%%%%%%%%%%%%%%%%%%%%%%%%%%%%%%%%%%%%%
\begin{definition}[Normal Conditional Rewriting]
\label{def-norm-rew}
%%%%%%%%%%%%%%%%%%%%%%%%%%%%%%%%%%%%%%%%%%%%%%%%%%%%%%%%%%%%%%%%%%%%%%%%%%%
Let $\R$ be a conditional rewrite system.
If for every rule $\vd = \vc \sgt l \at_\R r$, the conditions
$\vc$ are closed terms in $\a_{\R}$-normal form, then we say that
$\a_{\R}$ is a
{\em normal conditional rewrite relation}.
\end{definition}

\noindent
In general, for a given conditional system the normal forms
w.r.t.\ join and semi-equational rewriting are not the same
(this is a by-product of the fact that
semi-equational orthogonal rewriting is confluent,
while join orthogonal rewriting is not~\cite{bk86jcss,ohlebusch02book},
see also Theorem~\ref{thm-orth-cond}).
The notion of normal conditional rewriting presented in
Definition~\ref{def-norm-rew} is thus specific to join conditional rewriting
(it is easy to see that it coincides with normality for oriented rewriting).

An important point with conditional rewriting
is the possible undecidability of a rewriting step.
This impacts of effectiveness of the notion
of normal conditional rewriting.

%%%%%%%%%%%%%%%%%%%%%%%%%%%%%%%%%%%%%%%%%%%%%%%%%%%%%%%%%%%%%%%%%%%%%%%%%%%
\begin{remark}[Decidability]
\label{rem-dec}
%%%%%%%%%%%%%%%%%%%%%%%%%%%%%%%%%%%%%%%%%%%%%%%%%%%%%%%%%%%%%%%%%%%%%%%%%%%
One-step conditional rewrite relations are in general {\em not} decidable.
Consider a rule $\vd = \vc \sgt l \at r$.
Because of the recursive definition of $\a_\R$, to know if $l\s \a_\R r\s$,
we need to reduce the terms $\vd\s$ and $\vc\s$.
This is in general undecidable,
even for terminating systems~\cite{kaplan84tcs}
(see also~\cite{ohlebusch02book}).

These facts have consequences on normal rewriting.
Given a set of conditional rules, to determine whether it can 
generate a normal relation, we have to check that a part of the conditions
is in normal form.
This is in general undecidable, even when the rewrite relation terminates.

We therefore focus on join rewriting because it seems to be 
a more easily and generally applicable theory than normal rewriting,
even if the implementation of conditional rewriting is easier
when we {\em already know} that the conditional rewrite relation is normal.
\end{remark}

Our results on the preservation of confluence impose restrictions
on rewrite rules. Some of them concern the terms
which can appear in different parts of the rules.
This motivates the following definition.
Recall from Definition~\ref{def-cond-rules}
that left-hand sides are always assumed to be algebraic.

%%%%%%%%%%%%%%%%%%%%%%%%%%%%%%%%%%%%%%%%%%%%%%%%%%%%%%%%%%%%%%%%%%%%%%%%%%%
\begin{definition}[Applicative and Algebraic Conditional Rewrite Rules]
%%%%%%%%%%%%%%%%%%%%%%%%%%%%%%%%%%%%%%%%%%%%%%%%%%%%%%%%%%%%%%%%%%%%%%%%%%%
A conditional rewrite rule $\vd = \vc \sgt l \at r$ is
\begin{itemize}
\item {\em right-applicative} if $r$ is an applicative term,

\item {\em applicative} if it is right-applicative
and if moreover the terms $\vd,\vc$ are applicative,

\item {\em right-algebraic} if $r$ is an algebraic term,

\item {\em algebraic} if it is right-algebraic
and if moreover the terms $\vd,\vc$ are algebraic.
\end{itemize}
A rewrite system $\R$ is {\em right-applicative}
(resp. {\em applicative}, {\em right-algebraic}, {\em algebraic})
if all its rules are right-applicative
(resp. applicative, right-algebraic, algebraic).
\end{definition}

In the conditional rewrite system~(\ref{eqn-filter}), 
the first rule
$\filter \esp p \esp \nil \ \at\ \nil$ is algebraic.
The two other rules are right-algebraic.
They both use the term $p \esp x$ in their conditions, where $p$
is a variable. This term is applicative but not algebraic.

%% file: ex.tex
%%%%%%%%%%%%%%%%%%%%%%%%%%%%%%%%%%%%%%%%%%%%%%%%%%%%%%%%%%%%%%%%%%%%%%%%%%%
\subsection{Examples}
\label{sec-ex}
%%%%%%%%%%%%%%%%%%%%%%%%%%%%%%%%%%%%%%%%%%%%%%%%%%%%%%%%%%%%%%%%%%%%%%%%%%%

We now give some examples of conditional rewrite systems.

%%%%%%%%%%%%%%%%%%%%%%%%%%%%%%%%%%%%%%%%%%%%%%%%%%%%%%%%%%%%%%%%%%%%%%%%%%%
\subsubsection{Coherence of lambda-calculus with surjective pairing}
\label{sec-ex-sp}
%%%%%%%%%%%%%%%%%%%%%%%%%%%%%%%%%%%%%%%%%%%%%%%%%%%%%%%%%%%%%%%%%%%%%%%%%%%

We begin by recalling one use of conditional rewriting
in the study of $\la$-calculus with surjective pairing.
We use $\pair \esp t_1 \esp t_2$ to denote the pairing of $t_1$
and $t_2$.
The rewrite rules for binary products are the following:
\[
  \fst \esp (\pair \esp x_1 \esp x_2) \quad\at_\pi\quad x_1
  \qquad\qquad
  \snd \esp (\pair \esp x_1 \esp x_2) \quad\at_\pi\quad x_2
~.
\]
It is well-known that the combination of these rules with
$\b$-reduction is confluent (see Theorem~\ref{thm-muller},
proved in~\cite{muller92ipl}).
This follows from the {\em left-linearity} of the 
rewrite rules.
However, when we add the rule for surjective pairing
\[
  \pair \esp (\fst \esp x) \esp (\snd \esp x)
  \quad\at_{\SP}\quad x
\]
then the combination of the resulting rewrite relation with
$\b$-reduction is not confluent~\cite{klop80phd}.
Note that the rule $\at_\SP$ is not left-linear:
the variable $x$ appears twice in the left-hand side.
However, the corresponding conversion is coherent:
there are two terms that are not convertible.
This has been first shown using semantic methods~\cite{scott75scandiv}.

In~\cite{vrijer89lics}, de Vrijer uses semi-equational
$\b$-conditional rewriting to give a syntactic proof
of the coherence of $\b$-reduction combined with surjective pairing.
His rules are those of $\at_\pi$ plus
\[
  \snd \esp x \esp=\esp y
  \esp\sgt\esp \pair \esp (\fst \esp x) \esp y \dbesp\at_{lr}\dbesp x 
  \qquad\qquad
  \fst \esp x \esp=\esp y
  \esp\sgt\esp \pair \esp y \esp (\snd \esp x) \dbesp\at_{lr}\dbesp x 
  ~.
\]
The resulting relation is confluent modulo an equivalence relation,
and this allows de Vrijer to show that $\la$-calculus
plus pairs and surjective pairing is a conservative extension
of the pure $\la$-calculus:
for any two {\em pure} $\la$-terms $t,u \in \Te$,
\[
  t \alr_\b u
\qquad\text{if and only if}\qquad
  t \alr_{\b \cup \pi \cup \SP} u
~.
\]

%%%%%%%%%%%%%%%%%%%%%%%%%%%%%%%%%%%%%%%%%%%%%%%%%%%%%%%%%%%%%%%%%%%%%%%%%%%
\subsubsection{A term manipulation system}
\label{sec-ex-cond-term}
%%%%%%%%%%%%%%%%%%%%%%%%%%%%%%%%%%%%%%%%%%%%%%%%%%%%%%%%%%%%%%%%%%%%%%%%%%%

Our main example is an adaptation of a \textsc{CAML} program
of~\cite{huet86notes}.
It defines functions that perform in a term
the replacement of the subterm at a given occurrence by another term.
Terms are represented by trees whose nodes contain a label
and the list of their successor nodes.

This system must be read having in mind the combination of
$\la$-calculus with (join) $\b$-conditional rewriting.

We begin by some basic functions on lists.
\[
\begin{array}{c !{\qquad} c}
	\begin{array}{l !{\esp \at \esp} l}
	{\car}\esp (\cons \esp x \esp l) 	& x		\\
	{\car}\esp \nil		& {\err}
	\end{array}
&
	\begin{array}{l !{\esp \at \esp} l}
	{\cdr}\esp (\cons \esp x \esp l) 	& l		\\
	{\cdr}\esp \nil		& {\err}
	\end{array}
\\
\\
	\begin{array}{l !{\esp \at \esp} l}
	{\get}\esp l\esp  \zero 	& {\car}\esp l				\\
	{\get}\esp l\esp (\suc \esp n)	& {\get}\esp ({\cdr}\esp l)\esp n
	\end{array}
&
	\begin{array}{l !{\esp \at \esp} l}
	{\length}\esp \nil	& \zero	\\
	{\length}\esp (\cons \esp x \esp l)& \suc \esp (\length \esp l)
	\end{array}
\\
\\
  \multicolumn{2}{c}
  {
    \begin{array}{l c l c l c l}
	        &   &          &      & {\filter}\esp p \esp \nil
      & \at & \nil \\
	  p\esp x & = & {\true}  & \sgt & {\filter}\esp p \esp (\cons \esp x \esp l)
	  	& \at & \cons \esp x \esp ({\filter}\esp p \esp l) \\
	  p\esp x & = & {\false} & \sgt & {\filter}\esp p \esp (\cons \esp x \esp l)
		  & \at & {\filter}\esp p \esp l
	  \end{array}
  }
\end{array}
\]

Let us define $\apply$ such that
$\apply \ptesp f \ptesp n \ptesp l$ applies $f$ to the $n$th element of $l$.
It uses $\appaux$
as an auxiliary function:
\[
\begin{array}{l c l c l c l}
> ({\length}\esp l)\esp n & = & {\true}  & \sgt & 
  {\apply}\esp f\esp n\esp l & \at & {\appaux}\esp f\esp n\esp l \\
> ({\length}\esp l)\esp n & = & {\false} & \sgt &
  {\apply}\esp f\esp n\esp l & \at & {\err} \\
\\
                      &     &          &        &
  {\appaux}\esp f\esp \zero            \esp l
& \at
& \cons \esp (f\esp ({\car}\esp l)) \esp ({\cdr}\esp l)
\\
                      &     &          &        &
  {\appaux}\esp f\esp (\suc \esp n)\esp l
& \at
& \cons \esp ({\car}\esp l) \esp ({\appaux}\esp f \esp n\esp ({\cdr}\esp l))
\end{array}
\]
We represent
first-order terms by trees with nodes $\node \ptesp y \ptesp l$ where
$y$ is intended to be a label and $l$ the list of sons.
Positions are lists of integers
and $\occ \ptesp u \ptesp t$ tests if $u$ is an occurrence of $t$. 
We define it as follows:
\[
\begin{array}{l c l c l c l}
                        &     &           &         &
  {\occ}\esp \nil\esp t
& \at
& \true
\\
>  ({\length}\esp l)\esp x  & = & {\false}  & \sgt  &	 
	{\occ}\esp  (\cons \esp x \esp o)\esp ({\node}\esp y\esp l)
& \at
& \false
\\
> ({\length}\esp l)\esp x   &= & {\true}   & \sgt  &
	{\occ}\esp (\cons \esp x \esp o)\esp ({\node}\esp y\esp l)
& \at
& \occ\esp o\esp ({\get}\esp l\esp x)
\end{array}
\]
To finish, $\replace \ptesp t \ptesp o \ptesp s$
replaces by $s$ the subterm of $t$ at occurrence $o$.
\[
\begin{array}{c}
  \begin{array}{l c l c l c l}
    {\occ}\esp u \esp t & = & \true     & \sgt & 
      {\replace}\esp t \esp o \esp s
& \at
& {\rep}\esp t \esp o \esp s
\\
  {\occ}\esp u \esp t & = & {\false}  & \sgt &
  {\replace}\esp t \esp o \esp s
& \at
& {\err}
\\
  \end{array}
\\
\\
  \begin{array}{l c l}
  {\rep}\esp t\esp \nil \esp s
& \at
& s \\
  {\rep}\esp ({\node}\esp y\esp l) \esp (\cons \esp x \esp o)\esp s
& \at
& {\node}\esp y \esp 
  ({\apply}\esp (\la z.{\rep}\esp z \esp o\esp s)\esp x \esp l)
  \end{array}
\end{array}
\]

The system {\sf Tree} that consists of the rules defining 
$\car$, $\cdr$, $\get$, $\length$ and $\occ$ is algebraic.
The rules of $\apply$ and $\appaux$ 
are right-applicative and those for $\filter$
contain in their conditions the variable $p$ in active position.  
This definition of $\rep$ involves a $\la$-abstraction in a right
hand side. In Section~\ref{sec-exclus}, we prove confluence of the
relation $\a_{\b \cup \R(\b)}$ induced by the whole system.

%%%%%%%%%%%%%%%%%%%%%%%%%%%%%%%%%%%%%%%%%%%%%%%%%%%%%%%%%%%%%%%%%%%%%%%%%%%
%\subsection{Realizability interpretation of the negative translation
%  of the axiom of choice}
%%%%%%%%%%%%%%%%%%%%%%%%%%%%%%%%%%%%%%%%%%%%%%%%%%%%%%%%%%%%%%%%%%%%%%%%%%%

%% file: confl.tex
%%%%%%%%%%%%%%%%%%%%%%%%%%%%%%%%%%%%%%%%%%%%%%%%%%%%%%%%%%%%%%%%%%%%%%%%%%%
\subsection{Confluence}
\label{sec-confl}
%%%%%%%%%%%%%%%%%%%%%%%%%%%%%%%%%%%%%%%%%%%%%%%%%%%%%%%%%%%%%%%%%%%%%%%%%%%

%%%%%%%%%%%%%%%%%%%%%%%%%%%%%%%%%%%%%%%%%%%%%%%%%%%%%%%%%%%%%%%%%%%%%%%%%%%
%\paragraph{General notions.}
%%%%%%%%%%%%%%%%%%%%%%%%%%%%%%%%%%%%%%%%%%%%%%%%%%%%%%%%%%%%%%%%%%%%%%%%%%%
The main property on rewrite relations studied in this paper
is {\em confluence}.
The confluence of a relation $\a$ which has at least two
distinct normal forms entails the coherence of
the conversion $\alr$.
Moreover, it allows to evaluate terms in a modular way:
the choice of the subterm to be evaluated first
has no impact on the result of the evaluation.

In this section we recall some well-known facts about confluence
which will be useful in the following.

A sufficient condition for confluence is the {\em diamond property}.

%%%%%%%%%%%%%%%%%%%%%%%%%%%%%%%%%%%%%%%%%%%%%%%%%%%%%%%%%%%%%%%%%%%%%%%%%%%
\begin{definition}[Confluence]
%%%%%%%%%%%%%%%%%%%%%%%%%%%%%%%%%%%%%%%%%%%%%%%%%%%%%%%%%%%%%%%%%%%%%%%%%%%
A rewrite relation $\a$ is {\em confluent} if
${\al^* \a^*} \sle {\a^* \al^*}$
and has the {\em diamond property} if
${\al\a} \sle {\a\al}$.
\comment{ % Long version
Let $\a$ be a rewrite relation on $\Te(\Si)$.
\begin{enumerate}
\item
We say that $\a$ is {\em confluent} if
%$\al^* \a^* \sle \a^* \al^*$.
for all $t,u,v \in \Te(\Si)$ such that $u \al^* t \a^* v$,
there exists $w \in \Te(\Si)$ such that $u \a^* w \al^* v$.

\item
We say that $\a$ has the {\em diamond property} if
%$\al \a \sle \a \al$.
for all $t,u,v \in \Te(\Si)$ such that $u \al t \a v$,
there exists $w \in \Te(\Si)$ such that $u \a w \al v$.
\end{enumerate}
} % Long Version

In diagrammatic form:
\[
\begin{array}{c !{\qquad\qquad} c}
% Confluence
\xymatrix@R=\xyS@C=\xyS{
  & \cdot \ar@{->}[dr]^{}_{*}
          \ar@{->}[dl]_{}^{*}
  & \\
    \cdot \ar@{..>}[dr]_{}^{*}
  & 
  & \cdot \ar@{..>}[dl]^{}_{*}
    \\
  & \cdot
  & \\
}
&
% Diamond proerty
\xymatrix@R=\xyS@C=\xyS{
  & \cdot \ar@{->}[dr]^{}
          \ar@{->}[dl]_{}
  & \\
    \cdot \ar@{..>}[dr]_{}
  & 
  & \cdot \ar@{..>}[dl]^{}
    \\
  & \cdot
  & \\
}
\\
  \text{Confluence}
& \text{Diamond property}
\end{array}
\]
\end{definition}

\noindent
The stratification of conditional rewrite relations
leads to fine-grained notions of confluence.

%%%%%%%%%%%%%%%%%%%%%%%%%%%%%%%%%%%%%%%%%%%%%%%%%%%%%%%%%%%%%%%%%%%%%%%%%%%
\begin{definition}[Stratified Confluences]
\label{def-strat-confl}
%%%%%%%%%%%%%%%%%%%%%%%%%%%%%%%%%%%%%%%%%%%%%%%%%%%%%%%%%%%%%%%%%%%%%%%%%%%
Assume that $(\a_i)_{i \in \N}$ are rewrite relations
and let ${\a} \ptesp\deq\ptesp {\bigcup_{i \in \N} \a_i}$.
We say that $\a$
is {\em level confluent} if for all $i \geq 0$ we have
${\al^*_{i} \a^*_{i}} \sle {\a^*_{i} \al^*_{i}}$;
and {\em shallow confluent} if for all $i,j \geq 0$ we have
${\al^*_{i} \a^*_{j}} \sle {\a^*_{j} \al^*_{i}}$.

In diagrammatic form:
\[
\begin{array}{c !{\qquad\qquad} c}
% Level Confluence
\xymatrix@R=\xyS@C=\xyS{
  & \cdot \ar@{->}[dr]^{i}_{*}
          \ar@{->}[dl]_{i}^{*}
  & \\
    \cdot \ar@{..>}[dr]_{i}^{*}
  & 
  & \cdot \ar@{..>}[dl]^{i}_{*}
    \\
  & \cdot
  & \\
}
&
% Shallow Confluence
\xymatrix@R=\xyS@C=\xyS{
  & \cdot \ar@{->}[dr]^{i}_{*}
          \ar@{->}[dl]_{j}^{*}
  & \\
    \cdot \ar@{..>}[dr]_{i}^{*}
  & 
  & \cdot \ar@{..>}[dl]^{j}_{*}
    \\
  & \cdot
  & \\
}
\\
\text{Level Confluence} & \text{Shallow Confluence}
\end{array}
\]
\end{definition}
%These usual notions are depicted in the following diagrams.

\noindent
Note that shallow confluence implies level confluence
which in turns implies confluence.
For instance, in Section~\ref{sec-exclus}
we show that $\a_{\b \cup \R(\b)}$ is shallow confluent
for some conditional rewrite systems $\R$ called orthonormal.
This entails their confluence.

%%%%%%%%%%%%%%%%%%%%%%%%%%%%%%%%%%%%%%%%%%%%%%%%%%%%%%%%%%%%%%%%%%%%%%%%%%%
\paragraph{Combinations of rewrite relations.}
%%%%%%%%%%%%%%%%%%%%%%%%%%%%%%%%%%%%%%%%%%%%%%%%%%%%%%%%%%%%%%%%%%%%%%%%%%%
Since we are interested in the confluence of the combination
of two rewrite relations (conditional rewriting and $\la$-calculus),
we will use the following notions.

%%%%%%%%%%%%%%%%%%%%%%%%%%%%%%%%%%%%%%%%%%%%%%%%%%%%%%%%%%%%%%%%%%%%%%%%%%%
\begin{definition}[Commutation]
%%%%%%%%%%%%%%%%%%%%%%%%%%%%%%%%%%%%%%%%%%%%%%%%%%%%%%%%%%%%%%%%%%%%%%%%%%%
A rewrite relation $\a_A$ {\em commutes} with a rewrite relation $\a_B$
if ${\al^*_A \a^*_B} \sle {\a^*_B \al^*_A}$.
\end{definition}

\noindent
The Hindley-Rosen Lemma is a simple but important tool
to prove the confluence of the combination of two rewrite relations.

%%%%%%%%%%%%%%%%%%%%%%%%%%%%%%%%%%%%%%%%%%%%%%%%%%%%%%%%%%%%%%%%%%%%%%%%%%%
\begin{lemma}[Hindley-Rosen]
\label{lem-hr}
%%%%%%%%%%%%%%%%%%%%%%%%%%%%%%%%%%%%%%%%%%%%%%%%%%%%%%%%%%%%%%%%%%%%%%%%%%%
If $\a_A$ and $\a_B$ are two confluent rewrite relations
that commute then $\a_{A \cup B}$ is confluent.
\end{lemma}

\noindent
The next simple lemma is useful to prove the commutation
of two relations.

%%%%%%%%%%%%%%%%%%%%%%%%%%%%%%%%%%%%%%%%%%%%%%%%%%%%%%%%%%%%%%%%%%%%%%%%%%%
\begin{lemma}
\label{lem-commut}
%%%%%%%%%%%%%%%%%%%%%%%%%%%%%%%%%%%%%%%%%%%%%%%%%%%%%%%%%%%%%%%%%%%%%%%%%%%
Let $\a_A$ and $\a_B$ be
two rewrite relations such that for all $t,u,v \in \Te(\Si)$,
if $u \al_A t \a_B v$ then there is a term $w$
such that $u \a^*_B w \al_A v$.
Then $\a_A$ commutes with $\a_B$.
\end{lemma}

\begin{proof}
We show (i) by induction on $\a^*_B$ and deduce (ii)
by induction on $\a^*_A$.
\[
\begin{array}{c !{\qquad} c}
\xymatrix@R=\xyTS@C=\xyS{
    \cdot \ar@{->}[rr]^{B}_{*}
          \ar@{->}[dd]_{A}
  & & \cdot \ar@{..>}[dd]^{A}
  \\
  \\
    \cdot \ar@{..>}[rr]_{B}^{*}
  & & \cdot
  \\
}
&
\xymatrix@R=\xyTS@C=\xyS{
    \cdot \ar@{->}[rr]^{B}_{*}
          \ar@{->}[dd]_{A}^{*}
  & & \cdot \ar@{..>}[dd]^{A}_{*}
  \\
  \\
    \cdot \ar@{..>}[rr]_{B}^{*}
  & & \cdot
  \\
}
\\
\text{(i)} & \text{(ii)}
\end{array}
\]
\end{proof}

%% file: over.tex
%%%%%%%%%%%%%%%%%%%%%%%%%%%%%%%%%%%%%%%%%%%%%%%%%%%%%%%%%%%%%%%%%%%%%%%%%%%
\section{Confluence: from unconditional to conditional rewriting}
\label{sec-over-gen}
%%%%%%%%%%%%%%%%%%%%%%%%%%%%%%%%%%%%%%%%%%%%%%%%%%%%%%%%%%%%%%%%%%%%%%%%%%%

In this section we state precisely
the known results from which this paper starts
and give a short overview of our results.
In Section~\ref{sec-rew} we review the results on the combination
of $\la$-calculus with {\em unconditional} rewriting that we
extend to conditional and $\b$-conditional rewriting in Section~\ref{sec-Aconfl}
and Section~\ref{sec-Bconfl} respectively.
In Section~\ref{sec-orth-cond} we recall a result on the confluence of
orthogonal normal rewrite relations, that we generalize
to orthonormal $\b$-conditional rewriting in Section~\ref{sec-exclus}.
We then give a short
overview of our results in Section~\ref{sec-overview}.

%%%%%%%%%%%%%%%%%%%%%%%%%%%%%%%%%%%%%%%%%%%%%%%%%%%%%%%%%%%%%%%%%%%%%%%%%%%
\subsection{Confluence of beta-reduction with unconditional rewriting}
\label{sec-rew}
%%%%%%%%%%%%%%%%%%%%%%%%%%%%%%%%%%%%%%%%%%%%%%%%%%%%%%%%%%%%%%%%%%%%%%%%%%%

Our results of Section~\ref{sec-Aconfl} and~\ref{sec-Bconfl}
on the preservation of confluence for the combination of $\la$-calculus
with (left-algebraic) {\em conditional} rewriting are extensions of similar results
on the combination of $\la$-calculus with (left-algebraic) 
{\em unconditional} rewriting.
We concentrate of two cases, both untyped, that we review in this section:
\begin{itemize}
\item In Section~\ref{sec-muller} we discuss
M{\"u}ller's result~\cite{muller92ipl} (stated in Theorem~\ref{thm-muller})
on left-linear rewriting.

\item In Section~\ref{sec-dough} we discuss
Dougherty's result~\cite{dougherty92ic}
(stated in Theorem~\ref{thm-dough})
on strongly $\b$-normalizing terms with arity conditions.
\end{itemize}

%%%%%%%%%%%%%%%%%%%%%%%%%%%%%%%%%%%%%%%%%%%%%%%%%%%%%%%%%%%%%%%%%%%%%%%%%%%
\subsubsection{Left-linear rewriting}
\label{sec-muller}
%%%%%%%%%%%%%%%%%%%%%%%%%%%%%%%%%%%%%%%%%%%%%%%%%%%%%%%%%%%%%%%%%%%%%%%%%%%

Using the example of surjective pairing~\cite{klop80phd},
we have recalled in Section~\ref{sec-ex-sp}
that the combination of a confluent non left-linear rewrite
system with $\b$-reduction may not be confluent.
An example has also been presented in~\cite{bm87popl},
which can be seen as an adaptation of an example due to Huet~\cite{huet80jacm}
concerning first-order rewriting.

%%%%%%%%%%%%%%%%%%%%%%%%%%%%%%%%%%%%%%%%%%%%%%%%%%%%%%%%%%%%%%%%%%%%%%%%%%%
\begin{example}[\cite{bm87popl}]
\label{ex-minus}
%%%%%%%%%%%%%%%%%%%%%%%%%%%%%%%%%%%%%%%%%%%%%%%%%%%%%%%%%%%%%%%%%%%%%%%%%%%
Consider the confluent rewrite system
\[
  \minus \esp x \esp x \quad\at_\minus\quad \zero
\qquad\qquad
  \minus \esp (\suc \esp x) \esp x \quad\at_\minus\quad (\suc \esp \zero)
~,
\]
and let
\[
  Y_\suc \quad\deq\quad
  (\la x. \suc \esp (x \esp x)) \esp (\la x. \suc \esp (x \esp x))
~.
\]
Since $Y_\suc \a_\b \suc \ptesp Y_\suc$,
we have the following unjoinable peak:
\[
  \xymatrix@C=\xyS@R=\xyS{
    & \minus \esp Y_\suc \esp Y_\suc
      \ar@{>}[dl]
      \ar@{>}[dr]
    \\
    \minus \esp (\suc \esp Y_\suc) \esp Y_\suc \ar@{>}[d]
    & & \zero
    \\
      \suc \esp \zero
  }
\]
\end{example}

%%%%%%%%%%%%%%%%%%%%%%%%%%%%%%%%%%%%%%%%%%%%%%%%%%%%%%%%%%%%%%%%%%%%%%%%%%%
\begin{remark}[Interpretation with B{\"o}hm trees]
%%%%%%%%%%%%%%%%%%%%%%%%%%%%%%%%%%%%%%%%%%%%%%%%%%%%%%%%%%%%%%%%%%%%%%%%%%%
A simple interpretation of this system is to see $Y_\suc$
as representing the "infinite integer" $\infty$,
the term $\minus \ptesp Y_\suc \ptesp Y_\suc$
representing the undefined operation $\infty - \infty$.
This interpretation can be made concrete by using
B{\"o}hm trees~\cite{barendregt84book}.
The B{\"o}hm tree of the term $Y_\suc$ is the infinite term
\[
\begin{array}{c}
  \suc \\
  \vline \\
  \suc \\
  \vline \\
  \vdots
\end{array}
\]
Intuitively,
Ex.~\ref{ex-minus} can be seen as an instance of the fact that
confluence of non left-linear systems is not preserved
when we extend the term algebra (in this case by infinite terms).
\end{remark}

As shown in~\cite{muller92ipl}, confluence is preserved when
rewriting is left-linear.
The original result concerns only algebraic systems, but 
can easily be extended to unconditional rewrite systems
with arbitrary right-hand sides.

%%%%%%%%%%%%%%%%%%%%%%%%%%%%%%%%%%%%%%%%%%%%%%%%%%%%%%%%%%%%%%%%%%%%%%%%%%%
\begin{theorem}[\cite{muller92ipl}]
\label{thm-muller}
%%%%%%%%%%%%%%%%%%%%%%%%%%%%%%%%%%%%%%%%%%%%%%%%%%%%%%%%%%%%%%%%%%%%%%%%%%%
If $\R$ is a left-linear {\em unconditional} rewrite system such that
$\a_{\R}$ is confluent then $\a_{\b \cup \R}$ is confluent.
\end{theorem}

\noindent
This result has been generalized to the case of
Higher-Order Rewrite Systems~\cite{or94lfcs}.

%%%%%%%%%%%%%%%%%%%%%%%%%%%%%%%%%%%%%%%%%%%%%%%%%%%%%%%%%%%%%%%%%%%%%%%%%%%
\subsubsection{Strongly beta-normalizing terms}
\label{sec-dough}
%%%%%%%%%%%%%%%%%%%%%%%%%%%%%%%%%%%%%%%%%%%%%%%%%%%%%%%%%%%%%%%%%%%%%%%%%%%

To handle non left-linear systems, as seen in Example~\ref{ex-minus}
we have to forbid infinite terms.
This is possible for example by focusing on algebraic
rewriting on typed terms.
Confluence is preserved for the combination of algebraic
rewriting with simply typed $\la$-calculus~\cite{breazu88lics}.
This result has been then extended to the polymorphic
$\la$-calculus~\cite{bg89icalp,bg94ic,okada89issac}.

A question arises from these results: besides strong normalization
of $\b$-reduction, what is the role of typing in the preservation
of confluence ?
This is studied in~\cite{dougherty92ic},
which shows that for algebraic rewriting
terms must satisfy some {\em arity conditions}.

%%%%%%%%%%%%%%%%%%%%%%%%%%%%%%%%%%%%%%%%%%%%%%%%%%%%%%%%%%%%%%%%%%%%%%%%%%%
\begin{example}
\label{ex-dough}
%%%%%%%%%%%%%%%%%%%%%%%%%%%%%%%%%%%%%%%%%%%%%%%%%%%%%%%%%%%%%%%%%%%%%%%%%%%
Consider the rewrite system
\[
  \id \esp x \quad\at_\id\quad x
~,
\]
and let $\W_\suc \esp\deq\esp \la x.\suc \esp (x \esp x)$.
The term
\[
t \quad\deq\quad
\minus
\esp
  (\id \esp \W_\suc \esp \W_\suc)
\esp
  (\id \esp \W_\suc \esp \W_\suc)
\]
is in $\b$-normal form, hence strongly $\b$-normalizing.
Moreover, we can check that the rewrite system
$\at_{\minus \cup \id}$ is confluent.
However, $\at_{\b \cup \minus \cup \id}$
is not confluent since $t$ rewrites to the unjoinable peak
of Example~\ref{ex-minus}:
\[
\minus
\ptesp
  (\id \esp \W_\suc \esp \W_\suc)
\ptesp
  (\id \esp \W_\suc \esp \W_\suc)
\quad\a^2_\id\quad  
\minus \esp Y_\suc \esp Y_\suc
~.
\]
\end{example}

\noindent
The problem is that reducing $\id$ in the $\b$-normal form
$\id \ptesp \W_\suc \ptesp \W_\suc$ leads to a term
which is no longer in $\b$-normal form:
rewriting does not preserve $\b$-normal forms.
The approach taken in~\cite{dougherty92ic} is to find
{\em arity conditions} on terms for rewriting to preserve $\b$-normal forms.
Consider a symbol $\sff \in \Si$
such that for all $\sff \vl \at_\R r$, we have $|\vl| \leq n$.
Then we discard terms of the form $\sff \vt$
with $|\vt| > n$.
For example, the term
$\id \esp \W_\suc \esp \W_\suc$
is not allowed since $\id$ takes two arguments
whereas its rewrite rule takes only one.

%%%%%%%%%%%%%%%%%%%%%%%%%%%%%%%%%%%%%%%%%%%%%%%%%%%%%%%%%%%%%%%%%%%%%%%%%%%
\begin{definition}[Applicative Arity]
\label{def-arity}
%%%%%%%%%%%%%%%%%%%%%%%%%%%%%%%%%%%%%%%%%%%%%%%%%%%%%%%%%%%%%%%%%%%%%%%%%%%
An {\em arity} is a function $\ap : \Si \a \N$.
\begin{enumerate}[(i)]
\item
\label{def-arity-term}
A term $t$ {\em respects $\ap$}
if it contains no subterm $\sff \vt$
with $|\vt| > \ap(\sff)$.

\item
\label{def-arity-rule}
A rewrite system $\R$ respects $\ap$ if 
for all $\sff \vl \at_\R r$,
$\sff \vl$ and $r$ respect $\ap$
and moreover $|\vl| = \ap(\sff)$.
\end{enumerate}
\end{definition}

However, the respect of an arity is not stable by
$\b$-reduction.
For instance,
with $\ap(\id) = 1$ the term
$(\la x.x \ptesp \W_\suc \ptesp \W_\suc) \ptesp \id$
respects $\ap$ but it $\b$-reduces to 
$\id \ptesp \W_\suc \ptesp \W_\suc$ which does not respect $\ap$.
In order to work with terms which respect an arity $\ap$
and whose $\b\R$-reducts respect $\ap$ too, it is convenient
to consider sets of terms respecting $\ap$ and which are stable by
reduction. This motivates the following definition.

%%%%%%%%%%%%%%%%%%%%%%%%%%%%%%%%%%%%%%%%%%%%%%%%%%%%%%%%%%%%%%%%%%%%%%%%%%%
\begin{definition}[$(\R,\ap)$-Stable Terms]
\label{def-rstable}
%%%%%%%%%%%%%%%%%%%%%%%%%%%%%%%%%%%%%%%%%%%%%%%%%%%%%%%%%%%%%%%%%%%%%%%%%%%
Given an arity $\ap$ and a rewrite system $\R$,
a set of terms $S$ is $(\R,\ap)$-stable if
\begin{enumerate}[(i)]
\item for all $t \in S$, $t$ respects $\ap$,
\item for all $t \in S$, if $t \a_{\b \cup \R} u$ then $u \in S$,
\item for all $t \in S$, if $u$ is a subterm of $t$
  then $u \in S$.
\end{enumerate}
\end{definition}

Dougherty~\cite{dougherty92ic}
obtain the preservation of confluence on $(\R,\ap)$-stable
sets of strongly $\b$-normalizing terms.
%The result of~\cite{dougherty92ic} is the following.

%%%%%%%%%%%%%%%%%%%%%%%%%%%%%%%%%%%%%%%%%%%%%%%%%%%%%%%%%%%%%%%%%%%%%%%%%%%
\begin{theorem}[\cite{dougherty92ic}]
\label{thm-dough}
%%%%%%%%%%%%%%%%%%%%%%%%%%%%%%%%%%%%%%%%%%%%%%%%%%%%%%%%%%%%%%%%%%%%%%%%%%%
If $\R$ is an algebraic confluent {\em unconditional} rewrite system
that respects an arity $\ap$,
then $\a_{\b \cup \R}$ is confluent on every 
$(\R,\ap)$-stable set $S \sle \SN_{\b}$.
\end{theorem}

%%%%%%%%%%%%%%%%%%%%%%%%%%%%%%%%%%%%%%%%%%%%%%%%%%%%%%%%%%%%%%%%%%%%%%%%%%%
\begin{remark}
\label{rem-alg-rhs}
%%%%%%%%%%%%%%%%%%%%%%%%%%%%%%%%%%%%%%%%%%%%%%%%%%%%%%%%%%%%%%%%%%%%%%%%%%%
To get the preservation $\b$-normal forms by rewriting it is necessary
to restrict to algebraic right-hand sides, since
in contrast with algebraic terms, substituting a variable
in an applicative term may produce a $\b$-redex.
For instance $(x \ptesp z)\wth{x}{\la y.y} = (\la y.y) \ptesp z$.
%The method of~\cite{dougherty92ic} to prove the preservation of confluence
%is to show that $\b \cup \R$-steps can be project to $\R$-steps
%on $\b$-normal forms. This suppose that rewriting preserves $\b$-normal forms.
%This is need arity assumptions as shown in Example~\ref{ex-dough}.
%This also needs rewriting to be algebraic.
%In particular, it is not sufficient to restrict to right-applicative
%rewriting since
%in contrast with algebraic terms, substituting a variable
%in an applicative term may produce a $\b$-redex.
%For instance $(x \ptesp z)\wth{x}{\la y.y} = (\la y.y) \ptesp z$.
\end{remark}

%%%%%%%%%%%%%%%%%%%%%%%%%%%%%%%%%%%%%%%%%%%%%%%%%%%%%%%%%%%%%%%%%%%%%%%%%%%
\subsection{Orthogonal conditional rewriting}
\label{sec-orth-cond}
%%%%%%%%%%%%%%%%%%%%%%%%%%%%%%%%%%%%%%%%%%%%%%%%%%%%%%%%%%%%%%%%%%%%%%%%%%%

Orthogonality is a sufficient condition for the confluence
of some kinds of conditional rewriting.
In this section we recall some known results about the confluence
of algebraic orthogonal conditional rewrite systems.
They were initially formulated in the framework of first-order
conditional rewriting, of which algebraic rewriting
is an instance.
The main result is the shallow confluence of orthogonal normal
conditional rewriting. We generalize it in Section~\ref{sec-exclus}
to orthonormal $\b$-conditional rewriting.

For {\em unconditional} rewriting, orthogonality is a simple syntactic 
criterion:
it entails the confluence of left-linear systems with no
critical pairs~\cite{huet80jacm}.
With conditional rewriting, things get more complicated
since
the notion of critical pairs has to take into account the conditions
of rewrite rules.

%%%%%%%%%%%%%%%%%%%%%%%%%%%%%%%%%%%%%%%%%%%%%%%%%%%%%%%%%%%%%%%%%%%%%%%%%%%
\begin{definition}[Conditional Critical Pairs]
\label{def-cond-cp}
%%%%%%%%%%%%%%%%%%%%%%%%%%%%%%%%%%%%%%%%%%%%%%%%%%%%%%%%%%%%%%%%%%%%%%%%%%%
Let $\R$ be a set of conditional rules and
suppose that $\r_1 : \vd = \vc \sgt l\at r$ and 
$\r_2 : \vd' =  \vc' \sgt l' \at r'$ 
are two renaming of rules in $\R$ such that they have no
variable in common.
If $p$ is a non-variable occurrence of $l$ and $\s$ is a
most general unifier of $l|_p$ and $l'$, then
\[
  \vd\s \esp=\esp \vc\s
\quad\land\quad
  \vd'\s \esp=\esp \vc'\s
\quad\sgt\quad
  \left(l[p \gets r']\s ~,~ r\s \right)
\]
is a {\em conditional critical pair} of $\R$.
If $\rho_1$ and $\rho_2$ are renaming of the same rule,
we assume that $p$ is not the root position of $l$.
A critical pair of the form $\vd = \vc \sgt (s,s)$ is called {\em trivial}.
\end{definition}

The important point is that in a conditional critical pair
$\vd = \vc \sgt (s,t)$,
it is possible that there is no substitution $\s$ such that $\vd\s = \vc\s$.
Thus, critical pairs can be {\em feasible} or {\em unfeasible}.
According to the kind of conditional rewriting considered
(with and without $\b$-reduction in the evaluation of conditions),
the satisfaction of conditions is done differently.
Therefore, we consider two notions of feasibility.

%%%%%%%%%%%%%%%%%%%%%%%%%%%%%%%%%%%%%%%%%%%%%%%%%%%%%%%%%%%%%%%%%%%%%%%%%%%
\begin{definition}[Feasibility of Conditional Critical Pairs]
%%%%%%%%%%%%%%%%%%%%%%%%%%%%%%%%%%%%%%%%%%%%%%%%%%%%%%%%%%%%%%%%%%%%%%%%%%%
A critical pair $\vd = \vc \sgt (s,t)$ of a conditional system $\R$ is
\begin{itemize}
\item {\em feasible} if there is a substitution $\s$ such that
  $\vd\s \ad_{\R} \vc\s$;

\item {\em $\b$-feasible} if there is a substitution $\s$ such that
  $\vd\s \ad_{\b \cup \R(\b)} \vc\s$;
\end{itemize}
A critical pair which is not feasible (resp. $\b$-feasible)
is said {\em unfeasible} (resp. {\em $\b$-unfeasible}).
\end{definition}

The easiest way to prove unfeasibility of critical pairs is often to 
use confluence.
We come back on this question in Section~\ref{sec-exclus}.
Both notions of feasibility induce a notion of orthogonality.

%%%%%%%%%%%%%%%%%%%%%%%%%%%%%%%%%%%%%%%%%%%%%%%%%%%%%%%%%%%%%%%%%%%%%%%%%%%
\begin{definition}[Orthogonality]
%%%%%%%%%%%%%%%%%%%%%%%%%%%%%%%%%%%%%%%%%%%%%%%%%%%%%%%%%%%%%%%%%%%%%%%%%%%
A set $\R$ of left-linear conditional rewrite rules is
\begin{itemize}
\item {\em orthogonal}
(resp. {\em $\b$-orthogonal}) if all its critical pairs are unfeasible
(resp. {\em $\b$-unfeasible});

\item {\em weakly orthogonal} (resp. {\em weakly $\b$-orthogonal})
if all its critical pairs are either trivial or
unfeasible (resp. {\em $\b$-unfeasible}).
\end{itemize}
%%  and result from a root superposition (i.e. an overlay).
\end{definition}

Hence, to test the orthogonality of a conditional system,
we have to evaluate the conditions of its critical pairs.
According to Remark~\ref{rem-dec}, this is in general undecidable.

%%%%%%%%%%%%%%%%%%%%%%%%%%%%%%%%%%%%%%%%%%%%%%%%%%%%%%%%%%%%%%%%%%%%%%%%%%%
%\renewcommand\fntext{This is also the case of semi-equational conditional rewriting,
%not considered in this paper~\cite{bk86jcss,ohlebusch02book}}
%%%%%%%%%%%%%%%%%%%%%%%%%%%%%%%%%%%%%%%%%%%%%%%%%%%%%%%%%%%%%%%%%%%%%%%%%%%
It is well-known that for normal (and semi-equational) rewriting,
weak orthogonality implies confluence.
This in general {\em not} the case for join conditional rewriting,
as shown in~\cite{bk86jcss}.

%%%%%%%%%%%%%%%%%%%%%%%%%%%%%%%%%%%%%%%%%%%%%%%%%%%%%%%%%%%%%%%%%%%%%%%%%%%
\begin{theorem}[\cite{bk86jcss,ohlebusch02book}]
\label{thm-orth-cond}
%%%%%%%%%%%%%%%%%%%%%%%%%%%%%%%%%%%%%%%%%%%%%%%%%%%%%%%%%%%%%%%%%%%%%%%%%%%
Let $\R$ be a conditional rewrite system.
If $\R$ is weakly orthogonal, and moreover 
is a normal system, then $\a_{\R}$ is shallow confluent.
\end{theorem}

%%%%%%%%%%%%%%%%%%%%%%%%%%%%%%%%%%%%%%%%%%%%%%%%%%%%%%%%%%%%%%%%%%%%%%%%%%%
\subsection{Overview of the results}
\label{sec-overview}
%%%%%%%%%%%%%%%%%%%%%%%%%%%%%%%%%%%%%%%%%%%%%%%%%%%%%%%%%%%%%%%%%%%%%%%%%%%

The goal of this paper is to give sufficient conditions for the 
confluence of $\beta$-reduction with $\b$-conditional rewriting
(i.e.\ with $\b$-steps allowed in the evaluation of conditions).

More precisely, we seek to obtain results on the
{\em preservation} of confluence, that is to get the confluence
of $\a_{\b \cup \R(\b)}$ assuming the confluence of $\a_\R$.
Our approach is to generalize the results summarized in Section~\ref{sec-rew}
on the combination of $\b$-reduction with unconditional rewriting.
%to its combination with $\b$-conditional rewriting.
We thus consider two different cases:
\begin{itemize}
\item First, the extension of M{\"u}ller's result~\cite{muller92ipl},
when $\b$-reduction is not restricted
(we thus need to assume left-linearity, and to extend this notion to
conditional rewriting).

\item Second, the extension of Dougherty's result~\cite{dougherty92ic},
when we restrict to $\b$-normalizing terms
(we thus need some arity conditions on terms).
In fact, we improve~\cite{dougherty92ic} from strongly $\b$-normalizing
terms to weakly $\b$-normalizing terms.
\end{itemize}

In each case, we proceed in two steps.
We first consider in Section~\ref{sec-Aconfl}
the case of $\b$-reduction with conditional rewriting
$\a_{\R}$ (when $\b$-reduction is not allowed in the evaluation of conditions).
We then extend these results to $\b$-conditional rewriting $\a_{\R(\b)}$ in
Section~\ref{sec-Bconfl}.

As discussed at the beginning of Section~\ref{sec-Bconfl}
(see Example~\ref{ex-Bconfl}), for the extension of both~\cite{muller92ipl}
and~\cite{dougherty92ic} to $\b$-conditional rewriting,
rewrite rules must be algebraic
and respect arity conditions.
In contrast, the extension of~\cite{muller92ipl} to the simpler
case of conditional rewriting holds without these restrictions.
This motivates the definition of criteria for the confluence
of $\b$-reduction with $\b$-conditional rewriting when rules need
not be algebraic nor to respect an arity
(recall from Definition~\ref{def-cond-rules} that left-hand sides are always
algebraic in this paper).
We propose such a criterion in Section~\ref{sec-exclus}, which defines
{\em orthonormal} conditional rewriting, an extension of orthogonal rewriting.
We show the {\em shallow} confluence of $\b$-reduction with
$\b$-conditional rewriting for these systems, hence extending
Theorem~\ref{thm-orth-cond}.

Our results are summarized in 
Figure~\ref{fig-oversimple} page~\pageref{fig-oversimple}.
%See Figure~\ref{fig-summary} page~\pageref{fig-summary} for precise statements.

%% file: Aconfl.tex
%%%%%%%%%%%%%%%%%%%%%%%%%%%%%%%%%%%%%%%%%%%%%%%%%%%%%%%%%%%%%%%%%%%%%%%%%%%
\section{Confluence of beta-reduction with conditional rewriting}
  \label{sec-Aconfl}
  %\secttoc
%%%%%%%%%%%%%%%%%%%%%%%%%%%%%%%%%%%%%%%%%%%%%%%%%%%%%%%%%%%%%%%%%%%%%%%%%%%

We now turn to conditional rewriting. 
In this section we focus on the combination of
join conditional rewriting $\a_\R$ with $\b$-reduction:
we do not allow the use of $\b$-reduction in the evaluation of conditions.

%%%%%%%%%%%%%%%%%%%%%%%%%%%%%%%%%%%%%%%%%%%%%%%%%%%%%%%%%%%%%%%%%%%%%%%%%%%
%\begin{notation}[Join Conditional Rewriting]
%%%%%%%%%%%%%%%%%%%%%%%%%%%%%%%%%%%%%%%%%%%%%%%%%%%%%%%%%%%%%%%%%%%%%%%%%%%
%We denote by $\a_\R$ the join conditional rewrite relation
%issued from a conditional rewrite system $\R$.
%\end{notation}

The important point of left-linearity 
is to prevent {\em unconditional} rewriting from comparing arbitrary terms.
It forbids in particular comparisons of infinite terms such as $Y_\suc$.
But with conditional rewriting, rewrite rules can make
this comparison in their conditions while being left-linear.
Hence, starting from Example~\ref{ex-minus}, we can define a left-linear
conditional rewrite system which makes the commutation of
rewriting with $\b$-reduction fail.

%%%%%%%%%%%%%%%%%%%%%%%%%%%%%%%%%%%%%%%%%%%%%%%%%%%%%%%%%%%%%%%%%%%%%%%%%%%
\begin{example}
\label{ex-minus-cond}
%%%%%%%%%%%%%%%%%%%%%%%%%%%%%%%%%%%%%%%%%%%%%%%%%%%%%%%%%%%%%%%%%%%%%%%%%%%
Consider the conditional system
\[
  x = y \esp\sgt\esp \minus \esp x \esp y
  \esp\at\esp \zero
\qquad\qquad
  x = (\suc \esp y) \esp\sgt\esp \minus \esp x \esp y
  \esp\at\esp (\suc \esp \zero)
~.
\]
This system is left-linear, but the conditions
can test the equality of open terms.
The join conditional rewrite 
relation issued from this system forms with $\a_\b$
the following unjoinable peak:
\[
\xymatrix@R=\xyS@C=\xyTS{
%\xymatrix{
  &
  & \minus \esp Y_{\suc} \esp Y_{\suc} 
    \ar@{>}[dl]
    \ar@{>}[dr]
  & (Y_\suc \esp\ad\esp Y_\suc) 
  \\
    ((\suc \esp Y_\suc) \esp\ad\esp (\suc \esp Y_\suc))
  & \minus \esp (\suc \esp Y_{\suc}) \esp Y_{\suc} \ar@{>}[d]
  & & \zero
  \\
  & (\suc \esp \zero)
}
\]
\end{example}

As for unconditional rewriting in Section~\ref{sec-rew},
we consider two ways to overcome the problem:
to restrict rewriting (Section~\ref{sec-Aconfl-sc})
or to restrict $\b$-reduction (Section~\ref{sec-Aconfl-wn}).

%%%%%%%%%%%%%%%%%%%%%%%%%%%%%%%%%%%%%%%%%%%%%%%%%%%%%%%%%%%%%%%%%%%%%%%%%%%
\subsection{Confluence for left-linear semi-closed systems}
\label{sec-Aconfl-sc}
%%%%%%%%%%%%%%%%%%%%%%%%%%%%%%%%%%%%%%%%%%%%%%%%%%%%%%%%%%%%%%%%%%%%%%%%%%%

In this section we are interested in the extension of
Theorem~\ref{thm-muller} (\cite{muller92ipl}) to conditional
rewriting: we want sufficient conditions on rewrite rules
for the preservation of confluence on {\em all untyped terms of $\Te(\Si)$}.
As seen in Example~\ref{ex-minus-cond}, for conditional rewriting
we have to extend the notion of left-linearity
in order to forbid comparison of open terms in the conditions
of rewrite rules.
To this end we restrict to {\em semi-closed} conditional rewrite rules.

%%%%%%%%%%%%%%%%%%%%%%%%%%%%%%%%%%%%%%%%%%%%%%%%%%%%%%%%%%%%%%%%%%%%%%%%%%%
\begin{definition}[Semi-Closed Conditional Rewrite Rules]
\label{def-sc-cond}
%%%%%%%%%%%%%%%%%%%%%%%%%%%%%%%%%%%%%%%%%%%%%%%%%%%%%%%%%%%%%%%%%%%%%%%%%%%
A conditional rewrite system $\R$ is {\em semi-closed}
if for all rules
\[
  d_1 \esp=\esp c_1
  \esp\land\esp \dots \esp\land\esp
  d_n \esp=\esp c_n
  \esp\sgt\esp l \dbesp\at_\R\dbesp r
  ~,
\]
the terms $c_1,\dots,c_n$ are applicative and closed.
\end{definition}

For example, the system {\sf Tree} of Section~\ref{sec-ex}
is left-linear and semi-closed.
In a semi-closed rule $\vd = \vc \sgt l \at r$, since $\vc$ are closed terms, 
it is tempting to normalize them and obtain a normal rewrite relation,
but as noted in Remark~\ref{rem-dec},
results on join rewriting seem more easily applicable.

We show that the confluence of $\a_\R$ implies the confluence of $\a_{\b\cup\R}$
for semi-closed left-linear 
right-applicative systems (Theorem~\ref{thm-Aconfl-sc}).
Using Hindley-Rosen lemma (Lemma~\ref{lem-hr}), this follows
from the commutation of conditional rewriting with $\b$-reduction.
As in~\cite{muller92ipl}, we obtain this commutation as a consequence
of the commutation of conditional rewriting with a relation $\rpr_\b$ of
{\em parallel} $\b$-reduction (see Definition~\ref{def-par-rew}).
This is shown in Lemma~\ref{lem-commut-cond}, which relies
on Property~\ref{prop-par-rew}
($\s \rpr_\b \s'$ implies $t\s \rpr_\b t\s'$).
This property holds for parallel rewrite relations but fails with $\a_\b$.

The parallel $\b$-reduction $\rpr_\b$ we use is
Tait and Martin-L{\"o}f's relation~\cite{barendregt84book,takahashi95ic}.
It is defined as follows.

%%%%%%%%%%%%%%%%%%%%%%%%%%%%%%%%%%%%%%%%%%%%%%%%%%%%%%%%%%%%%%%%%%%%%%%%%%%
\begin{definition}[Parallel $\b$-Reduction]
%%%%%%%%%%%%%%%%%%%%%%%%%%%%%%%%%%%%%%%%%%%%%%%%%%%%%%%%%%%%%%%%%%%%%%%%%%%
We let $\rpr_\b$ be the smallest parallel rewrite relation closed under the rule
\[
(\pBeta)~
  \dfrac{t_1 ~\rpr_\b~ u_1 \qquad\qquad t_2 ~\rpr_\b~ u_2}
        {(\la x.t_1)t_2 ~\rpr_\b~ u_1\wth{x}{u_2}}
\]
\end{definition}

We will use some well-known properties of $\rpr_\b$. If
$\s \rpr_\b \s'$ then $s\s \rpr_\b s\s'$; this is the one-step
reduction of parallel redexes. We can also simulate $\b$-reduction:
$\a_\b \sle \rpr_\b \sle \a^*_\b$. And third, $\rpr_\b$ enjoys the
diamond property:
$\rpl_\b \rpr_\b \sle \rpr_\b \rpl_\b$.
%This corresponds to the fact that any complete development of 
%$\a_\b$ can be done in {\em one} $\rpr_\b$-step.

The relation $\rpr_\b$ is stronger than the one used
in~\cite{muller92ipl}: it can reduce in one step
nested $\b$-redexes, while the relation of~\cite{muller92ipl}
is simply the smallest
parallel rewrite relation containing $\b$-reduction
(i.e.\ the {\em parallel closure} of $\a_\b$).
The diamond property (which holds for $\rpr_\b$)
fails for the parallel closure of $\b$-reduction precisely
because it cannot reduce nested $\b$-redexes in one step.
The parallel closure of $\a_\b$ 
would have been sufficient to obtain Lemma~\ref{lem-commut-cond},
but we use the nested relation $\rpr_\b$ because we rely on the
diamond property in Section~\ref{sec-Bconfl-sc}.

Nested parallelizations (corresponding to complete developments) are already
used in~\cite{or94lfcs} for their confluence proof of HORSs.
However, our method inherits more from~\cite{muller92ipl}
than from~\cite{or94lfcs},
as we use complete developments of $\a_\b$ only,
whereas complete developments of $\a_\b$ {\em and} of $\a_{\R}$ are used 
for the modularity result of~\cite{or94lfcs}.
%For more information on $\rpr_\b$, the interested reader may 
%consult~\cite{takahashi95ic}.

The left-linearity assumption is used in the proof of Lemma~\ref{lem-commut-cond}
via the following property of linear algebraic terms.

%%%%%%%%%%%%%%%%%%%%%%%%%%%%%%%%%%%%%%%%%%%%%%%%%%%%%%%%%%%%%%%%%%%%%%%%%%%
\begin{proposition}
\label{prop-subst-lin-rhd}
%%%%%%%%%%%%%%%%%%%%%%%%%%%%%%%%%%%%%%%%%%%%%%%%%%%%%%%%%%%%%%%%%%%%%%%%%%%
Let $t$ be an algebraic linear term and $\s$ be a substitution
such that $t\s \rpr_\b u$.
There is a substitution $\s'$ such that $u = t\s'$
with $\s \rhd_\b \s'$
and $\s'(x) = \s(x)$ for all $x \notin \FV(t)$.
\end{proposition}

\begin{proof}
By induction on $t$.
\begin{description}
\item[$t = x \in \Vte$.]
  In this case $t\s = \s(x)$.
  Take $\s'$ such that $\s'(x) = u$ and $\s'(y) = \s(y)$
  for all $y \neq x$.

\item[$t = \sff \in \Si$.]
  In this case $t\s = t = u$, hence $\s' = \s$ fits
  (recall that $\rpr_\b$ is reflexive).

\item[$t = t_1 t_2$.]
  Since $t$ is algebraic, $t_1\s \esp t_2\s$ is not a $\b$-redex.
  It follows that $u$ is of the form $u_1 \esp u_2$
  with $(t_1\s,t_2\s) \rpr_\b (u_1,u_2)$.
  By induction hypothesis, there are two substitutions
  $\s'_1$ and $\s'_2$ such that for each $i \in \{1,2\}$ we have
  $\s \rhd_\b \s'_i$, $u_i = t_i\s'_i$, 
  and $\s'_i(x) = \s(x)$ for all $x \notin \FV(t_i)$.
  Since $t$ is linear, $\FV(t_1) \cap \FV(t_2) = \emptyset$,
  hence with $\s' \deq \s'_1 \uplus \s'_2$, we have
  $u = u_1 u_2 = t_1\s' \esp t_2\s' = t\s'$,
  $\s \rhd_\b \s'$
  and $\s(y) = \s'(y)$ for all $y \notin \FV(t)$.
  \qedhere
\end{description}
\end{proof}

We are now ready to prove the commutation of
$\a_{\R}$ and $\rpr_\b$.
In fact we prove a slightly stronger statement,
which can be termed as the "level commutation" of
$\a_{\R}$ and $\rpr_\b$.

%%%%%%%%%%%%%%%%%%%%%%%%%%%%%%%%%%%%%%%%%%%%%%%%%%%%%%%%%%%%%%%%%%%%%%%%%%%
\begin{lemma}[Commutation of $\a_{\R_i}$ with $\rpr_\b$]
\label{lem-commut-cond}
%%%%%%%%%%%%%%%%%%%%%%%%%%%%%%%%%%%%%%%%%%%%%%%%%%%%%%%%%%%%%%%%%%%%%%%%%%%
If $\R$ is a conditional rewrite system which is
semi-closed, left-linear and right-applicative,
then $\rpr_\b$ commutes with $\a_{\R_i}$ for all $i \in \N$:
%In diagrammatic from:
\[
\xymatrix@R=\xyS@C=\xyL{
    \cdot \ar@{->}[r]^{\R_{i}}_{*}
          \ar@{->}[d]_{\rpr_\b}^{*}
  & \cdot \ar@{.>}[d]^{\rpr_\b}_{*} \\
    \cdot \ar@{.>}[r]_{\R_{i}}^{*}
  & \cdot
}
\]
\end{lemma}

\begin{proof}
We reason by induction on $i \in \N$.
The base case $i = 0$ is trivial.
Let $i \geq 0$ and assume that $\a_{\R_i}$ commutes with $\rpr_\b$.
We show that $\a_{\R_{i+1}}$ commutes with $\rpr_\b$.
%Let $i \geq 0$ and assume that $\a_{\R_i}$ commutes with $\a_\b$.
%We show that $\a_{\R_{i+1}}$ commutes with $\a_\b$.

We begin by showing that for all
$t,u,v \in \Te(\Si)$,
if $u \rpl_\b t \a_{\R_{i+1}} v$
then there is a term $w$ such that
$u \a^*_{\R_{i+1}} w \rpl_\b v$.
In diagrammatic form:
\begin{equation}
\label{eq-lem-commut-cond-base}
\begin{aligned}
\xymatrix@R=\xyS@C=\xyL{
    t \ar@{->}[r]^{\R_{i+1}}
      \ar@{->}[d]_{\rpr_\b}
  & v \ar@{.>}[d]^{\rpr_\b} \\
    u \ar@{.>}[r]_{\R_{i+1}}^{*}
  & w
}
\end{aligned}
\end{equation}
%Since ${\a^*_\b} = {\rpr^*_\b}$,
We deduce from~(\ref{eq-lem-commut-cond-base})
%that $\a_{\R_{i+1}}$ commutes with $\a_\b$
that $\a_{\R_{i+1}}$ commutes with $\rpr_\b$
by applying Lemma~\ref{lem-commut}.

We now show~(\ref{eq-lem-commut-cond-base})
by induction on $t$.
If both reductions $t \rpr_\b u$ and $t \a_{\R_{i+1}} v$
occur in a proper subterm of $t$ then we conclude by induction hypothesis.
Otherwise there are two cases.
\begin{enumerate}
\item
$t = (\la x.t_1)t_2$ with
$u = u_1\wth{x}{u_2}$ and $v = (\la x.v_1)v_2$
where $(u_1,u_2) \rpl_\b (t_1,t_2) \a_{\R_{i+1}} (v_1,v_2)$.
By induction hypothesis, there are terms $w_1$ and $w_2$
such that $u_i \a^*_{\R_{i+1}} w_i \rpl_\b v_i$.
We deduce that
$u_1\wth{x}{u_2} \a^*_{\R_{i+1}} w_1\wth{x}{w_2}$
and that
$(\la x.v_1)v_2 \rpr_\b w_1\wth{x}{w_2}$.

\item
$t$ is the $\R$-redex contracted in the step $t \a_{\R_{i+1}} v$.
In this case, there is a conditional rule $\vd = \vc \sgt l \at_\R r$
and a substitution $\s$ s.t.\ $t = l\s$ and $v = r\s$.
Moreover, there are terms $\vv$ such that
$\vd\s \a^*_{\R_{i}} \vv \al^*_{\R_{i}} \vc\s$.
Since $\R$ is semi-closed, the terms $\vc$ are closed and applicative,
hence $\vc\s = \vc$ and the terms $\vv$ are applicative since
$\R$ is right-applicative.
Since $l$ is left-linear and algebraic,
we deduce from Proposition~\ref{prop-subst-lin-rhd}
that there is a substitution $\s'$ such that $\s \rhd_\b \s'$ and $u=l\s'$.
It follows Proposition~\ref{prop-par-rew} that
$r\s \rhd_\b r\s'$ and $\vd\s \rhd_\b \vd\s'$.
%hence $\vd\s \a^*_\b \vd\s'$.

To conclude that $w \deq r\s'$ fits, it remains to show that
$l\s' \a_{\R_{i+1}} r\s'$,
that is $\vd\s' \ad_{\R_{i}} \vc$.
%We have $\vd\s' \al^*_\b \vd\s \a^*_{\R_{i}} \vv$,
We have $\vd\s' \rpl_\b \vd\s \a^*_{\R_{i}} \vv$,
hence by induction hypothesis there are $\vw$ s.t.\
%$\vd\s' \a^*_{\R_{i}} \vw \al^*_{\b} \vv$.
$\vd\s' \a^*_{\R_{i}} \vw \rpl^*_{\b} \vv$.
It follows that $\vd\s' \a^*_{\R_{i}} \vv$, the terms $\vv$
being applicative hence in $\b$-normal form.
We thus have $\vd\s' \ad_{\R_{i}} \vc$,
and conclude that $l\s' \a_{\R_{i+1}} r\s'$.
\qedhere
\end{enumerate}
\end{proof}

\noindent
A direct application of Hindley-Rosen's Lemma (Lem.~\ref{lem-hr})
then offers the preservation of confluence.

%%%%%%%%%%%%%%%%%%%%%%%%%%%%%%%%%%%%%%%%%%%%%%%%%%%%%%%%%%%%%%%%%%%%%%%%%%%
\begin{theorem}[Confluence of $\a_{\b\cup\R}$]
\label{thm-Aconfl-sc}
%%%%%%%%%%%%%%%%%%%%%%%%%%%%%%%%%%%%%%%%%%%%%%%%%%%%%%%%%%%%%%%%%%%%%%%%%%%
Let $\R$ be a semi-closed left-linear right-applicative system.
If $\a_\R$ is confluent then so is $\a_{\b\cup\R}$.
\end{theorem}

%For the system {\sf Tree} of Section~\ref{sec-ex}, the relation $\a_\R$ is
%confluent. As the rules are left-linear, semi-closed
%and right-applicative,
%Theorem~\ref{thm-Aconfl-sc} applies and $\a_{\b\cup\R}$ is confluent.

%%%%%%%%%%%%%%%%%%%%%%%%%%%%%%%%%%%%%%%%%%%%%%%%%%%%%%%%%%%%%%%%%%%%%%%%%%%
\paragraph{Comparison with M{\"u}ller's work.}
%%%%%%%%%%%%%%%%%%%%%%%%%%%%%%%%%%%%%%%%%%%%%%%%%%%%%%%%%%%%%%%%%%%%%%%%%%%
Our main result on the confluence of $\b$-reduction with conditional rewriting
for left-linear semi-closed system (Theorem~\ref{thm-Aconfl-sc})
is not a true extension of Theorem~\ref{thm-muller}.
Indeed, Theorem~\ref{thm-muller} applies to unconditional systems
with arbitrary right-hand sides, while Theorem~\ref{thm-Aconfl-sc}
requires right-hand sides to be applicative.

The problem is that Lemma~\ref{lem-commut-cond} may fail with
non-applicative right-hand sides.
Consider the system:
\[
  \sfh \esp\at\esp \la x.x
\qquad\qquad
  x = \sfh\ptesp \sfa \esp\sgt\esp \sfg\ptesp x \esp\at\esp \sfa
\]
We have
\(
  \sfg\ptesp \sfa
\esp\al_\b\esp
  \sfg\ptesp ((\la x.x) \sfa)
\esp\a_\R\esp
  \sfa
\).
But since the term $(\la x.x)\sfa$ is a $\R$-normal form,
the condition $x \ad_{\R} \sfh \ptesp \sfa$ is not satisfied,
and $\sfg\ptesp \sfa$ is a $\R$-normal form.

We can extend Theorem~\ref{thm-muller} to normal conditional rewriting,
i.e.\ to systems $\R$ such that in all rules $\vd = \vc \sgt l \at_\R r$,
the closed algebraic terms $\vc$ are required to be in normal form
w.r.t. $\a_\R$
(recall from Remark~\ref{rem-dec} that this is undecidable).
The proof follows exactly the same scheme as for Theorem~\ref{thm-Aconfl-sc}.
The only difference lies in the commutation of $\a_{\R_i}$
with $\a_\b$, for which Lemma~\ref{lem-commut-cond} does not apply.

%%%%%%%%%%%%%%%%%%%%%%%%%%%%%%%%%%%%%%%%%%%%%%%%%%%%%%%%%%%%%%%%%%%%%%%%%%%
\begin{theorem}[Extension of~\cite{muller92ipl}]
\label{thm-ext-mull}
%%%%%%%%%%%%%%%%%%%%%%%%%%%%%%%%%%%%%%%%%%%%%%%%%%%%%%%%%%%%%%%%%%%%%%%%%%%
Let $\R$ be a left-linear semi-closed system
such that $\a_\R$ is a normal conditional rewrite relation.
If $\a_\R$ is confluent then so is $\a_{\b\cup\R}$.
\end{theorem}

\begin{proof}
As in Theorem~\ref{thm-Aconfl-sc}, the proof
relies on the commutation of $\a_{\R_i}$ with $\a_\b$.
Since right-hand sides are not applicative,
we can not use Lemma~\ref{lem-commut-cond}.
However,
the commutation of $\a_{\R_i}$ with $\rpr_\b$ is proved
using the same general reasoning, excepted for the following point.
%The only difference is the following.
Assume that $\a_{\R_i}$ commutes with $\rpr_\b$
and that for a rule $\vd = \vc \sgt l \at_\R r$ and a substitution $\s$
we have
$l\s' \rpl_\b l\s \a_{\R_{i+1}} r\s$.
As $\a_{\R}$ is normal, we have $\vd\s \a^*_{\R_i} \vc$,
and by induction hypothesis there are $\vc'$ such that
$\vd\s' \a^*_{\R_i} \vc' \al^*_\b \vc$.
Since $\R$ is semi-closed, the terms $\vc$
are algebraic hence $\b$-normals.
It follows that
$\vd\s' \a^*_{\R_i} \vc$, hence $l\s' \a_{\R_{i+1}} r\s'$.
\end{proof}

%%%%%%%%%%%%%%%%%%%%%%%%%%%%%%%%%%%%%%%%%%%%%%%%%%%%%%%%%%%%%%%%%%%%%%%%%%%
\subsection{Confluence on weakly beta-normalizing terms}
\label{sec-Aconfl-wn}
%%%%%%%%%%%%%%%%%%%%%%%%%%%%%%%%%%%%%%%%%%%%%%%%%%%%%%%%%%%%%%%%%%%%%%%%%%%

We now turn to the problem of dropping the left-linearity and
semi-closure conditions.
We generalize 
Theorem~\ref{thm-dough}~\cite{dougherty92ic}
in two ways.
First, we adapt it to conditional rewriting.
%Second, we allow nested symbols in left-hand sides
%to be applied to less arguments than their arity.
Second, we use weakly $\b$-normalizing terms whose $\b$-normal forms 
respect an arity,
whereas Dougherty uses sets of strongly normalizing
arity compliant terms closed under reduction.

%%%%%%%%%%%%%%%%%%%%%%%%%%%%%%%%%%%%%%%%%%%%%%%%%%%%%%%%%%%%%%%%%%%%%%%%%%%
% Preliminaries
%%%%%%%%%%%%%%%%%%%%%%%%%%%%%%%%%%%%%%%%%%%%%%%%%%%%%%%%%%%%%%%%%%%%%%%%%%%

As seen in Example~\ref{ex-minus-cond},
fixpoint combinators make the commutation of $\a^*_\b$
and $\a^*_{\R}$ fail when rewriting involves equality tests
between open terms. When using weakly $\b$-normalizing terms, we can
project rewriting on $\b$-normal forms ($\bnf$), thus eliminating
fixpoints as soon as they are not significant for the reduction.

Hence, we seek to obtain 
\begin{equation}
\begin{aligned}
\label{diag-projA-bnf}
\xymatrix@R=\xyS@C=\xyL{
  s \ar@{->}[r]^{\b \cup \R}_{*}
    \ar@{.>}[d]_{\b}^{*}
  & t \ar@{.>}[d]^{\b}_{*} \\
  \bnf(s) \ar@{.>}[r]_{\R}^{*}
  & \bnf(t)
}
\end{aligned}
\end{equation}

We rely on the following:
\begin{enumerate}
\item
\label{enum-Aconfl-wn-alg-rhs}
First, $\b$-normal forms should be stable by
rewriting. By Remark~\ref{rem-alg-rhs} we must assume that right-hand sides are
algebraic, and as seen in Example~\ref{ex-dough},
we must use the notion of applicative arity
(Definition~\ref{def-arity}).

\item
We need normalizing $\b$-derivations to commute with rewriting.
This follows from using the leftmost-outermost strategy of
$\la$-calculus.

\item
Finally, we assume that conditions are algebraic.
Since left-hand sides and right-hand sides are algebraic
(by Definition~\ref{def-cond-rules} and item~\ref{enum-Aconfl-wn-alg-rhs} respectively),
this entails that
for all rules $\vd = \vc \sgt l \at_\R r$
and all substitutions $\s$, we have
$\bnf(\vd\s) = \vd \ptesp \bnf(\s)$,
$\bnf(\vc\s) = \vc \ptesp \bnf(\s)$,
$\bnf(l\s) = l \ptesp \bnf(\s)$,
$\bnf(r\s) = r \ptesp \bnf(\s)$.
\end{enumerate}

We now have to extend to conditional rewriting
the condition of arity on rewrite rules
stated in Definition~\ref{def-arity}.(\ref{def-arity-rule}).

%%%%%%%%%%%%%%%%%%%%%%%%%%%%%%%%%%%%%%%%%%%%%%%%%%%%%%%%%%%%%%%%%%%%%%%%%%%
\begin{definition}[Applicative Arity for Conditional Rules]
\label{def-arity-cond}
%%%%%%%%%%%%%%%%%%%%%%%%%%%%%%%%%%%%%%%%%%%%%%%%%%%%%%%%%%%%%%%%%%%%%%%%%%%
A rule $\vd=\vc\sgt l\at_\R r$ 
{\em respects} an arity $\ap : \Si \a \N$
if the terms $\vd,\vc$ respect $\ap$ and
the unconditional rule $l \at r$ respects $\ap$.
\end{definition}

As seen in Example~\ref{ex-dough}, terms and rewrite systems
respecting an arity prevent collapsing rules
from creating $\b$-redexes.

%%%%%%%%%%%%%%%%%%%%%%%%%%%%%%%%%%%%%%%%%%%%%%%%%%%%%%%%%%%%%%%%%%%%%%%%%%%
% weak beta-normalization
%%%%%%%%%%%%%%%%%%%%%%%%%%%%%%%%%%%%%%%%%%%%%%%%%%%%%%%%%%%%%%%%%%%%%%%%%%%

However, we do not assume that every term at hand respects an arity.
If a term has a $\b$-normal form, the leftmost-outermost
strategy for $\b$-reduction
never evaluates non-normalizing subterms.
It follows that such subterms may not respect any arity
without disturbing the projection on $\b$-normal forms.
Therefore it is sufficient to require that terms
have a $\b$-normal form that respects an arity.

%%%%%%%%%%%%%%%%%%%%%%%%%%%%%%%%%%%%%%%%%%%%%%%%%%%%%%%%%%%%%%%%%%%%%%%%%%%
\begin{definition}
\label{def-an}
%%%%%%%%%%%%%%%%%%%%%%%%%%%%%%%%%%%%%%%%%%%%%%%%%%%%%%%%%%%%%%%%%%%%%%%%%%%
Given an arity $\ap : \Si \a \N$,
we let $\AN$ be the set of terms having 
a $\b$-normal form, and whose $\b$-normal form respects $\ap$.
\end{definition}

The proof goes through essentially thanks to two points:
the well-foundedness of the leftmost-outermost strategy
for $\a_\b$ on weakly $\b$-normalizing terms~\cite{barendregt84book};
and the fact that this strategy can be described by means of
{\em head} $\b$-reductions, that are easily shown to commute with
(parallel) conditional rewriting.

%%%%%%%%%%%%%%%%%%%%%%%%%%%%%%%%%%%%%%%%%%%%%%%%%%%%%%%%%%%%%%%%%%%%%%%%%%%
% head beta-reduction
%%%%%%%%%%%%%%%%%%%%%%%%%%%%%%%%%%%%%%%%%%%%%%%%%%%%%%%%%%%%%%%%%%%%%%%%%%%
We use
a well-founded relation containing head $\b$-reductions.
Recall that by Lemma~\ref{lem-wadsworth},
any $\la$-term can be written either
\begin{align}
\tag{a}                 \la \vx .v\, &a_0 \, a_1 \dots a_n  \\
\tag{b} \text{or} \quad \la \vx .(\la y.b) &a_0 \, a_1 \dots a_n 
\end{align}
where $v \in \Vte \cup \Si$.
We denote {\em head} $\b$-reductions by $\a_h$.
They consist of head $\b$-steps:
\[
\la \vx.(\la y.b) a_0 \, a_1 \dots a_n ~\a_h~ \la \vx.b\wth{y}{a_0} a_1 \dots a_n ~. 
\]
We use the relation $\succ$ defined as:
\begin{align}
\tag{a} \la \vx.v \, a_0 \, a_1 \dots a_n         &\succ a_i        \\
\tag{b} \la \vx.(\la y.b) a_0 \, a_1 \dots a_n 
  &\succ \la \vx.b\wth{y}{a_0} \, a_1 \dots a_n
\end{align}
where $v \in \Vte \cup \Si$ and $0 \leq i \leq n$ and $n > 0$.
Note that in the case (a), $a_i$ can have free variables among $\vx$,
hence it can also be a subterm of a term
$\alpha$-equivalent to $\la \vx.v \va$;
for instance $\la x.{\sf f}x \succ y$ for all $y \in \Vte$.
Recall that $\WN_\b$ is the set 
of weakly $\b$-normalizing terms.

%%%%%%%%%%%%%%%%%%%%%%%%%%%%%%%%%%%%%%%%%%%%%%%%%%%%%%%%%%%%%%%%%%%%%%%%%%%
\begin{lemma}
\label{lem-succ}
%%%%%%%%%%%%%%%%%%%%%%%%%%%%%%%%%%%%%%%%%%%%%%%%%%%%%%%%%%%%%%%%%%%%%%%%%%%
If $s \in \WN_\b$ and $s \succ t$ then $t \in \WN_\b$. Moreover, 
$\succ$ is well-founded on $\WN_\b$.
\end{lemma}

\begin{proof}
For the first part, let $s \in \WN_\b$ and $s \succ t$.
If $s$ is of the form (b), the first step of the leftmost-outermost derivation normalizing
$s$ is $t$. Hence $t \in \WN_\b$.
Otherwise, if $t$ has no $\b$-normal form,
then $s$ has no $\b$-normal form.

For the second part, we write $\#(s)$ for the number
of $\a_h$-steps in the leftmost-outermost derivation
starting from $s$ and $|s|$ for the size of $s$.
We show that if $s \succ t$, then $(\#(s),|s|) >_{lex} (\#(t),|t|)$.
If $s$ is of the form (b), by the first point $t \in \WN_\b$.
Since $s \a_h t$, we have $\#(s) > \#(t)$.
Otherwise, the leftmost-outermost strategy starting from $s$
reduces each $a_i$
%($1 \leq i \leq n$)
by leftmost-outermost reductions.
Hence $\#(s) \ge \#(t)$.
But in this case, $t$ is a proper subterm of $s$, hence $|s| > |t|$.
\end{proof}

%%%%%%%%%%%%%%%%%%%%%%%%%%%%%%%%%%%%%%%%%%%%%%%%%%%%%%%%%%%%%%%%%%%%%%%%%%%
% Parallelization of conditional rewriting
%%%%%%%%%%%%%%%%%%%%%%%%%%%%%%%%%%%%%%%%%%%%%%%%%%%%%%%%%%%%%%%%%%%%%%%%%%%

It follows that we can reason by well-founded induction on $\succ$. 
For all $i \geq 0$, we
use a {\em nested} parallelization of $\a_{\R_i}$.
It corresponds to the one used in~\cite{or94lfcs},
that can be seen as a generalization
of Tait and Martin-L{\"o}f parallel relation.
As for $\rpr_\b$ and $\a_{\b}$, in the orthogonal case,
a complete development of $\a_{\R_i}$ can be simulated
by {\em one step} $\rpr_{\R_i}$-reduction.
This relation is also
an adaptation to conditional rewriting of the
parallelization used in~\cite{dougherty92ic}.

%%%%%%%%%%%%%%%%%%%%%%%%%%%%%%%%%%%%%%%%%%%%%%%%%%%%%%%%%%%%%%%%%%%%%%%%%%%
\begin{definition}[Conditional Nested Parallel Relations]
\label{def-walkA}
%%%%%%%%%%%%%%%%%%%%%%%%%%%%%%%%%%%%%%%%%%%%%%%%%%%%%%%%%%%%%%%%%%%%%%%%%%%
For all $i \geq 0$, let $\rpr_{\R_i}$ be the smallest parallel
rewrite relation closed under the rule
\[
(\pRule)~
\dfrac{\vd = \vc \sgt l \at_\R r
        \qquad
       l\s \a_{\R_i} r\s
        \qquad
       \s \rpr_{\R_i} \t}
	    { l\s \rpr_{\R_i} r\t }
\]
\end{definition}

Recall that $l\s \a_{\R_i} r\s$ is ensured by $\vd\s \ad_{\R_{i-1}} \vc\s$. 
These relations enjoy some nice
properties:

%%%%%%%%%%%%%%%%%%%%%%%%%%%%%%%%%%%%%%%%%%%%%%%%%%%%%%%%%%%%%%%%%%%%%%%%%%%
\begin{proposition}
\label{prop-tgtA}
%%%%%%%%%%%%%%%%%%%%%%%%%%%%%%%%%%%%%%%%%%%%%%%%%%%%%%%%%%%%%%%%%%%%%%%%%%%
For all $i \geq 0$, 
\begin{enumerate}
\item
\label{subset-tgtA}
$\a_{\R_i}  ~\sle~  \rpr_{\R_i} ~\sle~ \a_{\R_i}^*$;

\item
\label{tgtA-subst}
${s \rpr_{\R_i} t} ~\imp~ {u\wth{x}{s} \rpr_{\R_i} u\wth{x}{t}}$;

\item
\label{tgtA-subst2}
\(
  {\lbrack s \rpr_{\R_i} t ~\&~ u \rpr_{\R_i} v \rbrack}
  ~\imp~
  {u\wth{x}{s} \rpr_{\R_i} v\wth{x}{t}}
\).
\end{enumerate}
\end{proposition}

\begin{proof}
The first point is shown by induction on the definition of $\rpr_{\R_i}$ ; 
the second follows from Proposition~\ref{prop-par-rew}
and the fact that $\rpr_{\R_i}$ is a parallel rewrite relation.
For the last one, we use an induction on
$\rpr_{\R_i}$ in $u \rpr_{\R_i} v$.
If $u$ is $v$, the result is trivial.
If $u \rpr_{\R_i} v$ was obtained by $(\papp)$ or $(\abs)$,
the result follows from induction hypothesis.
Otherwise, $u \rpr_{\R_i} v$ is obtained by $\pRule$. That is, there is 
a rule $\vd = \vc \sgt l \at_\R r$ such that $u = l\s$, $v = r\t$,
$\s \rpr_{\R_i} \t$ and $l\s \a_{\R_i} r\s$.
Since $\a_{\R_i}$ is a rewrite relation, we have
$l\s\wth{x}{s} \rpr_{\R_i} r\s\wth{x}{s}$.
By induction hypothesis, we have $\s\wth{x}{s} \rpr_{\R_i} \t\wth{x}{t}$.
Therefore $l\s\wth{x}{s} \rpr_{\R_i} r\t\wth{x}{t}$.
\end{proof}

We now turn to the {\em one step} commutation of $\rpr_{\R_i}$ and $\a_h$.
This is a direct consequence of the third point of Proposition~\ref{prop-tgtA}.
Commutation of 
$\a_h$ with (unconditional) rewriting has already be coined in~\cite{bfg97jfp}.

%%%%%%%%%%%%%%%%%%%%%%%%%%%%%%%%%%%%%%%%%%%%%%%%%%%%%%%%%%%%%%%%%%%%%%%%%%%
\begin{lemma}
\label{lem-commut-h}
%%%%%%%%%%%%%%%%%%%%%%%%%%%%%%%%%%%%%%%%%%%%%%%%%%%%%%%%%%%%%%%%%%%%%%%%%%%
Let $i \geq 0$.
%For all $s$, $t$, $u$,
%If $u ~\al_h~ s ~\rpr_{\R_i}~ t$ then there exists $v$ such that
%$u ~\rpr_{\R_i}~ v ~\al_h~ t$.
%In diagrammatic form:
If $u \al_h s \rpr_{\R_i} t$ then there exists $v$ such that
$u \rpr_{\R_i} v \al_h t$~:
\begin{equation*}
\begin{aligned}
\xymatrix@R=\xyS@C=\xyL{
    s \ar@{>}[r]^{\rpr_{\R_i}}
          \ar@{>}[d]_{h}
  & t \ar@{..>}[d]^{h} \\
    u \ar@{..>}[r]_{\rpr_{\R_i}}
  & v
}
\end{aligned}
\end{equation*}
\end{lemma}

\begin{proof}
Assume that 
$s \al_h \la \vx. (\la y.a_0)a_1 \dots a_p \rpr_{\R_i} t$.
Because rules have non-variable algebraic left-hand sides, 
$t= \la \vx. (\la y.b_0)b_1 \dots b_p$ with
for all $k \in \{0,\dots,p \}$, $a_k
\rpr_{\R_i} b_k$.
On the other hand, $s = \la \vx. a_0\wth{y}{a_1}a_2 \dots a_p$.
It follows from Proposition~\ref{prop-tgtA}.\ref{tgtA-subst2} that
$a_0\wth{x}{a_1} \rpr_{\R_i} b_0\wth{x}{b_1}$
(in {\em one} step). 
Hence we have 
$s \rpr_{\R_i} \la \vx. b_0\wth{y}{b_1}b_2 \dots b_p \al_h t$.
\end{proof}

The main lemma is the projection of rewriting on $\b$-normal forms,
that is, the commutation of diagram (\ref{diag-projA-bnf}).

%%%%%%%%%%%%%%%%%%%%%%%%%%%%%%%%%%%%%%%%%%%%%%%%%%%%%%%%%%%%%%%%%%%%%%%%%%%
\begin{lemma}
\label{lem-projA-bnf}
%%%%%%%%%%%%%%%%%%%%%%%%%%%%%%%%%%%%%%%%%%%%%%%%%%%%%%%%%%%%%%%%%%%%%%%%%%%
Let $\ap : \Si \a \N$ be an arity
and $\R$ be an algebraic conditional rewrite system
which respects $\ap$.
For all $i \in \N$, if $t \in \AN$ and $t\a_{\b\cup\R_i}^* u$,
then $u\in \AN$ and $\bnf(t)\a_{\R_i}^* \bnf(u)$.
\end{lemma}

\begin{proof}
We reason by induction on $i \in \N$.  The base case $i = 0$ is trivial.
We assume that the property holds for $i \geq 0$ and show it for $i+1$.
The proof is in two steps.

\begin{enumerate}
\item
\label{lem-projA-bnf-base}
We begin by showing that for all $t \in \AN$ we have
\begin{equation}
\begin{aligned}
\label{diag-projAbnf-par}
\xymatrix@R=\xyS@C=\xyL{
    t \ar@{->}[r]^{\rpr_{\R_{i+1}}}
      \ar@{.>}[d]_{\b}^{*}
  & u \ar@{.>}[d]^{\b}_{*} \\
    \bnf(t) \ar@{.>}[r]_{\rpr_{\R_{i+1}}}
  & \bnf(u)
}
\end{aligned}
\end{equation}
We reason by induction on $\succ$ using Lemma~\ref{lem-wadsworth}.
\begin{description}
\item[$t = \la \vx.x t_1 \dots t_n$.]
In this case,
$\bnf(t) = \la \vx.x \bnf(t_1) \dots \bnf(t_n)$
and $u = \la \vx. x u_1 \dots u_n$
with $t_k \rpr_{\R_{i+1}} u_k$ for all $k \in \{1,\dots,n\}$.
As $t \succ t_k$, for all $k \in \{1,\dots,n\}$,
by induction hypothesis on $\succ$ we have $u_k \in \AN$
and $\bnf(t_k) \rpr_{\R_{i+1}} \bnf(u_k)$.
Since $\bnf(u) = \la \vx.x \bnf(u_1) \dots \bnf(u_n)$,
we have $\bnf(u) \in \AN$ and
$\bnf(t) \rpr_{\R_{i+1}} \bnf(u)$.

\item[$t = \la \vx.f t_1 \dots t_n$.]
If no rule is reduced at the head of $t$, the
result follows from induction hypothesis on $\succ$.
Otherwise, there is a rule $\vd = \vc \sgt l \at r$ such that
$t = \la \vx.l\s\va$ and $u = \la \vx.r\t\vb$
with
$l\s \rpr_{\R_{i+1}} r\t$ and $\vd\s \ad_{\R_{i}} \vc\s$.
Since $l$ is algebraic, $\bnf(t)$ is of the form $\la \vx. l\s' \va'$
where $\s' = \bnf(\s)$ and $\va' = \bnf(\va)$.
Since $\bnf(t)$ respects $\ap$,
$\va' = \emptyset$, hence $\va = \emptyset$ and $t = \la \vx . l\s$.
Therefore, because $l\s \rpr_{\R_{i+1}} r\t$, we have $\vb = \emptyset$ and 
$u = \la \vx.r\t$.
It remains to show that $u \in \AN$ 
and that $\bnf(t) = \la \vx.l\s' \rpr_{\R_{i+1}} \bnf(u)$.
Because $l$ is algebraic, its variables are $\prec^+ l$.
We can then apply induction hypothesis on $\s \rpr_{\R_{i+1}} \t$.
It follows that $\t$ has a $\b$-normal form $\t'$,
which respects $\ap$ and moreover such that $\s' \rpr_{\R_{i+1}} \t'$.
Since $r$ is algebraic, $\la \vx.r\t'$ is the 
$\b$-normal form of $u$ (which respects $\ap$). 
Hence it remains to show that
$l\s' \rpr_{\R_{i+1}} r\t'$.
Because $\s' \rpr_{\R_{i+1}} \t'$, it suffices to prove that
$l\s' \a_{\R_{i+1}} r\s'$.
Thus, we are done if we show that $\vd\s' \ad_{\R_{i}} \vc\s'$.
Since $\vd$ and $\vc$ are algebraic, $\bnf(\vd\s) = \vd\s'$ and
$\bnf(\vc\s) = \vc\s'$.
Now, since $\vd$ is algebraic and respects $\ap$,
and since $\s'$ respects $\ap$, it follows that
$\vd\s'$ respects $\ap$. The same holds for $\vc\s'$.
Hence we conclude by applying on
$\vd\s \ad_{\R_{i}} \vc\s$ the induction hypothesis on $i$.

\item[$t = \la \vx.(\la x.v)w t_1 \dots t_n$.]
In this case, we head $\b$-normalize $t$ and
obtain a term $t' \in \AN$.
Using the commutation of $\rpr_{\R_{i+1}}$ and $\a_h$, 
we obtain a term $u'$ such that $t' \rpr_{\R_{i+1}} u'$.
Since $t \succ^+ t'$, we can reason as in the preceding cases.
\end{description}

\item
We now show that $t \a^*_{\b \cup \R_i} u$
implies $\bnf(t) \a^*_{\R_i} \bnf(u)$.
We reason by induction on $t \a^*_{\b \cup \R_i} u$,
using Proposition~\ref{prop-tgtA}.\ref{subset-tgtA}.
The base case $t = u$ is trivial.
Assume that $t \a_{\b\cup\R_i} v \a^*_{\b\cup\R_i} u$.
By induction hypothesis we have $\bnf(v) \a^*_{\R_i} \bnf(u)$.
There are two cases. If $t \a_\b v$,
then $\bnf(t) = \bnf(v)$ and we are done.
Otherwise we have $t \a_{\R_i} v$, hence
$\bnf(t) \a^*_{\R_i} \bnf(v) \a^*_{\R_i} \bnf(u)$
by~\ref{lem-projA-bnf-base}.
\qedhere
\end{enumerate}
\end{proof}

%%%%%%%%%%%%%%%%%%%%%%%%%%%%%%%%%%%%%%%%%%%%%%%%%%%%%%%%%%%%%%%%%%%%%%%%%%%
% confluence out of terms having an AC \bNF
%%%%%%%%%%%%%%%%%%%%%%%%%%%%%%%%%%%%%%%%%%%%%%%%%%%%%%%%%%%%%%%%%%%%%%%%%%%

Preservation of confluence is a direct consequence of the
projection on $\b$-normal forms.

%%%%%%%%%%%%%%%%%%%%%%%%%%%%%%%%%%%%%%%%%%%%%%%%%%%%%%%%%%%%%%%%%%%%%%%%%%%
\begin{theorem}
\label{thm-Aconfl-bnf}
%%%%%%%%%%%%%%%%%%%%%%%%%%%%%%%%%%%%%%%%%%%%%%%%%%%%%%%%%%%%%%%%%%%%%%%%%%%
Let $\ap : \Si \a \N$ be an arity
and $\R$ be an algebraic conditional rewrite system which respects $\ap$.
If $\a_\R$ is confluent on $\AN$, then $\a_{\b\cup\R}$ is confluent on $\AN$.
\end{theorem}

\begin{proof}
Let $s,t,u$ such that $s \in \AN$ and 
$u ~\al^*_{\b \cup \R}~ s ~\a^*_{\b \cup \R}~ t$.
By two applications of Lemma~\ref{lem-projA-bnf} we get 
that $\bnf(u) ~\al^*_\R~ \bnf(s) ~\a^*_\R~ \bnf(t)$, with moreover
$\bnf(s)$, $\bnf(t)$ and $\bnf(u) \in \AN$.
Hence we conclude by $\a_\R$-confluence on $\AN$.
In diagrammatic form,
\[
\xymatrix{
  & s \ar@{->}[dr]^{\b \cup \R}_{*}
      \ar@{->}[dl]_{\b \cup \R}^{*}
      \ar@{..>}[d]_{\b}^{*} \\
  u   \ar@{..>}[d]_{\b}^{*}
  & \bnf(s) \ar@{..>}[dr]_{\R}^{*}
            \ar@{..>}[dl]^{\R}_{*} 
  & t \ar@{..>}[d]^{\b}_{*} \\
  \bnf(u)   \ar@{..>}[dr]^{\R}_{*}
  &
  & \bnf(t) \ar@{..>}[dl]_{\R}^{*} \\
  & v & \\
}
\]
\end{proof}

%%%%%%%%%%%%%%%%%%%%%%%%%%%%%%%%%%%%%%%%%%%%%%%%%%%%%%%%%%%%%%%%%%%%%%%%%%%
% related work
%%%%%%%%%%%%%%%%%%%%%%%%%%%%%%%%%%%%%%%%%%%%%%%%%%%%%%%%%%%%%%%%%%%%%%%%%%%

\comment{
In that paper, Dougherty proves, in the non-conditional case, that
left-linearity is not needed when considering so called $\R$-stable
sets of terms. A set of terms is $\R$-stable if it is made of
$\b$-strongly normalizing arity compliant terms and is closed under
$\a_\b$, $\a_\R$ and subterm. Then, $\a_{\b\cup\R}$ is confluent on
any $\R$-stable set of terms whenever $\a_\R$ is confluent and $\R$
is first-order.

We extend this result in three ways. First, we adapt it to conditional
rewriting. The main difference is that we need a more subtle induction
relation. Second, we do not limit ourselves to curried versions of
first-order rewrite rules. We allow nested symbols in rules to be
applied to less arguments that their arity. Third, and that is the
main point, we show that confluence is preserved out of terms having
an arity compliant $\b$-normal form. The amelioration is
two-fold. First we do not need to assume that {\em every}
$\b$-reduction starting from {\em any} $\a_{\b\cup\R}$-reduct of a
term terminates. We simply need that {\em there exists} a terminating
$\b$-reduction starting from the term. Second, we do not need to
assume that {\em every} $\a_{\b\cup\R}$-reduct of a term is
arity compliant but only that its $\b$-normal form is.
}

%% file: Bconfl.tex
%%%%%%%%%%%%%%%%%%%%%%%%%%%%%%%%%%%%%%%%%%%%%%%%%%%%%%%%%%%%%%%%%%%%%%%%%%%
\section{Using beta-reduction in the evaluation of conditions}
  \label{sec-Bconfl}
%%%%%%%%%%%%%%%%%%%%%%%%%%%%%%%%%%%%%%%%%%%%%%%%%%%%%%%%%%%%%%%%%%%%%%%%%%%

In this section we focus on the combination of $\b$-reduction
with the {\em join $\b$-conditional} rewrite relation $\a_{\R(\b)}$
issued from a conditional rewrite system $\R$
(see Definition~\ref{def-cond-rew}).

%%%%%%%%%%%%%%%%%%%%%%%%%%%%%%%%%%%%%%%%%%%%%%%%%%%%%%%%%%%%%%%%%%%%%%%%%%%
%\begin{notation}[Join $\b$-Conditional Rewriting]
%%%%%%%%%%%%%%%%%%%%%%%%%%%%%%%%%%%%%%%%%%%%%%%%%%%%%%%%%%%%%%%%%%%%%%%%%%%
%We denote by $\a_{\R(\b)}$ the join $\b$-conditional rewrite relation
%issued from a conditional rewrite system $\R$.
%\end{notation}

%%%%%%%%%%%%%%%%%%%%%%%%%%%%%%%%%%%%%%%%%%%%%%%%%%%%%%%%%%%%%%%%%%%%%%%%%%%
%\paragraph{General methodology.}
%%%%%%%%%%%%%%%%%%%%%%%%%%%%%%%%%%%%%%%%%%%%%%%%%%%%%%%%%%%%%%%%%%%%%%%%%%%
We give sufficient conditions on $\R$ to deduce
the confluence of $\a_{\b\cup\R(\b)}$ from the confluence of $\a_{\R}$.
We achieve this by exhibiting two different criteria ensuring that
derivations combining $\b$-reduction and $\b$-conditional rewriting
can be projected, via $\b$-reductions, to derivations made 
of conditional rewriting only
(hence without using $\b$-reduction in the evaluation of conditions):
\begin{equation}
\begin{aligned}
%{\a_{\b\cup\R(\b)}^*} \sle {\a_\b^*\a_\R^*\al_\b^*} ~.
\label{diag-BsleA}
\xymatrix@R=\xyS@C=\xyL{
  s \ar@{->}[r]^{\b \cup \R(\b)}_{*}
    \ar@{.>}[d]_{\b}^{*}
& t \ar@{.>}[d]^{\b}_{*} \\
  s' \ar@{.>}[r]_{\R}^{*}
& t'
}
\end{aligned}
\end{equation}

It is easy to see that property~(\ref{diag-BsleA}) combined to the confluence
of $\a_{\b\cup\R}$ entails the confluence of $\a_{\R}$.
%As explained in Section~\ref{},
We can actually prove property~(\ref{diag-BsleA})
on some subsets of $\Te(\Si)$ only.
This motivates the assumptions on the following proposition.

%%%%%%%%%%%%%%%%%%%%%%%%%%%%%%%%%%%%%%%%%%%%%%%%%%%%%%%%%%%%%%%%%%%%%%%%%%%
\begin{proposition}
\label{prop-Bconfl}
%%%%%%%%%%%%%%%%%%%%%%%%%%%%%%%%%%%%%%%%%%%%%%%%%%%%%%%%%%%%%%%%%%%%%%%%%%%
Let $\R$ be a conditional rewrite system and $S \sle \Te(\Si)$
be a set of terms closed under $\a_{\b \cup \R(\b)}$.
Assume that $\a_{\b\cup\R}$ is confluent on $S$.
If property~(\ref{diag-BsleA}) is satisfied for all $s,t \in S$
then $\a_{\b\cup\R(\b)}$ is confluent on $S$.
\end{proposition}

\begin{proof}
Let $t \in S$ and $u,v$ such that
\[
  u \quad\al^*_{\b \cup \R(\b)}\quad
  t \quad\a^*_{\b \cup \R(\b)}\quad
  v
~.
\]
Note that $u,v \in S$ since $S$ is closed under $\a_{\b\cup\R(\b)}$.
By property~(\ref{diag-BsleA}) applied twice and by confluence of 
$\a_{\b \cup \R}$ on $S$, there is $w$ such that
$u \a^*_{\b \cup \R} w \al^*_{\b \cup \R} v$.
We conclude by the fact that $\a_{\R} \sle \a_{\R(\b)}$.
In diagrammatic form,
\begin{equation*}
\begin{aligned}
\xymatrix@C=\xyS@R=\xyS{
  & u     \ar@{..>}[dl]_{\b}^{*}
  &
  & t     \ar@{>}[ll]_{\b \cup \R(\b)}^{*}
          \ar@{>}[rr]^{\b \cup \R(\b)}_{*}
          \ar@{..>}[dr]_{\b}^{*}
          \ar@{..>}[dl]^{\b}_{*}
  &
  & v     \ar@{..>}[dr]^{\b}_{*}
  &       \\
    \cdot \ar@{..>}[drrr]_{\b \cup \R}^{*}
  & 
  & \cdot \ar@{..>}[ll]_{\R}^{*}
  & 
  & \cdot \ar@{..>}[rr]^{\R}_{*}
  & 
  & \cdot \ar@{..>}[dlll]^{\b \cup \R}_{*} \\
  &
  &
  & w
  &
  &
  &
}
\end{aligned}
\end{equation*}
\end{proof}

\noindent
Our two different criteria to obtain~(\ref{diag-BsleA})
are the extensions to $\b$-conditional rewriting of the two criteria
studied for conditional rewriting in Section~\ref{sec-Aconfl}.
\begin{itemize}
\item The first one concerns left-linear (and semi-closed)
rewriting, with no termination assumption on $\b$-reduction.

\item The second one concerns arity-preserving algebraic rewriting,
with a weak-normalization assumption on $\b$-reduction.
\end{itemize}

%%%%%%%%%%%%%%%%%%%%%%%%%%%%%%%%%%%%%%%%%%%%%%%%%%%%%%%%%%%%%%%%%%%%%%%%%%%
%\paragraph{Difficulties with $\b$-conditional rewriting.}
%%%%%%%%%%%%%%%%%%%%%%%%%%%%%%%%%%%%%%%%%%%%%%%%%%%%%%%%%%%%%%%%%%%%%%%%%%%
In the left-linear and semi-closed case,
allowing $\b$-reduction in the evaluation of conditions
imposes us to put stronger assumptions on $\R$ than for
conditional rewriting in Section~\ref{sec-Aconfl-sc}:
rewrite rules need to be algebraic and to respect and arity.
Recall that these assumptions were already made in Section~\ref{sec-Aconfl-wn}
when considering possibly non left-linear rewriting on weakly $\b$-normalizing
terms.

The following example presents rules which either
are not algebraic or do not respect
the arity prescribed by left-hand sides.
With these rules property~(\ref{diag-BsleA})
fails and $\a_{\b\cup\R(\b)}$
is not confluent whereas $\a_{\R}$ and $\a_{\b\cup\R}$ are confluent.

%%%%%%%%%%%%%%%%%%%%%%%%%%%%%%%%%%%%%%%%%%%%%%%%%%%%%%%%%%%%%%%%%%%%%%%%%%%
\begin{example}
\label{ex-Bconfl}
%%%%%%%%%%%%%%%%%%%%%%%%%%%%%%%%%%%%%%%%%%%%%%%%%%%%%%%%%%%%%%%%%%%%%%%%%%%
With the conditional rewrite systems~(\ref{eq-ex1}), (\ref{eq-ex2}),
(\ref{eq-ex3}) and (\ref{eq-ex4}) below,
\begin{enumerate}
\item the relations $\a_\R$ and $\a_{\b\cup\R}$ are confluent,
\item property~(\ref{diag-BsleA}) is not satisfied and the relation
  $\a_{\b\cup\R(\b)}$ is not confluent.
\end{enumerate}
\begin{align}
% First line
      \label{eq-ex1}
        {\sfg} \esp x \esp y &\dbesp\at\dbesp x \esp y
		  & {\sfg} \esp x \esp \sfc \esp=\esp {\sfd}
        \esp\sgt\esp \sff \esp x &\dbesp\at\dbesp \sfa \esp x
		  & \sff \esp x &\dbesp\at\dbesp \sfb \esp x
      \\	
% Second line
      \label{eq-ex2}
      &
      & x \esp \sfc  \esp=\esp {\sfd}
        \esp\sgt\esp {\sff} \esp x &\dbesp\at\dbesp \sfa \esp x
		  & \sff \esp x &\dbesp\at\dbesp \sfb \esp x
      \\
% Third line
      \label{eq-ex3}
      &
		  & {\id} \esp x \esp {\sfc} \esp=\esp {\sfd}
        \esp\sgt\esp {\sff} \esp x &\dbesp\at\dbesp \sfa \esp x
		  & \sff \esp x &\dbesp\at\dbesp \sfb \esp x
      \\
% Fourth line
      \label{eq-ex4}
        {\sfh} \esp x \esp y &\dbesp\at\dbesp {\id} \esp x \esp y 
		  & {\sfh} \esp x \esp \sfc \esp=\esp \sfd
        \esp\sgt\esp \sff \esp x &\dbesp\at\dbesp \sfa \esp x 
		  & \sff \esp x &\dbesp\at\dbesp \sfb \esp x	
\end{align}
where $\id$ is defined by $\id \esp x \at x$.
\end{example}

\begin{proof}\item
\begin{enumerate}
\item
Since the symbol $\sfd$ is not defined,
these systems lead to normal conditional rewrite relations.
Since they are left-linear and semi-closed,
we can apply Theorem~\ref{thm-ext-mull},
and deduce the confluence of $\a_{\b \cup \R}$ from the confluence
of $\a_{\R}$.
Since they are left-linear systems,
if their critical pairs are unfeasible,
we can obtain the confluence of $\a_{\R}$
by Theorem~\ref{thm-orth-cond}.
Each system has a unique conditional rule and a unique critical
pair, issued from the root superposition of this rule with
$\sff \ptesp x \esp\at\esp \sfb \ptesp x$.
In each case, the number of occurrences of the symbol $\sfc$
in a term is preserved by $\a_{\R}$.
Moreover, for each instantiation of the conditional rule,
the instantiated left-hand side of the condition
contains at least one occurrence of $\sfc$.
It follows that it cannot reduce to $\sfd$,
and that the critical pair is unfeasible.
Therefore, we obtain the confluence of $\a_{\R}$
by Theorem~\ref{thm-orth-cond}
and we deduce the confluence of $\a_{\b \cup \R}$
thanks to Theorem~\ref{thm-ext-mull}.

\item
In each case, the step
$\sff \esp \la x.\sfd \a_{\R(\b)} \sfa \esp \la x.\sfd$
is not in $\a^*_{\b} \a^*_{\R} \al^*_\b$
and the following peak is unjoinable
\[
  \sfa \esp \la x.\sfd
  \quad\al_{\R(\b)}\quad
  \sff \esp \la x.\sfd
  \quad\a_{\R(\b)}\quad
  \sfb \esp \la x.\sfd
~.
\]
\qedhere
\end{enumerate}
\end{proof}

\noindent
Note that systems~(\ref{eq-ex1}) and~(\ref{eq-ex2})
contain respectively a right-hand side and a condition
which are not algebraic,
and that systems~(\ref{eq-ex3}) and~(\ref{eq-ex4})
contain respectively a right-hand side and a condition
that do not respect the arity of $\id$ imposed by the rewrite rule
$\id \ptesp x \at x$.

Note also that~(\ref{diag-BsleA}) is reminiscent of
a property required on the substitution calculus
used in~\cite{or94lfcs}.
This would require to see $\a_\b$ as the substitution calculus.
But this does not fit in our framework, in particular because we consider 
$\a_\b$ and rewriting at the same level.
Moreover, the substitution calculus used in~\cite{or94lfcs} is required
to be complete (i.e. strongly normalizing and confluent),
which is not the case here for $\a_\b$.

%%%%%%%%%%%%%%%%%%%%%%%%%%%%%%%%%%%%%%%%%%%%%%%%%%%%%%%%%%%%%%%%%%%%%%%%%%%
\paragraph{Outline.}
%%%%%%%%%%%%%%%%%%%%%%%%%%%%%%%%%%%%%%%%%%%%%%%%%%%%%%%%%%%%%%%%%%%%%%%%%%%
We begin in Section~\ref{sec-Bconfl-sc} by the extension
of Theorem~\ref{thm-Aconfl-sc} to $\b$-conditional rewriting for
left-linear and semi-closed systems.
In this case,
preservation of confluence only holds on terms respecting an arity
(namely {\em conditionally $(\R,\ap)$-stable terms}, 
see Definition~\ref{def-rstable-cond}).
This is an
extra hypothesis compared to the results of Section~\ref{sec-Aconfl-sc}.
Then, in Section~\ref{sec-Bconfl-wn}
we consider the case of Theorem~\ref{thm-Aconfl-bnf}.
It directly extends to $\b$-conditional rewriting.
In both cases, we assume that rules are algebraic
and respect an arity.
In each case our assumptions ensure that the
results of Section~\ref{sec-Aconfl} apply, hence that $\a_{\b\cup\R}$
is confluent whenever $\a_\R$ is confluent.
Hence, using Proposition~\ref{prop-Bconfl}
we deduce the confluence of $\a_{\b \cup \R(\b)}$
from the confluence of $\a_\R$ and property~(\ref{diag-BsleA}).

%%%%%%%%%%%%%%%%%%%%%%%%%%%%%%%%%%%%%%%%%%%%%%%%%%%%%%%%%%%%%%%%%%%%%%%%%%%
\paragraph{Remarks.}
%%%%%%%%%%%%%%%%%%%%%%%%%%%%%%%%%%%%%%%%%%%%%%%%%%%%%%%%%%%%%%%%%%%%%%%%%%%
In~\cite{bkr06fossacs}, we have shown~(\ref{diag-BsleA})
by using a stratification of $\a_{\R(\b)}$ in which,
instead of having $\a_{\R(\b)_0} = \emptyset$ as in
Definition~\ref{def-cond-rew}, we had $\a_{\R(\b)_0} = \a_\b$
(it is easy to show that these two base cases induce the same
relation $\a_{\R(\b)}$).

We proceed here in a slightly different and more general way.
We show~(\ref{diag-BsleA}) with $\a_{\R(\b)_0} = \emptyset$
and use the following intermediate property:
for all $i \in \N$,
\begin{equation}
\begin{aligned}
%{\a_{\b\cup\cB}^*} \sle {\a_\b^*\a_\cA^*\al_\b^*} ~.
\label{diag-BsleA-strat}
\xymatrix@R=\xyS@C=\xyL{
  t \ar@{->}[r]^{\b \cup \R(\b)_i}_{*}
    \ar@{.>}[d]_{\b}^{*}
  & u \ar@{.>}[d]^{\b}_{*} \\
  t' \ar@{.>}[r]_{\R_i}^{*}
  & u'
}
\end{aligned}
\end{equation}

%%%%%%%%%%%%%%%%%%%%%%%%%%%%%%%%%%%%%%%%%%%%%%%%%%%%%%%%%%%%%%%%%%%%%%%%%%%
\subsection{Confluence for left-linear semi-closed systems}
\label{sec-Bconfl-sc}
%%%%%%%%%%%%%%%%%%%%%%%%%%%%%%%%%%%%%%%%%%%%%%%%%%%%%%%%%%%%%%%%%%%%%%%%%%%

This section is devoted to the proof of (\ref{diag-BsleA-strat})
for left-linear semi-closed systems.
Using Proposition~\ref{prop-Bconfl} and Theorem~\ref{thm-Aconfl-sc},
we then easily deduce the confluence of $\a_{\b \cup \R(\b)}$
when $\a_\R$ is confluent.
We postpone the proof of~(\ref{diag-BsleA-strat}) until Lemma~\ref{lem-BtoA},
Section~\ref{sec-Bconfl-sc-confl}.
%The proof of~(\ref{diag-BsleA-strat}) is postponed until Lemma~\ref{lem-BtoA},
%Section~\ref{sec-Bconfl-sc-confl}.
The material used in the proof
is presented and motivated in
Section~\ref{sec-Bconfl-sc-prelim} below.

%%%%%%%%%%%%%%%%%%%%%%%%%%%%%%%%%%%%%%%%%%%%%%%%%%%%%%%%%%%%%%%%%%%%%%%%%%%
\subsubsection{Preliminaries}
\label{sec-Bconfl-sc-prelim}
%%%%%%%%%%%%%%%%%%%%%%%%%%%%%%%%%%%%%%%%%%%%%%%%%%%%%%%%%%%%%%%%%%%%%%%%%%%

The proof of~(\ref{diag-BsleA-strat})
involves some intermediate lemmas and the extension
of $(\R,\ap)$-stable sets of terms to conditional rewriting.
In order to motivate them, we sketch some steps of the proof.
Property~(\ref{diag-BsleA-strat}) is proved by induction on $i \in \N$.
Assuming the property for $i \in \N$, we discuss it for $i+1$.
We reason by induction on 
the length of the derivation $t \a^*_{\b \cup \R(\b)_{i+1}} u$.
We present the ingredients used 
in the different steps of this induction.

%%%%%%%%%%%%%%%%%%%%%%%%%%%%%%%%%%%%%%%%%%%%%%%%%%%%%%%%%%%%%%%%%%%%%%%%%%%
\paragraph{The base case.}
%%%%%%%%%%%%%%%%%%%%%%%%%%%%%%%%%%%%%%%%%%%%%%%%%%%%%%%%%%%%%%%%%%%%%%%%%%%
In the base case, we have $t \a_{\b \cup \R(\b)_{i+1}} u$ in one step.
The case of $t \a_\b u$ is trivial: take $t' \deq u' \deq u$.
The case of $t \a_{\R(\b)_{i+1}} u$ is more involved.
We show 
\begin{equation}
\label{diag-BsleA-strat-base-rew}
\begin{aligned}
\xymatrix@R=\xyS@C=\xyL{
    t \ar@{->}[r]^{\R(\b)_{i+1}}
      \ar@{.>}[d]_{\b}^{*}
  & u \ar@{.>}[d]^{\b}_{*} \\
    t' \ar@{.>}[r]_{\R_{i+1}}
  & u'
}
\end{aligned}
\end{equation}
Consider a rule $\vd = \vc \sgt l \at_\R r$.
Recall from Example~\ref{ex-Bconfl} that it
must be algebraic and respect an arity.
Hence
%Given an algebraic rule $\vd=\vc\sgt l\at_\R r$ that respects an arity,
every $\b$-redex
occurring in $\vd\s$ or $\vc\s$ also occurs in $l\s$. Then,
property~(\ref{diag-BsleA-strat-base-rew})
means that there is a $\b$-reduction starting from $l\s$ that reduces
these redexes and produce a substitution $\s'$ such that
\[
  l\s \quad\a^*_\b\quad l\s' \quad\a_{\R_{i+1}}\quad r\s' \quad\al^*_\b\quad r\s
~.
\]
In other words, if the
conditions are satisfied with $\s$ and $\a_{\b\cup\R(\b)_i}$
(i.e.\ $\vd\s\ad_{\b\cup\R(\b)_i} \vc$, recall that $\vc$ are closed terms
since $\R$ is semi-closed),
then they are satisfied with $\s'$ and
$\a_{\R_i}$ (i.e.\ $\vd\s'\ad_{\R_i} \vc$).
Let us look at this more precisely.
Assume that $\vd\s\ad_{\b\cup\R(\b)_i} \vc$.
Hence there are terms $\vv$ such that
\[
\vd\s \quad\a^*_{\b\cup\R(\b)_i}\quad \vv \quad\al^*_{\b\cup\R(\b)_i}\quad \vc
~.
\]
By induction hypothesis on $i$, we get terms $\vw$ and $\vv'$ such that
\[
\xymatrix@R=\xyS@C=\xyL{
  \vd\s \ar@{->}[r]^{\b \cup \R(\b)_i}_{*} \ar@{.>}[d]_{\b}^{*}
& \vv \ar@{.>}[d]^{\b}_{*} 
& \vc \ar@{->}[l]_{\b\cup \R(\b)_i}^{*} \\
  \vw \ar@{.>}[r]_{\R_i}^{*}
& \vv'
}
\]
In order to conclude, we need a substitution $\s'$ such that
$\s \a^*_{\b} \s'$ and $\vw \a^*_{\b} \vd\s'$.
Using the algebraicity of $\vd$, this follows from 
Proposition~\ref{prop-beta-alg}, which is stated and proved below.
The remaining of the proof uses the commutation of $\a_\b$ and $\a_{\R_i}$
(Lemma~\ref{lem-commut-cond}) and
relies on the semi-closure and the right-applicativity
of $\R$ (which follows from its algebraicity).
See the proof of Lemma~\ref{lem-BtoA} in Section~\ref{sec-Bconfl-sc-confl}
for details.

We need to show that if $t$ is an algebraic term such that
$t\s \a^*_{\b} v$, then there is a substitution $\s'$
such that $\s \a^*_\b \s'$ and $v \a^*_\b t\s'$.
This is provided by the two following technical propositions.
%We begin by some technical properties.
The first one is a generalization of the diamond property of $\rpr_\b$.
It is a direct consequence of Lemma 3.2 in~\cite{takahashi95ic}.

%%%%%%%%%%%%%%%%%%%%%%%%%%%%%%%%%%%%%%%%%%%%%%%%%%%%%%%%%%%%%%%%%%%%%%%%%%%
\begin{proposition}
\label{prop-diamond}
%%%%%%%%%%%%%%%%%%%%%%%%%%%%%%%%%%%%%%%%%%%%%%%%%%%%%%%%%%%%%%%%%%%%%%%%%%%
Let $n \geq 0$ and assume that $s,s_1,\dots, s_n$ are terms such
that $s \rpr_\b s_i$ for all $i \in \{1,\dots,n\}$.
Then there is a term $s'$ such that $s \rpr_\b s'$ 
and $s_i \rpr_\b s'$ for all $i \in \{1,\dots,n\}$.
\end{proposition}

We deduce the following property.
The proof of case~\ref{prop-beta-alg-star} uses the diamond
property of $\rpr_\b$.

%%%%%%%%%%%%%%%%%%%%%%%%%%%%%%%%%%%%%%%%%%%%%%%%%%%%%%%%%%%%%%%%%%%%%%%%%%%
\begin{proposition}
\label{prop-beta-alg}
%%%%%%%%%%%%%%%%%%%%%%%%%%%%%%%%%%%%%%%%%%%%%%%%%%%%%%%%%%%%%%%%%%%%%%%%%%%
Let $t_1,\dots,t_n$ be algebraic terms and let $\s$ be a substitution.
\begin{enumerate}
\item \label{prop-beta-alg-base}
If $t_i\s \rpr_\b u_i$ for all $i \in \{1,\dots,n\}$,
then there is a substitution $\s'$ such that
$\s \rpr_\b \s'$ and $u_i \rpr_\b t_i\s'$
for all $i \in \{1,\dots,n\}$.

\item \label{prop-beta-alg-star}
If $t_i\s \a^*_\b u_i$ for all $i \in \{1,\dots,n\}$,
then there is a substitution $\s'$ such that
$\s \a^*_\b \s'$ and $u_i \a^*_\b t_i\s'$
for all $i \in \{1,\dots,n\}$.
\end{enumerate}
\end{proposition}

Note that the terms $t_1,\dots,t_n$ need not be linear.

\begin{proof}\item
\begin{enumerate}
\item 
Since $t_i$ is algebraic, every occurrence of a $\b$-redex
in $t_i$ is of the form $p.d$ where $p$
is an occurrence of a variable $x$ in $t_i$.
Since $\rpr_\b$ is reflexive, 
for each $i \in \{1,\dots,n\}$, each $x \in \FV(t_i)$
and each $p \in \Occ(x,t_i)$,
there is a term $s_{(i,x,p)}$
%there are terms
%\[
%  (s_{(i,x,p)})_{i \in \{1,\dots,n\}
%    \quad\land\quad x \in \FV(t_i) \quad\land\quad p \in \Occ(x,t_i)}
    %\esp\&\esp x \in \FV(t_i) \esp\&\esp p \in \Occ(x,t_i)}
%\]
such that
%for all $i \in \{1,\dots,n\}$, all $x \in \FV(t_i)$
%and all $p \in \Occ(x,t_i)$ we have
\[
  t_i\s|_p \quad=\quad \s(x) \quad\rpr_\b\quad s_{(i,x,p)}
\]
and for all $i \in \{1,\dots,n\}$,
\[
  u_i \quad=\quad
  t_i[p \gets s_{(i,x,p)}
      \esp\tq\esp x \in \FV(t_i) \quad\land\quad p \in \Occ(x,t_i)]
  ~.
\]
By Proposition~\ref{prop-diamond},
for all $x \in \FV(t_1,\dots,t_n)$,
there is a term $v_x$ such that
$\s(x) \rpr_\b v_x$ and
$s_{(i,x,p)} \rpr_\b v_x$
for all $i \in \{1,\dots,n\}$ and all $p \in \Occ(x,t_i)$.
Therefore, for all $i \in \{1,\dots,n\}$
we have
\[
  u_i \quad\rpr_\b\quad
  t_i[p \gets v_x \esp\tq\esp x \in \FV(t_i) \quad\land\quad p \in \Occ(x,t_i)]
  ~.
\]
Let $\s'$ be the substitution of same domain as $\s$
such that
$\s'(x) = v_x$ for all $x \in \FV(t_1,\dots,t_n)$ and
$\s'(x) = \s(x)$ for all $x \notin \FV(t_1,\dots,t_n)$.
Then we have $\s \rpr_\b \s'$ and $u_i \rpr_\b t_i\s'$
for all $i \in \{1,\dots,n\}$.

\item
By induction on $k \in \N$,
we show that if $t_i\s \rpr^k_\b u_i$ for all $i\in \{1,\dots,n\}$,
then there is $\s'$ such that $\s \rpr^*_\b \s'$
and $u_i \rpr^*_\b t_i\s'$ for all $i \in \{1,\dots,n\}$.

The base case $t_i\s \rpr^0_\b u_i$ for all $i \in \{1,\dots,n\}$
is trivial.
For the induction case, there are $u'_1,\dots,u'_n$
such that
${t_i\s} \esp\rpr^k_\b\esp {u'_i} \esp\rpr_\b\esp {u_i}$
for all $i \in \{1,\dots,n\}$.
Then, by~\ref{prop-beta-alg-base}, there is $\s'$
such that $\s \rpr_\b \s'$ and $u'_i \rpr_\b t_i\s'$
for all $i \in \{1,\dots,n\}$.
Since $\rpr_\b$ satisfies the diamond property
(Proposition~\ref{prop-diamond}),
for all $i \in \{1,\dots,n\}$ there is $u''_i$ such that
${t_i\s'} \esp\rpr^k_\b\esp {u''_i} \esp\rpl_\b\esp {u_i}$,
and by induction hypothesis on $k$,
there is $\s''$ such that $\s' \rpr^*_\b \s''$
and $u''_i \rpr^*_\b t_i\s''$ for all $i \in \{1,\dots,n\}$.
We deduce that $u_i \rpr_\b^* t_i\s''$ for all $i \in \{1,\dots,n\}$.

In diagrammatic form:
\[
\xymatrix@R=\xyS@C=\xyL{
    \vt\s \ar@{->}[r]^{\rpr_\b}
  & \vu'  \ar@{->}[r]^{\rpr_\b}_{k}
        \ar@{.>}[d]_{\rpr_\b}
  & \vu   \ar@{.>}[d]^{\rpr_\b}
\\
  & \vt\s'  \ar@{.>}[r]_{\rpr_\b}^{k}
  & {\vu''} \ar@{.>}[d]^{\rpr_\b}_{*}
\\
  &
  & \vt{\s''}
}
\]
\qedhere
\end{enumerate}
\end{proof}

%%%%%%%%%%%%%%%%%%%%%%%%%%%%%%%%%%%%%%%%%%%%%%%%%%%%%%%%%%%%%%%%%%%%%%%%%%%
\paragraph{The induction case.}
%%%%%%%%%%%%%%%%%%%%%%%%%%%%%%%%%%%%%%%%%%%%%%%%%%%%%%%%%%%%%%%%%%%%%%%%%%%
In the induction case, we have
$t \a^*_{\b \cup \R(\b)_{i+1}} u$ in more than one step.
Hence, this derivation can be written as
$t \a_{\b \cup \R(\b)_{i+1}} v \a^*_{\b \cup \R(\b)_{i+1}} u$
for some $v$.
%Hence there is some $v$ such that
%this derivation can be written as
%$t \a_{\b \cup \R(\b)_{i+1}} v \a^*_{\b \cup \R(\b)_{i+1}} u$.
If $t \a_\b v$, then we easily conclude by induction hypothesis on
$v \a^*_{\b \cup \R(\b)_{i+1}} u$.
Otherwise, we have $t \a_{\R(\b)_{i+1}} v$ and things get more involved.
Using the induction hypothesis on $v \a^*_{\b \cup \R(\b)_{i+1}} u$
and the discussion of the above paragraph for $t \a_{\R(\b)_{i+1}} v$,
we arrive at the following situation:
\[
\xymatrix@R=\xyS@C=\xyS{
  &&
    t \ar@{->}[rr]^{\R(\b)_{i+1}}
      \ar@{.>}[dl]_{\b}^{*}
  &&
    v \ar@{->}[rr]^{\b \cup \R(\b)_{i+1}}_{*}
      \ar@{.>}[dl]^{\b}_{*} 
      \ar@{.>}[dr]_{\b}^{*} 
  &&
    u \ar@{.>}[dr]^{\b}_{*} 
\\
  & t'  \ar@{.>}[rr]_{\R_{i+1}}
  &&
    v'' 
  &&
    v'  \ar@{.>}[rr]_{\R_{i+1}}^{*}
  &&
    u'  
}
\]
Using the confluence of $\a_\b$ and the commutation of $\a_\b$
with $\a_{\R_i}$ (Lemma~\ref{lem-commut-cond}),
we get
\[
\xymatrix@R=\xyS@C=\xyS{
  &&
    t \ar@{->}[rr]^{\R(\b)_{i+1}}
      \ar@{.>}[dl]_{\b}^{*}
  &&
    v \ar@{->}[rr]^{\b \cup \R(\b)_{i+1}}_{*}
      \ar@{.>}[dl]^{\b}_{*} 
      \ar@{.>}[dr]_{\b}^{*} 
  &&
    u \ar@{.>}[dr]^{\b}_{*} 
\\
  & t'  \ar@{.>}[rr]_{\R_{i+1}}
  &&
    v'' \ar@{.>}[dr]^{\b}_{*}
  &&
    v'  \ar@{.>}[rr]_{\R_{i+1}}^{*}
        \ar@{.>}[dl]_{\b}^{*} 
  &&
    u'  \ar@{.>}[dl]^{\b}_{*}
\\
  &&&&
  v'''  \ar@{.>}[rr]_{\R_{i+1}}^{*}
  &&
  u'' 
}
\]
In order to conclude we use the following property:
%More generally, we show
for all $i \in \N$,
\begin{equation}
\begin{aligned}
%{\a_{\b\cup\cB}^*} \sle {\a_\b^*\a_\cA^*\al_\b^*} ~.
\label{diag-AsleA-strat}
\xymatrix@R=\xyS@C=\xyL{
  t \ar@{->}[r]^{\b \cup \R_i}_{*}
    \ar@{.>}[d]_{\b}^{*}
  & u \ar@{.>}[d]^{\b}_{*} \\
  t' \ar@{.>}[r]_{\R_i}^{*}
  & u'
}
\end{aligned}
\end{equation}
The intricate case of property~(\ref{diag-AsleA-strat})
is when there is an $\R_i$-step followed by a $\b$-step:
\[
t \quad\a_{\R_i}\quad v \quad\a_\b\quad u
~.
\]
In this case, we have to make sure that the step
$t \a_{\R_i} v$ did not create the $\b$-redex contracted in $v \a_\b u$.
As seen in Section~\ref{sec-rew},
this follows from arity assumptions on terms.

%%%%%%%%%%%%%%%%%%%%%%%%%%%%%%%%%%%%%%%%%%%%%%%%%%%%%%%%%%%%%%%%%%%%%%%%%%%
% techincal properties
%%%%%%%%%%%%%%%%%%%%%%%%%%%%%%%%%%%%%%%%%%%%%%%%%%%%%%%%%%%%%%%%%%%%%%%%%%%

We therefore use terms whose arity is compatible with that of the rewrite system.
We need this property to be preserved by $\a_{\b \cup \R(\b)}$,
but also by the conditions of rewrite rules:
given a semi-closed rule $\vd = \vc \sgt l \at_\R r$ 
and a substitution $\s$, if $l\s$ respects $\ap:\Si \a \N$,
then the terms $r\s,\vd\s$ should also respect $\ap$.
This is the case when $r,\vd$ are algebraic and respect $\ap$.
Moreover, in the following we have to make sure that every term at hand
satisfy these properties.
In particular, if we have a rule $\vd = \vc \sgt l \at r$
such that $l\s$ and all its reducts respect an arity $\ap$,
this has to be the case of $\vd\s$ too
(the case of $\vc$ follows from semi-closure).
Hence, we consider sets of terms which are stable under the rewrite relation
$\a_{\o{\R}}$ issued from the rewrite system
\[
\begin{array}{r !{\quad} c !{\quad} l @{\tq} l}
  \at_{\o\R}
&\deq& \{(l,d_i) &
    d_1 = c_1 \land \ldots \land d_n = c_n \sgt l \at_\R r
    \quad\land\quad i \in \{1,\dots,n\}\}
\\
&\cup& \{(l,r) &
    d_1 = c_1 \land \ldots \land d_n = c_n \sgt l \at_\R r\}
~.
\end{array}
\]
This motivates the following definition,
which extends $(\R,\ap)$-stability (Definition~\ref{def-rstable})
to conditional rewriting.

%%%%%%%%%%%%%%%%%%%%%%%%%%%%%%%%%%%%%%%%%%%%%%%%%%%%%%%%%%%%%%%%%%%%%%%%%%%
\begin{definition}[Conditionally $(\R,\ap)$-Stable Terms]
%-- Extension of Def.~\ref{def-rstable}]
\label{def-rstable-cond}
%%%%%%%%%%%%%%%%%%%%%%%%%%%%%%%%%%%%%%%%%%%%%%%%%%%%%%%%%%%%%%%%%%%%%%%%%%%
Let $\ap : \Si \a \N$ be an arity and $\R$
be a conditional rewrite system.
A set of terms $S$ is {\em conditionally $(\R,\ap)$-stable} if
it is $(\o{\R},\ap)$-stable.
%\begin{enumerate}
%\item for all $t \in S$, $t$ respects $\ap$,
%\item for all $t \in S$, if $t \a_{\b \cup \o\R} u$ then $u \in S$,
%\item for all $t \in S$, if $u$ is a subterm of $t$ then $u \in S$.
%\end{enumerate}
\end{definition}

We now show~(\ref{diag-AsleA-strat}).
The proof of this property occupies Proposition~\ref{prop-post} and
Lemma~\ref{lem-post-star}.
Note that we prove Proposition~\ref{prop-post}
for systems whose conditions need not be algebraic.
However, this property may fail in presence of 
right-hand sides which either are not algebraic or do not respect
the arity prescribed by the left-hand sides.
Note also that we work on conditionally $(\R,\ap)$-stable terms.

%%%%%%%%%%%%%%%%%%%%%%%%%%%%%%%%%%%%%%%%%%%%%%%%%%%%%%%%%%%%%%%%%%%%%%%%%%%
\begin{proposition}
\label{prop-post}
%%%%%%%%%%%%%%%%%%%%%%%%%%%%%%%%%%%%%%%%%%%%%%%%%%%%%%%%%%%%%%%%%%%%%%%%%%%
Let $\R$ be a left-linear semi-closed system which is right-algebraic
and respects $\ap: \Si \a \N$,
and let $S \sle \Te(\Si)$ be conditionally $(\R,\ap)$-stable. 
For all $i \in \N$ and all $t,u,v \in S$,
if $t \a_{\R_i} u \rpr_\b v$,
then there are $t'$ and $v'$ such that
$t \rpr_\b t' \a^*_{\R_i} v' \rpl_\b v$~:
%In diagrammatic form:
\begin{equation*}
\begin{aligned}
\xymatrix@R=\xyS@C=\xyS{
    t \ar@{>}[r]^{\R_i}
          \ar@{..>}[d]_{\rpr_\b}
  & u \ar@{>}[r]^{\rpr_\b}  
  & v \ar@{..>}[d]^{\rpr_\b}  \\
    t' \ar@{..>}[rr]_{\R_i}^{*}
  & 
  & v'
}
\end{aligned}
\end{equation*}
\end{proposition}

\begin{proof}
The base case $i = 0$ is trivial, and we assume $i > 0$.
We reason by induction on $t$ using Lemma~\ref{lem-wadsworth}.
\begin{description}
\item[$t = \la \vx.x \esp t_1 \dots t_n$.]
In this case, $u = \la \vx.x u_1 \dots u_n$
with $(t_1,\dots,t_n) \a_{\R_i} (u_1,\dots,u_n)$.
Moreover, $v = \la \vx.x \esp v_1 \dots v_n$ with
$(u_1,\dots,u_n) \rpr_\b (v_1,\dots,v_n)$.
By induction hypothesis, there are 
$(t'_1,\dots,t'_n)$ and $(v'_1,\dots,v'_n)$ such that
\[
  (t_1,\dots,t_n)
  \quad\rpr_\b\quad
  (t'_1,\dots,t'_n)
  \quad\a^*_{\R_i}\quad
  (v'_1,\dots,v'_n)
  \quad\rpl_\b\quad
  (v_1,\dots,v_n)
  ~.
\]
It follows that
\[
  \la \vx.x \esp t_1 \dots t_n
  \quad\rpr_\b\quad
  \la \vx.x \esp t'_1 \dots t'_n
  \quad\a^*_{\R_i}\quad
  \la \vx.x \esp v'_1 \dots v'_n
  \quad\rpl_\b\quad
  \la \vx.x \esp v_1 \dots v_n
~.
\]

\item[$t = \la \vx.\sff t_{1} \dots t_{n}$.]
If
$u = \la \vx.\sff u_{1} \dots u_{n}$
with
\[
  (t_1,\dots,t_{n}) \quad\a_{\R_i}\quad (u_1,\dots,u_{n})
~,
\]
then
$v = \la \vx.\sff v_{1} \dots v_{n}$ with
$(u_1,\dots,u_{n}) \rpr_\b (v_1,\dots,v_{n})$
and we reason as in the previous case.

Otherwise, there is a rule $\vd = \vc \sgt l \at_\R r$,
a substitution $\s$ and $k \in \{1,\dots,n\}$ such that
$t = \la \vx.l\s t_{k+1} \dots t_{n}$
and $u = \la \vx.r\s t_{k+1} \dots t_{n}$.
As $t$ and $\R$ respect $\ap$, we have $n = k$,
hence $t = \la \vx.l\s$, $u = \la \vx.r\s$ and $v = \la \vx.w$
with $r\s \rpr_\b w$.

Since $r$ is algebraic, by Proposition~\ref{prop-beta-alg}.\ref{prop-beta-alg-base}
there is $\s'$ such that $\s \rpr_\b \s'$ and $w \rpr_\b r\s'$.
As $l$ is linear, by Proposition~\ref{prop-subst-lin-rhd} we have $l\s \rpr_\b l\s'$.
It remains to show that $l\s' \a_{\R_i} r\s'$.
Since $l\s \a_{\R_i} r\s$, there are terms $\vv$
such that
$\vd\s \a^*_{\R_{i-1}} \vv \al^*_{\R_{i-1}} \vc$.
Since $\vd\s \a^*_\b \vd\s'$, by Lemma~\ref{lem-commut-cond}, we obtain terms
$\vv'$ such that
\[
  \vd\s \quad\a^*_\b\quad \vd\s' \quad\a^*_{\R_{i-1}}\quad
  \vv' \quad\al^*_{\b}\quad \vv \quad\al^*_{\R_{i-1}}\quad \vc
~.
\]
Since terms $\vc$ are applicative and closed, they are algebraic,
and since $\R$ is right-algebraic, terms $\vv$ are also algebraic,
hence in $\b$-normal form.
It follows that $\vv = \vv'$, hence that
$\vd\s' \ad_{\R_{i-1}} \vc$, and we deduce that $l\s' \a_{\R_i} r\s'$.

\item[$t = \la \vx.(\la x.t_0)t_1 \dots t_n ~(n \geq 1)$.]
Then $u$ is of the form
$u = \la \vx.(\la x.u_0)u_1 \dots u_n$
and we have $(t_0,\dots,t_n) \a_{\R_i} (u_0,\dots,u_n)$.
If $v = \la \vx.(\la x.v_0)v_1 \dots v_n$
with
\[
(u_0,\dots,u_n) \quad\rpr_\b\quad (v_0,\dots,v_n)
~,
\]
then we conclude by induction hypothesis, as in the first case.

Otherwise, $v = \la \vx.v_0\wth{x}{v_1} v_2 \dots v_n$  
with $(u_0,\dots,u_n) \rpr_\b (v_0,\dots,v_n)$.
By induction hypothesis, we have
\[
  (t_0,\dots,t_n)
  \quad\rpr_\b\quad
  (t'_0,\dots,t'_n)
  \quad\a^*_{\R_i}\quad
  (v'_0,\dots,v'_n)
  \quad\rpl_\b\quad
  (v_0,\dots,v_n)
~.
\]
It follows that by using $(\pBeta)$, $(\papp)$ 
we have
\[
\begin{array}{c c c}
    \la \vx.(\la x.t_0)t_1 \dots t_n
  && 
    \la \vx.v_0\wth{x}{v_1} v_2 \dots v_n
  \\
    \triangledown_\b
  &&\triangledown_\b
  \\
    \la \vx.t'_0\wth{x}{t'_1} t'_2 \dots t'_n
  & \quad\a^*_{\R_i}\quad
  & \la \vx.v'_0\wth{x}{v'_1} v'_2 \dots v'_n
  ~.
\end{array}
\]
\qedhere
\end{description}
\end{proof}

%%%%%%%%%%%%%%%%%%%%%%%%%%%%%%%%%%%%%%%%%%%%%%%%%%%%%%%%%%%%%%%%%%%%%%%%%%%
\begin{lemma}
\label{lem-post-star}
%%%%%%%%%%%%%%%%%%%%%%%%%%%%%%%%%%%%%%%%%%%%%%%%%%%%%%%%%%%%%%%%%%%%%%%%%%%
Let $\R$ be a semi-closed left-linear right-algebraic system
which respects $\ap : \Si \a \N$, 
and let $S \sle \Te(\Si)$ be conditionally $(\R,\ap)$-stable.
For all $s,t \in S$,
if $s \a^*_{\b \cup {\R_i}} t$ then there are $s'$, $t'$ such that
%$s ~\a^*_\b~ s' \a^*_{\R_i} t' ~\al^*_\b~ t$.
%In diagrammatic form,
$s \a^*_\b s' \a^*_{\R_i} t' \al^*_\b t$~:
\begin{equation*}
\begin{aligned}
\xymatrix@R=\xyS@C=\xyL{
  s \ar@{->}[r]^{\b \cup {\R_i}}_{*}
    \ar@{.>}[d]_{\b}^{*}
  & t \ar@{.>}[d]^{\b}_{*} \\
  s' \ar@{.>}[r]_{{\R_i}}^{*}
  & t'
}
\end{aligned}
\end{equation*}
\end{lemma}

\comment{
\begin{proof}
The proof is in four steps.  We begin (1) to show that $\a_{\R_i} \rpr_\b
\sle \rpr_\b \a^*_{{\R_i}} \rpl_\b$, reasoning by cases on the step
$\rpr_\b$. This inclusion relies on an important fact of algebraic
terms: if $s$ is algebraic and $s\s\rpr_\b v$ then $v \rpr_\b s\s'$
with $\s \rpr^*_\b \s'$.  From (1), it follows that (2)
$\a_{\R_i}^*\rpr_\b\sle \rpr_\b\a_{\R_i}^*\rpl_\b^*$, by induction on the
number of $\a_{\R_i}$-steps. Then (3), we obtain $\a_{\R_i}^*\rpr_\b^*\sle
\rpr_\b^*\a_{\R_i}^*\rpl_\b^*$ using an induction on the number of
$\rpr_\b$-steps and the diamond property of $\rpr_\b$.
Finally (4), we deduce that $(\rpr_\b\cup\a_{\R_i})^*\sle
\rpr_\b^*\a_{\R_i}^*\rpl_\b^*$ by induction on the length of
$(\rpr_\b\cup\a_{\R_i})^*$.
\end{proof}}

\begin{proof}
The proof is in three steps.
\begin{enumerate}
\item
\label{lem-post-star-postpone}
We show $\a_{\R_i}^*\rpr_\b\sle \rpr_\b\a_{\R_i}^*\rpl_\b^*$ by
induction on the number of ${\R_i}$-steps. Assume that $s\a_{\R_i}^*
t'\a_{\R_i} t\rpr_\b u$. By Lemma~\ref{prop-post}, there are $v$ and $v'$
such that $t'\rpr_\b v\a_{\R_i}^* v'\rpl_\b u$. By induction hypothesis,
there are $s'$ and $s''$ such that $s\rpr_\b s'\a_{\R_i}^* s''\rpl_\b^* v$.
Then, by Lemma~\ref{lem-commut-cond}, there is $t''$ such that
$s''\a_{\R_i}^* t''\rpl_\b^* v'$. Thus, $s\rpr_\b s'\a_{\R_i}^* t''\rpl_\b^*
u$.

\item
\label{lem-post-star-postpone-star}
We show $\a_{\R_i}^*\rpr_\b^*\sle \rpr_\b^*\a_{\R_i}^*\rpl_\b^*$
by induction on the number of $\rpr_\b$-steps.
Assume that $s\a_{\R_i}^* t\rpr_\b u'\rpr_\b^* u$.
After~\ref{lem-post-star-postpone}, there are $s'$ and $t'$ such that
$s\rpr_\b s'\a_{\R_i}^* t'\rpl_\b^* u'$.
By the diamond property of $\rpr_\b$, there is $v$ such
that $t'\rpr_\b^* v\rpl_\b^* u$, where $t'\rpr_\b^* v$ is no longer
than $u'\rpr_\b^* u$. Hence, by induction hypothesis, there are $s''$
and $t''$ such that $s'\rpr_\b^* s''\a_{\R_i}^* t''\rpl_\b^*
v$. Therefore, $s\rpr_\b^* s''\a_{\R_i}^* t''\rpl_\b^* u$.

\item
We prove
$(\rpr_\b\cup\a_{\R_i})^*\sle \rpr_\b^*\a_{\R_i}^*\rpl_\b^*$
by induction on the length of $(\rpr_\b\cup\a_{\R_i})^*$.
Assume that $s \a_{\rpr_\b\cup{\R_i}} t \a_{\rpr_\b\cup{\R_i}}^* u$.
There are two cases. First, $s\rpr_\b t$.
This case follows directly from the induction hypothesis.
Second, $s\a_{\R_i} t$. By induction hypothesis, there are $t'$ and $u'$ such
that $t\rpr_\b^* t'\a_{\R_i}^* u'\rpl_\b^* u$.
After~\ref{lem-post-star-postpone-star}, there are $s'$
and $t''$ such that $s\rpr_\b^* s'\a_{\R_i}^* t''\rpl_\b^* t'$. Finally,
by Lemma~\ref{lem-commut-cond}, there is $u''$ such that
$t''\a_{\R_i}^* u'' \rpl_\b^* u'$.
Hence, $s\rpr_\b^* s'\a_{\R_i}^* u''\rpl_\b^* u$.
\end{enumerate}
We conclude by the fact that $\rpr_\b^* = \a_\b^*$.
\end{proof}

%%%%%%%%%%%%%%%%%%%%%%%%%%%%%%%%%%%%%%%%%%%%%%%%%%%%%%%%%%%%%%%%%%%%%%%%%%%
\paragraph{Remark.}
%%%%%%%%%%%%%%%%%%%%%%%%%%%%%%%%%%%%%%%%%%%%%%%%%%%%%%%%%%%%%%%%%%%%%%%%%%%
Note that $\b$-reduction is the only way 
to obtain a term not respecting $\ap$ from a term respecting it.
For instance, with $\ap(\id) = 1$ the term $(\la x.x\esp y\esp y)\id$
respects $\ap$ whereas $\id \esp y \esp y$ does not respect $\ap$.

%%%%%%%%%%%%%%%%%%%%%%%%%%%%%%%%%%%%%%%%%%%%%%%%%%%%%%%%%%%%%%%%%%%%%%%%%%%
\begin{proposition}
%%%%%%%%%%%%%%%%%%%%%%%%%%%%%%%%%%%%%%%%%%%%%%%%%%%%%%%%%%%%%%%%%%%%%%%%%%%
Let $\R$ be an algebraic conditional rewrite system
and $t \in \Te(\Si)$ that both respect $\ap : \Si \a \N$.
If $t \a_{\R(\b)} u$ then $u$ respects $\ap$.
\end{proposition}

\begin{proof}
We reason by induction on $t$, using Lemma~\ref{lem-wadsworth}.
The only case which does not directly follow from the induction hypothesis
is when
$t = \la \vx.\sff t_{1} \dots t_{n}$
and there is a rule
$\vd = \vc \sgt l \at_\R r$,
a substitution $\s$ and
$k \in \{1,\dots,n\}$
such that
$t = \la \vx.l\s t_{k+1} \dots t_{n}$.
Since $t$ and $\R$ respect $\ap$, we have $k = n$.
Hence $u = \la \vx.r\s$ and $u$ respects $\ap$ since $r$
is an algebraic term that respects $\ap$.
\end{proof}

%%%%%%%%%%%%%%%%%%%%%%%%%%%%%%%%%%%%%%%%%%%%%%%%%%%%%%%%%%%%%%%%%%%%%%%%%%%
\subsubsection{Confluence of beta-reduction with beta-conditional rewriting}
\label{sec-Bconfl-sc-confl}
%%%%%%%%%%%%%%%%%%%%%%%%%%%%%%%%%%%%%%%%%%%%%%%%%%%%%%%%%%%%%%%%%%%%%%%%%%%

We now have all we need to show property~(\ref{diag-BsleA-strat}). 
As seen in Example~\ref{ex-Bconfl},
rules have to be algebraic and arity compliant.
We reason by induction on $i \in \N$.

%%%%%%%%%%%%%%%%%%%%%%%%%%%%%%%%%%%%%%%%%%%%%%%%%%%%%%%%%%%%%%%%%%%%%%%%%%%
\begin{lemma}
\label{lem-BtoA}
%%%%%%%%%%%%%%%%%%%%%%%%%%%%%%%%%%%%%%%%%%%%%%%%%%%%%%%%%%%%%%%%%%%%%%%%%%
Let $\R$ be a semi-closed left-linear algebraic system
which respects $\ap : \Si \a \N$, and let $S \sle \Te(\Si)$
be conditionally $(\R,\ap)$-stable.
For all $t,u \in S$,
if $t \a^*_{\b \cup {\R(\b)_i}} u$ then there are $t'$, $u'$ such that
%$t ~\a^*_\b~ t' \a^*_{\R_i} u' ~\al^*_\b~ u$.
%In diagrammatic form,
$t \a^*_\b t' \a^*_{\R_i} u' \al^*_\b u$~:
\begin{equation}\label{eq-lem-BtoA}
\begin{aligned}
\xymatrix@R=\xyS@C=\xyL{
    t \ar@{->}[r]^{\b \cup \R(\b)_i}_{*}
      \ar@{.>}[d]_{\b}^{*}
  & u \ar@{.>}[d]^{\b}_{*} \\
    t' \ar@{.>}[r]_{\R_i}^{*}
  & u'
}
\end{aligned}
\end{equation}
\end{lemma}

\begin{proof}
We show~(\ref{eq-lem-BtoA}) by induction on $i \in \N$.
The base case $i = 0$ is trivial.
We assume that the property holds for $i \geq 0$ and show it for
$i+1$.  The proof is in two steps.
\begin{enumerate}
\item
\label{lem-BtoA-base}
We begin by showing that diagram~(\ref{eq-lem-BtoA-base})
commutes:
\begin{equation}\label{eq-lem-BtoA-base}
\begin{aligned}
\xymatrix@R=\xyS@C=\xyL{
    t \ar@{->}[r]^{\R(\b)_{i+1}}
      \ar@{.>}[d]_{\b}^{*}
  & u \ar@{.>}[d]^{\b}_{*} \\
    t' \ar@{.>}[r]_{\R_{i+1}}
  & u'
}
\end{aligned}
\end{equation}
We reason by induction on $t$, using Lemma~\ref{lem-wadsworth}.
The only case that does not directly follow from the induction
hypothesis is when $t = \la \vx.\sff t_1 \dots t_n$
and there is a rule $\vd = \vc \sgt l \at_\R r$, a substitution $\s$
and $k \in \{1,\dots,n\}$ such that
$t = \la \vx.l\s t_{k+1} \dots t_{n}$
and $u = \la \vx.r\s t_{k+1} \dots t_{n}$
with $l\s \a_{\R(\b)_{i+1}} r\s$.
Since $t$ and $\R$ respect $\ap$, we have $k = n$,
hence $u = \la \vx.r\s$.

To deduce~(\ref{eq-lem-BtoA-base}), it remains to show that 
there is a substitution $\s'$ such that
\[
  l\s \quad\a^*_\b\quad
  l\s' \quad\a_{\R_{i+1}}\quad
  r\s' \quad\al^*_\b\quad r\s
  ~.
\]
Since $l\s \a_{\R(\b)_{i+1}} r\s$, there are terms
$\vv$ such that
$\vd\s \a^*_{\b \cup \R(\b)_i} \vv \al^*_{\b \cup \R(\b)_i} \vc$.
By induction hypothesis on $i$, there are terms
$\vw$ and $\vv'$ such that
\[
  \vd\s \quad\a^*_\b\quad
  \vw \quad\a^*_{\R_i}\quad
  \vv' \quad\al^*_\b\quad \vv
  ~.
\]
By Proposition~\ref{prop-beta-alg}.\ref{prop-beta-alg-star},
as terms $\vd$ are algebraic there is a substitution $\s'$
such that $\s \a^*_\b \s'$ and $\vw \a^*_\b \vd\s'$.
By Lemma~\ref{lem-commut-cond} (commutation of $\a_{\R_i}$ with $\a_\b$),
we obtain terms $\vv'$ such that
$\vd\s' \a^*_{\R_i} \vv' \al^*_\b \vv$.
It follows that
\[
  \vd\s \quad\a^*_\b\quad \vd\s' 
        \quad\a^*_{\R_i}\quad \vv'
        \quad\al^*_\b\quad \vv
        \quad\al^*_{\b \cup \R(\b)_i}\quad \vc
  ~.
\]
Since terms $\vc$ are algebraic and $\R$ is right-applicative,
every reduct of $\vc$ by $\a_{\R(\b)}$
is $\b$-normal.
We thus have $\vv' = \vv$ and by induction hypothesis on $i$
we deduce that $\vc \a^*_{\R_i} \vv$.
It follows that $\vd\s' \ad_{\R_i} \vc$,
hence $l\s' \a_{\R_{i+1}} r\s'$.
We have $l\s \a^*_\b l\s'$ and $r\s \a^*_\b r\s'$
since $\s \a^* \s'$, hence
\[
  t \quad\a^*_\b\quad
  \la \vx.l\s' \quad\a_{\R_{i+1}}\quad \la \vx.r\s'
    \quad\al^*_\b\quad u
  ~.
\]
%This conclude the proof of~(\ref{eq-lem-BtoA-base}).

\item
We now show~(\ref{eq-lem-BtoA}) by induction on
the length of $t \a^*_{\b \cup \R(\b)_{i+1}} u$.
Assume that
\[
  t\a_{\b\cup\R(\b)_{i+1}} v\a_{\b\cup\R(\b)_{i+1}}^* u
~.
\]
By induction hypothesis, there are $v'$ and $u'$ such that
$v \a_\b^* v' \a_{\R_{i+1}}^* u' \al_\b^* u$.
and there are two cases.
If $t \a_\b v$, then we are done since $t \a_\b^* v'$.

Otherwise, we have $t\a_{\R(\b)_{i+1}} u$.
From~\ref{lem-BtoA-base}, there are $t'$ and $v''$
such that
\[
  t\a_\b^* t' \a_{\R_{i+1}}^* v'' \al_\b^* v
~.
\]
Now, by confluence of $\a_\b$, there is $v'''$ such that
$v''\a_\b^* v''' \al_\b^* v'$.
Commutation of $\a_\b$ and $\a_{\R_{i+1}}$ (Lemma~\ref{lem-commut-cond})
applied to $v''' \al^*_\b v' \a^*_{\R_{i+1}} u'$
gives us a term $u''$ such that $v''' \a_{\R_{i+1}}^* u''\al_\b^* u'$.
We thus have $t' \a_{\b \cup \R_{i+1}}^* u''$
and by Lemma~\ref{lem-post-star} there are $t''$ and $u'''$ such that
$t'' \a_\b^*\a_{\R_{i+1}}^*\al_\b^* u'''$.
Therefore, $t\a_\b^*\a_{\R_{i+1}}^*\al_\b^* u$.

In diagrammatic form,
\[
\xymatrix@R=\xyS@C=\xyS{
  &&
    t \ar@{->}[rr]^{\R(\b)_{i+1}}
      \ar@{.>}[dl]_{\b}^{*}
  &&
    v \ar@{->}[rr]^{\b \cup \R(\b)_{i+1}}_{*}
      \ar@{.>}[dl]^{\b}_{*} 
      \ar@{.>}[dr]_{\b}^{*} 
  &&
    u \ar@{.>}[dr]^{\b}_{*} 
\\
  & t'  \ar@{.>}[rr]_{\R_{i+1}}
        \ar@{.>}[dl]_{\b}^{*}
  &&
    v'' \ar@{.>}[dr]^{\b}_{*}
  &&
    v'  \ar@{.>}[rr]_{\R_{i+1}}^{*}
        \ar@{.>}[dl]_{\b}^{*} 
  &&
    u'  \ar@{.>}[dl]^{\b}_{*}
\\
  t''   \ar@{.>}[drrrrr]_{\R_{i+1}}^{*}
  &&&&
  v'''  \ar@{.>}[rr]^{\R_{i+1}}_{*}
  &&
  u''   \ar@{.>}[dl]^{\b}_{*}
\\
  &&&&&
  u'''
}
\]
\qedhere
\end{enumerate}
\end{proof}

We easily deduce~(\ref{diag-BsleA}) from~(\ref{eq-lem-BtoA}).
We get the confluence of $\a_{\b\cup\R}$ using Theorem~\ref{thm-Aconfl-sc}.
By Proposition~\ref{prop-Bconfl},
the confluence of $\a_{\b \cup \R(\b)}$ follows from 
the confluence of $\a_{\R}$ on conditionally $(\R,\ap)$-stable sets of terms.

%%%%%%%%%%%%%%%%%%%%%%%%%%%%%%%%%%%%%%%%%%%%%%%%%%%%%%%%%%%%%%%%%%%%%%%%%%%
\begin{theorem}
\label{thm-Bconfl-sc}
%%%%%%%%%%%%%%%%%%%%%%%%%%%%%%%%%%%%%%%%%%%%%%%%%%%%%%%%%%%%%%%%%%%%%%%%%%%
Let $\R$ be a semi-closed left-linear algebraic system
which respects $\ap : \Si \a \N$.
Then, on any conditionally $(\R,\ap)$-stable set of terms, 
if $\a_{\R}$ is confluent then so is $\a_{\b \cup \R(\b)}$.
\end{theorem}

\begin{proof}
Since $\R$ is semi-closed, left-linear and right-applicative,
confluence of $\a_{\b \cup \R}$ follows from confluence of $\a_{\R}$
by Theorem~\ref{thm-Aconfl-sc}.
We then conclude by Lemma~\ref{lem-BtoA} and Proposition~\ref{prop-Bconfl},
since conditionally $(\R,\ap)$-stable sets of terms are closed under $\a_{\b\cup\R(\b)}$.
\end{proof}

%%%%%%%%%%%%%%%%%%%%%%%%%%%%%%%%%%%%%%%%%%%%%%%%%%%%%%%%%%%%%%%%%%%%%%%%%%%
\subsection{Confluence on weakly beta-normalizing terms}
\label{sec-Bconfl-wn}
%%%%%%%%%%%%%%%%%%%%%%%%%%%%%%%%%%%%%%%%%%%%%%%%%%%%%%%%%%%%%%%%%%%%%%%%%%%

In this section, we extend to $\a_{\R(\b)}$ the results of 
Section~\ref{sec-Aconfl-wn}.
The main point is to obtain the lemma corresponding to
Lemma~\ref{lem-projA-bnf}.
Moreover, as in Section~\ref{sec-Bconfl-sc}, for all $i \in \N$
we project $\a_{\R(\b)_i}$ on $\a_{\R_i}$.
We thus want to obtain the following property, which
implies~(\ref{diag-BsleA}):
\begin{equation}
\begin{aligned}
\label{diag-projB-bnf}
\xymatrix@R=\xyS@C=\xyL{
  t \ar@{->}[r]^{\b \cup \R(\b)_i}_{*}
    \ar@{.>}[d]_{\b}^{*}
  & u \ar@{.>}[d]^{\b}_{*} \\
  \bnf(t) \ar@{.>}[r]_{\R_i}^{*}
  & \bnf(u)
}
\end{aligned}
\end{equation}
We use the same tools as in Section~\ref{sec-Aconfl-wn}.
We consider weakly $\b$-normalizing terms whose $\b$-normal form
respects the arity specified by rewrite rules,
and we reason by induction on $\succ$. We also assume that
rewrite rules are algebraic.

We denote by $\rpr_{\R(\b)}$ the nested parallelization
of join $\b$-conditional rewriting, defined similarly as in
Definition~\ref{def-walkA}. It satisfies Proposition~\ref{prop-tgtA}
and Lemma~\ref{lem-commut-h}.

We now show~(\ref{diag-projB-bnf})
using exactly the same method as for showing~(\ref{diag-projA-bnf})
in Lemma~\ref{lem-projA-bnf}.

%%%%%%%%%%%%%%%%%%%%%%%%%%%%%%%%%%%%%%%%%%%%%%%%%%%%%%%%%%%%%%%%%%%%%%%%%%%
\begin{lemma}
\label{lem-projB-bnf}
%%%%%%%%%%%%%%%%%%%%%%%%%%%%%%%%%%%%%%%%%%%%%%%%%%%%%%%%%%%%%%%%%%%%%%%%%%%
Let $\ap : \Si \a \N$ be an arity
and $\R$ be an algebraic conditional rewrite system
which respects $\ap$.
For all $i \in \N$,
if $t \in \AN$ and $t \a_{\b\cup\R(\b)_{i}}^* u$,
then $u\in \AN$ and $\bnf(t) \a_{\R_{i}}^* \bnf(u)$.
\end{lemma}

\begin{proof}
We reason exactly as in the proof of Lemma~\ref{lem-projA-bnf}.
We prove the property by induction on $i \in \N$.
In the induction case we show that for all $t \in \AN$,
\begin{equation}
\begin{aligned}
\label{diag-projBbnf-par}
\xymatrix@R=\xyS@C=\xyL{
    t \ar@{->}[r]^{\rpr_{\R(\b)_{i+1}}}
      \ar@{.>}[d]_{\b}^{*}
  & u \ar@{.>}[d]^{\b}_{*} \\
    \bnf(t) \ar@{.>}[r]_{\rpr_{\R_{i+1}}}
  & \bnf(u)
}
\end{aligned}
\end{equation}
We reason by induction on $\succ$ using Lemma~\ref{lem-wadsworth}.
The only difference with the proof of Lemma~\ref{lem-projA-bnf}
is the case where $t = \la \vx.f t_1 \dots t_n$
and there is a rule
$\vd = \vc \sgt l \at r$ such that
$t = \la \vx.l\s\va$ and $u = \la \vx.r\t\vb$
with
$l\s \rpr_{\R(\b)_{i+1}} r\t$ and $\vd\s \ad_{\b \cup \R(\b)_{i}} \vc\s$.
Exactly for the same reasons as in Lemma~\ref{lem-projA-bnf},
we have
$\va = \vb = \emptyset$, $t = \la \vx . l\s$ and $u = \la \vx.r\s$.
Moreover, $\bnf(t) = \la \vx. l \s'$ and $\bnf(u) = \la \vx.r\t'$
with $\s' \deq \bnf(\s)$ and $\t' = \bnf(\t)$, and 
by induction hypothesis on $\succ$ we have $\s' \rpr_{\R_{i+1}} \t'$.
It remains to show that $l\s' \rpr_{\R_{i+1}} r\t'$.
Because $\s' \rpr_{\R_{i+1}} \t'$, it suffices to prove that
$l\s' \a_{\R_{i+1}} r\s'$.
Thus, we are done if we show that $\vd\s' \ad_{\R_{i}} \vc\s'$.
Since $\vd$ and $\vc$ are algebraic, $\bnf(\vd\s) = \vd\s'$ and
$\bnf(\vc\s) = \vc\s'$.
Now, since $\vd$ is algebraic and respects $\ap$,
and since $\s'$ respects $\ap$, it follows that
$\vd\s'$ respects~$\ap$. The same holds for $\vc\s'$.
Hence we conclude by applying on
$\vd\s \ad_{\b \cup \R(\b)_{i}} \vc\s$ the induction hypothesis on $i$.
\end{proof}

We deduce the preservation of confluence.
%%%%%%%%%%%%%%%%%%%%%%%%%%%%%%%%%%%%%%%%%%%%%%%%%%%%%%%%%%%%%%%%%%%%%%%%%%%
\begin{theorem}
\label{thm-Bconfl-bnf}
%%%%%%%%%%%%%%%%%%%%%%%%%%%%%%%%%%%%%%%%%%%%%%%%%%%%%%%%%%%%%%%%%%%%%%%%%%%
Let $\ap : \Si \a \N$ be an arity
and $\R$ be an algebraic conditional rewrite system which respects $\ap$.
If $\a_\R$ is confluent on $\AN$,
then $\a_{\b\cup\R(\b)}$ is confluent on $\AN$.
\end{theorem}

\begin{proof}
We can reason as described at the beginning of this section,
using Proposition~\ref{prop-Bconfl}, Theorem~\ref{thm-Aconfl-bnf}
and Lemma~\ref{lem-projB-bnf}.
A direct proof is also possible, reasoning as for
Theorem~\ref{thm-Aconfl-bnf}.
\end{proof}

%% file: orth.tex
%%%%%%%%%%%%%%%%%%%%%%%%%%%%%%%%%%%%%%%%%%%%%%%%%%%%%%%%%%%%%%%%%%%%%%%%%%%
\section{Orthonormal systems}
\label{sec-exclus}
%%%%%%%%%%%%%%%%%%%%%%%%%%%%%%%%%%%%%%%%%%%%%%%%%%%%%%%%%%%%%%%%%%%%%%%%%%%

In this section, we give a criterion ensuring the confluence of
$\a_{\b\cup\R(\b)}$ when conditions and right-hand sides possibly contain
abstractions and active variables.

This criterion comes from peculiarities of orthogonality with
conditional rewriting. 
As remarked in Section~\ref{sec-orth-cond}, a conditional critical pair
can be {\em feasible} or not.
In~\cite{ohlebusch02book}, it is remarked that results on the
confluence of semi-equational and normal orthogonal conditional systems could
be extended to systems that have no feasible critical pair. But the
results obtained this way are not directly applicable, since proving
unfeasibility of critical pairs may require confluence.
%We give an example of this situation, using a rewrite system
An example of such situation is the following rewrite system.
%taken from Section~\ref{sec-ex-cond-term}.

%%%%%%%%%%%%%%%%%%%%%%%%%%%%%%%%%%%%%%%%%%%%%%%%%%%%%%%%%%%%%%%%%%%%%%%%%%%
\begin{example}
\label{ex-orthonormal}
%%%%%%%%%%%%%%%%%%%%%%%%%%%%%%%%%%%%%%%%%%%%%%%%%%%%%%%%%%%%%%%%%%%%%%%%%%%
Consider the following two rules, taken from the system
presented in Section~\ref{sec-ex-cond-term}:
\[
\begin{array}{l c l c l c l}
> ({\length}\esp l)\esp x  & = & {\false}  & ~\sgt~  &	 
    {\occ}\esp (\cons \esp x \esp o)\esp ({\node}\esp y\esp l)  & ~\at~  & {\false}  \\
> ({\length}\esp l)\esp x   & = & {\true}   & ~\sgt~  &
    {\occ}\esp (\cons \esp x \esp o)\esp ({\node}\esp y\esp l)   & ~\at~  &
{\occ}\esp o\esp ({\get}\esp l\esp x)
\end{array}
\]
The only conditional critical pair between them is
\[
  > (\length\esp l)\esp x
  \esp=\esp
  \true
\quad\land\quad
  > (\length\esp l)\esp x
  \esp=\esp
  \false
\quad\sgt\quad
  (\false ~,~ \occ \esp o\esp (\get\esp l \esp x))
\]
The condition of this pair cannot be satisfied by a confluent
relation. Hence, if $\a_{\b \cup \R(\b)_i}$ is confluent then
we can reason as in
Lemma~\ref{lem-commut-cond} and obtain the confluence of
$\a_{\b \cup \R(\b)_{i+1}}$.
\end{example}

In this section,
we define a class of systems, called {\em orthonormal},
that allows to generalize this reasoning.
As in Example~\ref{ex-orthonormal},
confluence can be shown stratified way:
the confluence of $\a_{\b \cup \R(\b)_i}$
implies the unfeasibility of critical pairs w.r.t.\ $\a_{\b \cup \R(\b)_i}$,
which in turn entails the confluence of the next stratum $\a_{\b \cup \R(\b)_{i+1}}$.
We thus obtain the level confluence of $\a_{\b \cup \R(\b)}$.
%Their confluence can be shown in a stratified way.
%For rewrite systems such as the one of Example~\ref{ex-orthonormal},
%we can reason in a stratified way.
%We can prove by induction on $i \in \N$ that the relation $\a_{\b \cup \R(\b)_i}$
%is confluent. This implies the $\b$-unfeasibility of critical pairs
%w.r.t.\ the next stratum $\a_{\b \cup \R(\b)_{i+1}}$, which
%in turn entails the confluence of $\a_{\b \cup \R(\b)_{i+1}}$.
%We thus obtain the level confluence of $\a_{\b \cup \R(\b)}$.

Rules of orthonormal systems can have $\la$-terms
in their right-hand sides and conditions.
Moreover, no arity assumption is made.
Hence, orthonormality ensures
the confluence of $\b$-conditional rewriting combined to $\b$-reduction
when we cannot deduce it from the confluence
of conditional rewriting (see Section~\ref{sec-Bconfl}).

Systems similar to orthonormal systems have already been studied
in the first-order case~\cite{gm87ctrs,kw97rta}.
It is worth relating orthonormal systems with approaches to conditional rewriting
in which conditions are arbitrary predicates on terms.
For first-order conditional rewriting
this approach has been taken in~\cite{bk86jcss}.
It has been applied to $\la$-calculus~\cite{takahashi93tlca},
and this is the way conditional rewrite rules are handled 
in the very expressive framework of CCERSs~\cite{gkk05ptc}.
Neither of these approaches can directly handle Example~\ref{ex-orthonormal}.
In each case, confluence is proved under the assumption that
the predicates used in conditions are stable by reduction,
while proving this property in the case of Example~\ref{ex-orthonormal}
requires confluence.

A symbol $\sff \in\Si$ is {\em defined}
if it is the head of the left-hand side of a rule.

%%%%%%%%%%%%%%%%%%%%%%%%%%%%%%%%%%%%%%%%%%%%%%%%%%%%%%%%%%%%%%%%%%%%%%%%%%%
\begin{definition}[Orthonormal Systems]
\label{def-orth}
%%%%%%%%%%%%%%%%%%%%%%%%%%%%%%%%%%%%%%%%%%%%%%%%%%%%%%%%%%%%%%%%%%%%%%%%%%%
A conditional rewrite system $\R$ is {\em orthonormal} if
\begin{enumerate}
\item\label{def-orth-ll}
it is left-linear; 
\item\label{def-orth-norm}
in every rule $\vd=\vc\sgt l\at_\R r$,
the terms in $\vc$ are closed $\b$-normal
forms not containing defined symbols;
\item\label{def-orth-orth}
for every critical pair
\[
  d_1 \esp=\esp c_1
\quad\land\quad\dots\quad\land\quad
  d_n \esp=\esp c_n
\quad\sgt\quad
  (s , t)
\]
there exist distinct $i,j \in \{1,\dots,n\}$ such that $d_i=d_j$ and $c_i\neq c_j$.
%$\vd=\vc\sgt (s,t)$, there exists $i\neq j$ such that
%$d_i=d_j$ and $c_i\neq c_j$.
\end{enumerate}
\end{definition}

\noindent
Condition~\ref{def-orth-norm} %of Definition~\ref{def-orth}
is a simple syntactic and decidable way to ensure
that orthonormal systems are normal
(recall from Remark~\ref{rem-dec} that normality is in general undecidable).
As explained in Example~\ref{ex-orthonormal},
assuming the confluence of $\a_{\b \cup \R(\b)_i}$,
condition~\ref{def-orth-orth} implies the unfeasibility
of critical pairs w.r.t.\ $\a_{\b \cup \R(\b)_i}$,
hence the confluence of the next stratum $\a_{\b \cup \R(\b)_{i+1}}$.
This entails the {\em level} confluence of $\a_{\b \cup \R(\b)}$.
We actually prove in Theorem~\ref{thm:lcr:beta:cond}
the {\em shallow} confluence of $\a_{\b \cup \R(\b)}$,
which is a stronger property (see Definition~\ref{def-strat-confl}).
Theorem~\ref{thm:lcr:beta:cond} is thus an extension of the shallow confluence
of orthogonal first-order normal conditional rewriting
(Theorem~\ref{thm-orth-cond})
to orthonormal $\b$-conditional rewriting.

%However, things are slightly more complex.
%Actually using condition~\ref{def-orth-orth} of Definition~\ref{def-orth} to get
%the unfeasibility of conditions for some stratum leads us to show
%the {\em shallow} confluence of $\a_{\b\cup \R(\b)}$, while the reasoning
%sketched above would only have implied its {\em level} confluence
%(see Definition~\ref{def-strat-confl}).

The most important point w.r.t. the results of Section~\ref{sec-Bconfl}
is that orthonormal systems
do not need to respect an arity nor to be algebraic.

%%%%%%%%%%%%%%%%%%%%%%%%%%%%%%%%%%%%%%%%%%%%%%%%%%%%%%%%%%%%%%%%%%%%%%%%%%%
\begin{example}
%%%%%%%%%%%%%%%%%%%%%%%%%%%%%%%%%%%%%%%%%%%%%%%%%%%%%%%%%%%%%%%%%%%%%%%%%%%
The system presented in Section~\ref{sec-ex-cond-term} is orthonormal.
\end{example}

%%%%%%%%%%%%%%%%%%%%%%%%%%%%%%%%%%%%%%%%%%%%%%%%%%%%%%%%%%%%%%%%%%%%%%%%%%%
% Shallow confluence
%%%%%%%%%%%%%%%%%%%%%%%%%%%%%%%%%%%%%%%%%%%%%%%%%%%%%%%%%%%%%%%%%%%%%%%%%%%

We now show that $\a_{\b\cup\R(\b)}$ is shallow confluent
%(i.e.\ $\a^*_{\b \cup \R(\b)_i}$ and $\a^*_{\b \cup \R(\b)_j}$
%commute for all $i,j\ge 0$)
when $\R$ is orthonormal.
This result is stated and proved in Theorem~\ref{thm:lcr:beta:cond}
below. We use some intermediate lemmas.
The parallel moves property occupies 
Lemmas~\ref{lem:par:moves:beta:cond} and~\ref{lem:scr:parallel:beta:cond}.
We begin by showing that the confluence of $\a_{\b \cup \R(\b)_i}$ implies
the commutation of $\a^*_\b$ and $\a^*_{\R(\b)_{i+1}}$.
%except that in a rule $\vd = \vc \sgt l \a r$,
%$\vc$ are closed $\a_{\b \cup \R(\b)}$-normal forms.

%%%%%%%%%%%%%%%%%%%%%%%%%%%%%%%%%%%%%%%%%%%%%%%%%%%%%%%%%%%%%%%%%%%%%%%%%%%
\begin{lemma}
\label{lem-commut-orth}
%%%%%%%%%%%%%%%%%%%%%%%%%%%%%%%%%%%%%%%%%%%%%%%%%%%%%%%%%%%%%%%%%%%%%%%%%%%
Let $\R$ be an orthonormal system.
For all $i \in \N$, if $\a_{\b\cup\R(\b)_i}$ is confluent
then $\a_{\R(\b)_{i+1}}$ commutes with $\a_\b$~:
%In diagrammatic form,
%$\a^*_{\R(\b)_{i+1}}$ and $\a^*_\b$ commute.
\begin{equation*}
\begin{aligned}
\xymatrix@R=\xyS@C=\xyL{
    \cdot \ar@{>}[r]^{\R(\b)_{i+1}}_{*}
          \ar@{>}[d]_{\b}^{*}
  & \cdot \ar@{..>}[d]^{\b}_{*} \\
    \cdot \ar@{..>}[r]_{\R(\b)_{i+1}}^{*}
  & \cdot
}
\end{aligned}
\end{equation*}
\end{lemma}

\begin{proof}
We reason as in Lemma~\ref{lem-commut-cond}.
We show property~(\ref{eq-lem-commut-orth-base}) below
and then deduce the commutation of $\a_{\R(\b)_{i+1}}$ and $\a_\b$
using Lemma~\ref{lem-commut} and the fact that $\a^*_{\b} \,=\, \rpr^*_\b$.
\begin{equation}
\label{eq-lem-commut-orth-base}
\begin{aligned}
\xymatrix@R=\xyS@C=\xyL{
    t \ar@{->}[r]^{\R(\b)_{i+1}}
      \ar@{->}[d]_{\rpr_\b}
  & v \ar@{.>}[d]^{\rpr_\b} \\
    u \ar@{.>}[r]_{\R(\b)_{i+1}}^{*}
  & w
}
\end{aligned}
\end{equation}
The only difference with the proof of Lemma~\ref{lem-commut-cond} is when
$t \a_{\R(\b)_{i+1}} v$ by contracting a rooted redex.
In this case, there is a rule $\vd = \vc \sgt l \at_\R r$
and a substitution $\s$ such that $t = l\s$ and $v = r\s$.
We show that there is a term $w$ such that
$u\a_{\R(\b)_{i+1}}^* w\rpl_\b r\s$. As $l$ is a non-variable linear
algebraic term, there is a substitution $\s'$ such that
$\s\rpr_\b \s'$ and $l\s\rpr_\b l\s'=u$.
Therefore we have $r\s\rpr_\b r\s'$.
It remains to show that $l\s'\a_{\R(\b)_{i+1}} r\s'$.
Recall that $\vd\s\a_{\b\cup\R(\b)_i}^* \vc$.
By assumption (confluence of $\a_{\b\cup \R(\b)_i}$), 
since $\vd\s \a^*_\b \vd\s'$
there are $\vv$ such that
$\vd\s'\a_{\b\cup\R(\b)_i}^* \vv\al_{\b\cup \R(\b)_i}^* \vc$.
But $\vc$ are $\b \cup \R(\b)$-normal forms, hence $\vv=\vc$.
We conclude that $l\s'\a_{\R(\b)_{i+1}} r\s'\rpl_\b r\s$.
\end{proof}

We follow the usual scheme of proofs of confluence of orthogonal conditional
rewrite systems~\cite{ohlebusch02book}.
For all $i \in \N$, we denote by $\a_{\parallel\R(\b)_i}$
the smallest parallel rewrite relation containing $\a_{\R(\b)_i}$
(see Definition~\ref{def-par-rew}).
Hence, $\a_{\parallel\R(\b)_i}$ is strictly included in the 
nested parallel relation $\rpr_{\R(\b)_i}$ used in Section~\ref{sec-Bconfl-wn}
(Definition~\ref{def-walkA}).
The main property is the commutation of 
$\a_{\parallel\R(\b)_i}$ and $\a_{\parallel\R(\b)_j}$ for all $i,j \in \N$,
which corresponds to the usual parallel moves property.
Let $<_{\mul}$ be the multiset extension of the usual
ordering on naturals numbers.
In our case, the parallel moves property is:
\begin{description}
\item[Parallel Moves.]
Given $i,j \in \N$, if 
$\a_{\b \cup \R(\b)_n}$ commutes with
$\a_{\b \cup \R(\b)_m}$ for all $n,m$ such that 
$\{n,m\} <_{\mul} \{i,j\}$,
then $\a_{\parallel\R(\b)_i}$ commutes with $\a_{\parallel\R(\b)_j}$.
\end{description}

The proof is decomposed into Lemma~\ref{lem:par:moves:beta:cond}
and Lemma~\ref{lem:scr:parallel:beta:cond}.
In Lemma~\ref{lem:par:moves:beta:cond}, assuming the commutation of
$\a_{\b\cup \R(\b)_n}$ and $\a_{\b\cup \R(\b)_m}$ for all $n,m$ such that
$\{n,m\} <_\mul \{i,j\}$, we consider, for the commutation of
$\a_{\parallel\R(\b)_i}$ and $\a_{\parallel\R(\b)_j}$,
the particular case of a rooted $\R(\b)_{i}$-reduction.

%%%%%%%%%%%%%%%%%%%%%%%%%%%%%%%%%%%%%%%%%%%%%%%%%%%%%%%%%%%%%%%%%%%%%%%%%%% 
\begin{lemma}
\label{lem:par:moves:beta:cond}
%%%%%%%%%%%%%%%%%%%%%%%%%%%%%%%%%%%%%%%%%%%%%%%%%%%%%%%%%%%%%%%%%%%%%%%%%%% 
Let $\R$ be an orthonormal system and $i,j\geq 0$.
Assume that $\a_{\b \cup \R(\b)_n}$ commutes with $\a_{\b \cup \R(\b)_m}$
for all $n$, $m$ such that $\{n,m\} <_{\mul} \{i,j\}$.
Then for all rules $\vd = \vc \sgt l \at_\R r$, we have
\[
\xymatrix@R=\xyS@C=\xyL{
  l\s \ar@{->}[r]^{\R(\b)_i} \ar@{->}[d]_{\parallel\R(\b)_j}
  & r\s \ar@{..>}[d]^{\parallel\R(\b)_j} \\
  u \ar@{..>}[r]_{\R(\b)_i}^{\re}
  & v
} 
\]
\end{lemma}

\begin{proof}
The result holds if $i=0$ since $\a_{\R(\b)_0}=\emptyset$.
If $j = 0$, then $u = l\s$ and take $v=r\s$.

Assume that $i,j >0$ and
write $q_1,\dots,q_n$ for the (disjoint) occurrences in
$l\s$ of the redexes 
contracted in $l\s \a_{\parallel\R(\b)_j} u$.
Therefore, for all $k$, $1 \leq k \leq n$, there exists a rule
$\rho_k: \vd_k = \vc_k \sgt l_k \at_\R r_k$
and a substitution $\t_k$ such that $l\s|_{q_k} =l_k\t_k$.
Thus,  $ u= l\s[r_1\t_1]_{q_1}\dots[r_n\t_n]_{q_n}$.
It is possible to rename variables and assume that $\rho$, $\rho_1, \dots
\rho_n$ have disjoint variables. 
Therefore, we can take $\s = \t_1 = \cdots = \t_n$.

Assume that there is a non-variable superposition, i.e.\ that a $q_k$
is a {\em non variable} occurrence in~$l$.
Hence rules $\rho$ and $\rho_k$ form an instance of a critical pair
$ \vd'\mu = \vc'\sgt ({l[r_k]_{q_k}}\mu,r\mu) $ and
there exists a substitution $\mu'$ such that $\s = \mu
\mu'$.
By definition of orthonormal systems, $|\vd'\mu| \geq 2$ and there is
$m\neq p$ such that $c'_m \neq c'_p$ and $d'_m\mu = d'_p\mu$.
Let us write $h$ for $max(i,j) - 1$.
As $d'_m\mu = d'_p\mu$ we have $d'_m\s = d'_p\s$ and it follows
that
\[
  c'_m \quad\al_{\b \cup \R(\b)_h}^*\quad
  d'_m\s \quad=\quad d'_p\s \quad\a_{\b \cup \R(\b)_h}^*\quad c'_p
~.
\]
But $\{h,h \} <_{\mul} \{i,j \}$ and by assumption $\a_{\b \cup \R(\b)_h}$ is
confluent. Therefore we must have $c'_m \ad_{\b \cup \R(\b)_h} c'_p $.
But it is not possible since $c'_m$ and $c'_p$ are distinct normal forms.
Hence, conditions of $\rho$ and $\rho_k$ cannot be both satisfied
by $\s$ and $\a_{\b \cup \R(\b)_{h}}$
and it follows that there is no non-variable superposition.

Therefore, each $q_k$ is of the form
$u_k.v_k$ where 
$l|_{u_k}$ is a variable $x_k$.
Let $\s'$ be such that $\s'(x_k) = \s(x_k)[r_k\s]_{v_k}$ 
and $\s'(y) = \s(y)$ if $y \neq x_k$ for all $1 \leq k \leq n$.
Then, $l\s \a_{\parallel\R(\b)_j} l\s'$ and by linearity of $l$, $u=l\s'$.
Furthermore, $r\s \a_{\parallel\R(\b)_j} r\s'$.
We now show that $l\s' \a_{\R(\b)_i} r\s'$.
We have $\vd\s  \a_{\b \cup \R(\b)_{i-1}}^* \vc$
and $\vd\s \a_{\R(\b)_j}^* \vd\s'$. 
As $i,j > 0$, we have $\{i-1,j\} <_{\mul} \{i,j\}$.
Therefore, by assumption $\a_{\b \cup \R(\b)_{i-1}}$ and 
$\a_{\b \cup \R(\b)_{j}}$
commute and there exist 
terms $\vc'$ such that
\[
  \vd\s' \quad\a_{\b \cup \R(\b)_{i-1}}^*\quad
  \vc'   \quad\al_{\b \cup \R(\b)_{j-1}}^*\quad  \vc
~.
\]
As terms $\vc$ are $\a_{\b\cup \R(\b)}$-normal forms,
we have $\vc' = \vc$ and it follows that $l\s' \a_{\R(\b)_i} r\s'$.
\end{proof}

Now, in Lemma~\ref{lem:scr:parallel:beta:cond}
we show that the commutation of 
$\a_{\parallel\R(\b)_i}$ and $\a_{\parallel\R(\b)_j}$
is ensured by the two particular cases of rooted $\R(\b)_i$-reduction
and $\R(\b)_j$-reduction.

%%%%%%%%%%%%%%%%%%%%%%%%%%%%%%%%%%%%%%%%%%%%%%%%%%%%%%%%%%%%%%%%%%%%%%%%%%%
\begin{lemma}
\label{lem:scr:parallel:beta:cond}
%%%%%%%%%%%%%%%%%%%%%%%%%%%%%%%%%%%%%%%%%%%%%%%%%%%%%%%%%%%%%%%%%%%%%%%%%%%
Let $\R$ be an orthonormal system and $i,j\geq 0$.
Property (i) below holds if and only if for all rules $\vd=\vc\sgt l\at_\R r$,
properties (ii) and (iii) hold.
\[
\begin{array}{c !{\quad} c !{\quad} c}
  % (i)
  \xymatrix@R=\xyS@C=\xyL{
    s \ar@{->}[r]^{\parallel\R(\b)_i}
      \ar@{->}[d]_{\parallel\R(\b)_j}
    & t \ar@{..>}[d]^{\parallel\R(\b)_j} \\
    u \ar@{..>}[r]_{\parallel\R(\b)_i}
    & v
  }
  &  % (ii)
  \xymatrix@R=\xyS@C=\xyL{
    l\s \ar@{->}[r]^{\R(\b)_i} \ar@{->}[d]_{\parallel\R(\b)_j}
    & r\s \ar@{..>}[d]^{\parallel\R(\b)_j} \\
    u \ar@{..>}[r]_{\R(\b)_i}^{\re}
    & v
  }
  &  % (iii)
  \xymatrix@R=\xyS@C=\xyL{
    l\s \ar@{->}[r]^{\R(\b)_j} \ar@{->}[d]_{\parallel\R(\b)_i}
    & r\s \ar@{..>}[d]^{\parallel\R(\b)_i} \\
    u \ar@{..>}[r]_{\R(\b)_j}^{\re}
    & v
  }
  \\
  \text{(i)} & \text{(ii)} & \text{(iii)}
\end{array}
\]
\end{lemma}

\begin{proof}
The ``only if'' statement is trivial.
For the ``if'' case, let $s,t,u$ be three terms such that
$ u  \al_{\parallel\R(\b)_j}  s  \a_{\parallel\R(\b)_i} t $.
If $s$ is $t$ (resp. $u$), then take $v = u$ (resp. $v = t$).
Otherwise, we reason by induction on the structure of $s$.
If there is a rooted reduction, we conclude by properties (ii) and (iii).
Now assume that both reductions are nested.
In this case $s$ cannot be a symbol $\sff \in \Si$ nor a variable.
If $s$ is an abstraction, we conclude by induction hypothesis.
Otherwise $s$ is an application $s_1 s_2$, and by assumption
$u = u_1 u_2$ and $t = t_1 t_2$ with
$u_k \al_{\parallel\R(\b)_j} s_k \a_{\parallel\R(\b)_i} t_k$.
In this case also we conclude by induction hypothesis. 
\end{proof}

Now, an induction on $<_{\mul}$ provides the commutation of
$\a_{\b\cup\R(\b)_i}$ and $\a_{\b\cup\R(\b)_j}$ for all $i,j\geq 0$,
i.e.\ the shallow confluence of $\a_{\b \cup \R(\b)}$.

%%%%%%%%%%%%%%%%%%%%%%%%%%%%%%%%%%%%%%%%%%%%%%%%%%%%%%%%%%%%%%%%%%%%%%%%%%%
\begin{theorem}
\label{thm:lcr:beta:cond}
%%%%%%%%%%%%%%%%%%%%%%%%%%%%%%%%%%%%%%%%%%%%%%%%%%%%%%%%%%%%%%%%%%%%%%%%%%%
If $\R$ is an orthonormal system, then $\a_{\b\cup\R(\b)}$ is shallow
confluent.
\end{theorem}

\begin{proof}
We reason by induction on unordered pairs $\{i,j\}$ seen as multisets
and compared with the well-founded relation $<_{\mul}$.
We show the commutation of $\a_{\b\cup\R(\b)_i}$
and $\a_{\b\cup\R(\b)_j}$ for all $i,j\geq 0$.
The least unordered pair $\{i,j\}$ (considered as a multiset)
with respect to $<_{\mul}$ is
$\{0,0\}$. As $\a_{\b\cup\R(\b)_0}=\a_\b$ by definition, this case holds by
confluence of $\b$.

Now, assume that $i>0$ and that the commutation of $\a_{\b\cup\R(\b)_n}$ and
$\a_{\b\cup\R(\b)_m}$ holds for all $n,m$ with $\{n,m\} <_{\mul} \{i,0\}$.  As
$\{i-1,i-1\} <_{\mul} \{i,0\}$, $\a_{\b\cup\R(\b)_{i-1}}$ is confluent and the
commutation of $\a_{\b\cup\R(\b)_i}$ with
$\a_{\b\cup\R(\b)_0}$ ($=\a_\b$) follows from Lemma~\ref{lem-commut-orth}.

The remaining case is when $i,j>0$. Using the induction hypothesis,
from Lemma~\ref{lem:par:moves:beta:cond} and
Lemma~\ref{lem:scr:parallel:beta:cond}, we obtain the commutation of
$\a_{\parallel\R(\b)_i}$ and $\a_{\parallel\R(\b)_j}$,
which in turn implies the commutation
of $\a^*_{\R(\b)_i}$ and $\a^*_{\R(\b)_j}$.
Now, as $\{i-1,i-1\} <_{\mul} \{i,j\}$, by
Lemma~\ref{lem-commut-orth}, $\a_\b$ and $\a_{\R(\b)_i}$ commute.
This way, we also
obtain the commutation of $\a_\b$ and $\a_{\R(\b)_j}$.
Then, the commutation of
$\a^*_{\b \cup \R(\b)_i}$ and $\a^*_{\b\cup\R(\b)_j}$ easily follows.
\end{proof}

%%%%%%%%%%%%%%%%%%%%%%%%%%%%%%%%%%%%%%%%%%%%%%%%%%%%%%%%%%%%%%%%%%%%%%%%%%%
\begin{example}
%%%%%%%%%%%%%%%%%%%%%%%%%%%%%%%%%%%%%%%%%%%%%%%%%%%%%%%%%%%%%%%%%%%%%%%%%%%
The relation $\a_{\b\cup\R(\b)}$ induced by the system
presented in Section~\ref{sec-ex-cond-term}
is shallow confluent and thus confluent.
\end{example}

%% file: conclu.tex
%%%%%%%%%%%%%%%%%%%%%%%%%%%%%%%%%%%%%%%%%%%%%%%%%%%%%%%%%%%%%%%%%%%%%%%%%%%
\section{Conclusion}
%%%%%%%%%%%%%%%%%%%%%%%%%%%%%%%%%%%%%%%%%%%%%%%%%%%%%%%%%%%%%%%%%%%%%%%%%%%

Our results are summarized in 
%Figure~\ref{fig-summary} page~\pageref{fig-summary}.
%See
Figure~\ref{fig-oversimple} page~\pageref{fig-oversimple}.
%for an informal overview.

%%%%%%%%%%%%%%%%%%%%%%%%%%%%%%%%%%%%%%%%%%%%%%%%%%%%%%%%%%%%%%%%%%%%%%%%%%%
%\begin{figure}
%%%%%%%%%%%%%%%%%%%%%%%%%%%%%%%%%%%%%%%%%%%%%%%%%%%%%%%%%%%%%%%%%%%%%%%%%%%
%\begin{center} \FigOverview \end{center}
%\caption{Summary of the results.~\label{fig-summary}}
%\end{figure}

We provide detailed conditions to ensure modularity of confluence when
combining $\b$-reduction and conditional rewriting, either when the
evaluation of conditions uses $\b$-reduction or when it does not.
This has useful applications on the high-level specification side and
for enriching the conversion used in logical frameworks or proof
assistants, while still preserving the confluence property.

These results lead us to the following remarks and further research points.
The results obtained in Section~\ref{sec-Aconfl} and~\ref{sec-Bconfl}
for the join conditional rewrite systems extend
to the case of oriented systems (hence to normal systems) and to
the case of level-confluent semi-equational systems.
For semi-equational systems, the proofs follow the same scheme,
provided that level-confluence of $\a_\R$ is assumed.
However, it would be
interesting to know if this restriction can be dropped.

Problems arising from non left-linear rewriting are directly transposed to left-linear
conditional rewriting. The semi-closure condition is sufficient to avoid this,
and it seems to provide the counterpart of left-linearity
for unconditional rewriting.
However, two remarks have to be made about this restriction.
First, it would be interesting to know if
it is a necessary condition and besides,
to characterize a class of non semi-closed
systems that can be translated into equivalent semi-closed ones.
Second, semi-closed terminating join systems behave like normal systems.
But normal systems can be easily
translated into equivalent non-conditional systems.
Moreover such a translation preserves
good properties such as left-linearity and non ambiguity.
As many practical uses of rewriting rely on terminating systems,
semi-closed join systems may be in practice
essentially an intuitive way to design rewrite systems
that can be then efficiently
implemented by non-conditional rewriting.

A wider interesting perspective 
would be to extend the results
to CCERSs~\cite{gkk05ptc}.